\documentstyle[12pt]{article}

\newcommand{\mysection}{\setcounter{equation}{0}\section}

\begin{document}
\vskip 0.2cm
\hfill{NIKHEF/96-027}
\vskip 0.2cm
\hfill{ITP-SB-96-66}
\vskip 0.2cm
\hfill{DESY 96-258}
\vskip 0.2cm
\hfill{INLO-PUB-22/96}
\vskip 0.2cm
\centerline{\large\bf {Charm electroproduction viewed in the
variable-flavour number }}
\centerline{\large\bf {scheme versus fixed-order perturbation theory}}
\vskip 0.2cm
\centerline {\sc M. Buza \footnote{supported by the Foundation
for Fundamental Research on Matter (FOM)}}
\centerline{\it NIKHEF/UVA,}
\centerline{\it POB 41882, NL-1009 DB Amsterdam,}
\centerline{\it The Netherlands.}
\vskip 0.2cm
\centerline {\sc Y. Matiounine }
\centerline{\it Institute for Theoretical Physics,}
\centerline{\it State University of New York at Stony Brook,}
\centerline{\it New York 11794-3840, USA.}
\vskip 0.2cm
\centerline {\sc  J. Smith \footnote{on leave from ITP, SUNY at Stony Brook,
New York 11794-3840, USA} }
\centerline{\it Deutsches Electronen-Synchrotron DESY}
\centerline{\it Notkestrasse 85, D-22603, Hamburg}
\centerline{\it Germany}
\vskip 0.2cm
\centerline {\sc W.L. van Neerven}
\centerline{\it Instituut-Lorentz,}
\centerline{\it University of Leiden,}
\centerline{\it PO Box 9506, 2300 RA Leiden,}
\centerline{\it The Netherlands.}
\vskip 0.2cm
\centerline{December 1996}
\vskip 0.2cm
\centerline{\bf Abstract}
\vskip 0.3cm
Starting from fixed-order perturbation theory (FOPT) we derive expressions
for the heavy-flavour components of the deep-inelastic structure functions
$(F_{i,H}(x,Q^2,m_H^2), i = 2,L;  H = c,b,t)$
in the variable-flavour number scheme (VFNS). These expressions are
valid in all orders of perturbation theory. 
This derivation establishes a relation between the parton densities
parametrized at $n_f$ and $n_f+1$ light flavours.
The consequences for the existing parametrizations of
the parton densities are discussed. Further we show that in charm
electroproduction the exact and asymptotic expressions
for the heavy-quark coefficient functions
yield identical results for $F_{2,c}(x,Q^2,m_c^2)$
when $Q^2 \ge 20$ (GeV$/c)^2$.
We also study the differences between the FOPT and the VFNS descriptions
for $F_{2,c}(x,Q^2,m_c^2)$. It turns out that the charm structure function
in the VFNS is larger than the one obtained in FOPT over the whole
$Q^2$-range. Furthermore inspection of the perturbation series reveals
that the higher order corrections in the VFNS
are smaller than those present in FOPT for $Q^2 \ge 10$ (GeV$/c)^2$. 
Therefore the VFNS gives a better prediction for the charm structure function
at large $Q^2$-values than FOPT.

\vfill
\newpage

\mysection{Introduction}
\newcommand{\be}{\begin{eqnarray}}
\newcommand{\ee}{\end{eqnarray}}


The study of charm production in deep-inelastic electron-proton scattering
and in photon-proton scattering provides us with important information
about heavy-quark-production mechanisms. In particular,
we can distinguish between intrinsic- and extrinsic-charm 
production. In the latter case the charm quark only appears in the final state
so the dominant subprocess is given by (virtual) photon-gluon fusion, 
which is the only reaction present on the Born level \cite{Wit}.
Hence one can measure the $x$-dependence of the gluon
density $G(x,\mu^2)$, where $x$ denotes the Bjorken scaling variable 
and $\mu$ stands for the factorization and renormalization scales. 
Next-to-leading (NLO) corrections 
\cite{ensn}, \cite{lrsn}, to which also other processes  
contribute, reveal that this picture remains unaltered. 

Besides extrinsic-charm production one can also consider
intrinsic-charm production \cite{bhmt}, which is given by the 
flavour-excitation process. In this case the charm quark also appears in 
the initial state and it is considered to be a part of the hadronic wave
function. Hence this quark is described by a parton density in the hadron, 
as in the cases of the other light flavours (u,d,s) and the gluon (g). 
Notice that in the context of
perturbative QCD the flavour-excitation mechanism is always accompanied by the
photon-gluon fusion process. However the latter can occur without the presence
of intrinsic-charm production because one can always assume that the
probability to find a charm quark inside the proton is equal to zero without
violating the principles of perturbative QCD.
The difference between
both production mechanisms becomes very clear when one looks at
differential distributions. In the case of extrinsic-charm production the
photon-gluon-fusion subprocess predicts that both the charm 
quark and the charm anti-quark appear back-to-back in the final state. 
For intrinsic-charm production, where the flavour-excitation mechanism 
dominates, it turns out
that either the charm quark or the charm anti-quark appears in the 
final state. 
Hence the transverse momentum of this heavy quark is not balanced by 
that of its antiparticle, in 
contrast to the situation in the photon-gluon fusion process.
Recent experiments carried out at HERA at $Q^2=0$ \cite{Aid} 
(photoproduction) or at
$10 < Q^2 < 100$ (GeV/$c)^2$ \cite{Adl} (electroproduction), where $Q^2$ 
denotes the virtual photon mass squared,
give strong evidence for
extrinsic-charm production. Therefore we only concentrate on the latter 
production mechanism in this paper.

As has been mentioned above the extrinsic-charm mechanism 
can be studied in photoproduction ($Q^2 = 0$) and in 
electroproduction ($Q^2 > 0$). The photoproduction reaction has the 
advantage that the production rate is much larger than in the case of 
electroproduction because the latter rate is suppressed by  
the photon propagator, which decreases when $Q^2$ gets larger. In the
context of perturbative QCD, however, the description of electroproduction 
is easier due to 
the absence of the hadronic (resolved) photon component which contributes
in photoproduction (for a discussion see \cite{rsn}). 
Moreover electroproduction
enables us to study the charm contribution to the deep-inelastic structure
functions $F_2(x,Q^2)$ and $F_L(x,Q^2)$. 

For the treatment of the charm component
of the structure functions $F_{k,c}(x,Q^2,m_c^2)$ ($k =2,L$)
one has adopted two different prescriptions for extrinsic-charm 
production
in the literature. The first one is advocated in \cite{ghr} where the charm 
quark is treated as a heavy quark and its contribution is given by fixed-order 
perturbation theory (FOPT). This involves
the computation of the photon-gluon-fusion process mentioned above and its
higher order corrections. 
The second approach is the so-called variable-flavour
number scheme (VFNS) \cite{acot}. 
Here charm is treated in a similar way to a massless quark 
and its contribution is described by a parton 
density in the hadron, denoted by $f_c^{\rm VFNS}(x,\mu^2)$.
At first sight this looks similar to intrinsic-charm production. However
there is one important difference. This becomes clear when one looks at
eq.(10) of \cite {acot} where there exists a relation between $f_c^{\rm VFNS}$
and the gluon density originally appearing in the 
photon-gluon process. In this way there is an intimate relation between
FOPT and the VFNS which imposes some conditions on the charm-quark density.
These conditions do not exist in the intrinsic-charm approach,
where this density is just an arbitrary function fitted to experimental data.
It is clear that the VFNS-procedure is not very well suited to describe charm
production in the threshold region because the photon-gluon fusion process
requires that the charm component of the structure function vanishes above
$x = Q^2/(Q^2+4~m_c^2)$, while this threshold condition gets lost in the
$x$-behaviour of $f_c^{\rm VFNS}(x,\mu^2)$.  
On the other hand FOPT also has its drawbacks when
$Q^2$ gets very large. The reason is that the heavy-quark 
coefficient functions calculated up to NLO in \cite{lrsn} 
are dominated by large logarithms of the type
$\ln^i(Q^2/m_c^2)\ln^j(\mu^2/m_c^2)$ when $Q^2 \gg m_c^2$
\cite{bmsmn}. Although it was 
shown in NLO \cite{grs} that these logarithms lead to  
rather stable charm structure functions $F_{i,c}(x,Q^2,m_c^2)$
with respect to variations in the factorization and  
renormalization scales $\mu$, their size still warrants some special 
treatment (for a discussion of the scale dependence of the charm contribution
in the FOPT and VFNS approaches see also \cite{Vogt},\cite{or},\cite{kls}). 
However as we will show in this paper the above criterion is not sufficient
to decide about the rate of convergence of the perturbation series.
It still may happen that these large logarithms vitiate the perturbation
series, in particular when corrections beyond NLO are included. Hence these
logarithms have to be resummed using the 
standard techniques of the renormalization group. 

This resummation proceeds as follows. First one has to add 
the light-parton component of the structure functions $F_k(x,Q^2)$, $k=2,L$ 
to the charm contributions $F_{k,c}(x,Q^2,m_c^2)$ both presented with
three light flavours. The light component consists of the light-parton 
(u,d,s,g) densities convoluted with the light-parton coefficient functions. 
The same
densities are used for the charm component where they are
convoluted with the heavy-quark coefficient functions. Then one
takes the limit $Q^2 \rightarrow \infty$ and performs mass factorization 
in order to remove the above mass singular 
logarithms in $m_c$ from the asymptotic heavy-quark coefficient functions.
The latter can be written as
a convolution of the heavy-quark operator-matrix elements (OME's), containing
these mass-singular logarithms, and the light-parton coefficient functions
which are finite in the limit $m_c \rightarrow 0$. 
The light-parton densities are then modified by multiplying
the original u,d,s and g densities by the heavy-quark OME's,
where one has to introduce a new
parton density to represent the charm quark. Notice that the latter is not
an arbitrary function since, via the heavy-quark OME's, it depends 
on the original densities represented by the
three light flavours (u,d,s) and the gluon (g) in the three-flavour scheme.
The whole procedure leads to the structure functions $F_i$ and $F_{i,c}$
which are now represented in the four-flavour scheme.  
The latter are expressed into four light-flavour densities 
(including charm) and the gluon density convoluted with the light-parton 
coefficient functions. Further the heavy-quark coefficient functions
due to the charm have completely disappeared.
Although these formulae are similar to those
obtained for intrinsic-charm production their origin is completely different.
Here we want to emphasize again that the VFNS is derived from FOPT, 
where, in the large-$Q^2$ limit, one has resummed the large 
logarithmic terms in 
the heavy-quark coefficient functions in all orders of perturbation theory. 
This procedure, which is only described 
up to leading order (LO) in \cite{acot}, will 
be generalized to all orders in QCD perturbation theory in Section 2.

In Section 3 we will study at which $Q^2$ the large logarithms in the 
heavy-quark coefficient functions actually dominate the charm component of 
the structure functions. 
We will investigate the $x$- and $\mu^2$-dependence 
of the charm density in the VFNS approach
and compare it with the existing charm densities in the literature. 
We will also study the differences between the FOPT approach 
and the VFNS approach for the charm component of the deep-inelastic structure
functions in the $x$- and $Q^2$-range explored by present experiments. Further
we investigate which approach is more stable with respect to higher order
corrections to the charm structure functions.
In Appendix A we present some heavy-quark OME's which were not previously
calculated in the literature. 
In Appendix B we list all renormalized heavy-quark
OME's which are needed for our analysis in Section 3.

\mysection{Derivation of the VFNS representation of the structure functions}

In this section we present the variable-flavour number scheme (VFNS)
representation of the structure functions 
$F_k(x,Q^2)$ in all orders of perturbation theory.
Our results hold for any species of heavy quark although at present
collider energies the VFNS is only interesting for the charm quark.
As mentioned in the previous section logarithms of the type
$\ln^i(Q^2/m^2)\ln^j(\mu^2/m^2)$ arise in the heavy-quark 
coefficient functions when $Q^2 \gg m^2$ and we work 
in fixed-order perturbation theory (FOPT). 
Here $m$ stands for the mass of the heavy quark, denoted by H in 
the subsequent part of this section. When going from 
the FOPT representation for the structure functions to that of the VFNS  
one has to remove
these mass-singular logarithms from the heavy-quark coefficient functions.
This is achieved using the technique of mass factorization, which 
is a generalization to all orders in perturbation theory of the procedure 
carried out up to lowest order (LO) in Sections II D and II E in \cite{acot}.  
Although this procedure resembles the usual mass factorization of the
collinear singularities, which appear in the partonic structure functions or
partonic cross sections, it is actually much more complicated. 
This is due to the presence of the light (u,d,s and g) partons, 
as well as of the heavy (c,b and t) quarks, in the Feynman diagrams
describing heavy-flavour production.

In the calculations of the heavy-flavour cross sections the light partons
are usually taken to be massless, whereas the heavy quarks get a mass 
$m \not= 0$. Note that the mass is defined by on-mass-shell renormalization.
When $Q^2 \gg m^2$ two types of collinear 
singularities appear in the partonic contributions. 
One type can be attributed to the light partons. In this
case the collinear divergences can be regularized using various techniques. 
The most well-known among them is $n$-dimensional regularization.
The second type can be traced back to the heavy quark and the singularity
manifests itself as $m \rightarrow 0$ in the large 
logarithmic terms mentioned above.
Beyond order $\alpha_s$ both types of singularities appear in the partonic
cross sections and in the heavy-quark operator matrix elements (OME's).
The latter show up in the mass-factorization formulae, and they are needed to 
remove the mass-singular terms (as $m\rightarrow 0$) from the 
heavy-quark coefficient functions.
Hence the mass factorization becomes much more complicated than 
when we only have to deal with collinear divergences due to massless
partons. Therefore we first present the mass factorization with
respect to the latter before we apply this technique to the mass singularities
related to the heavy-quark mass $m$.

Consider first deep-inelastic electron-proton scattering in which 
only light-partons show up in the calculation of the 
QCD corrections. The deep-inelastic
structure function $F_i(n_f,Q^2)$ can be expressed as follows
\begin{eqnarray}
F _i(n_f,Q^2)\!\! & = & \!\! \frac{1}{n_f} \sum\limits_{k=1}^{n_f} e_k^2 
\Biggl[
\hat\Sigma(n_f) \otimes 
\hat{\cal F}_{i,q}^{\rm S}\Big(n_f,\frac{Q^2}{p^2},\mu^2\Big)
+ \hat G(n_f) \otimes
\hat{\cal F}_{i,g}^{\rm S}\Big(n_f,\frac{Q^2}{p^2},\mu^2\Big)
\nonumber \\ &&
+n_f \hat \Delta_k(n_f) \otimes
\hat{\cal F}_{i,q}^{\rm NS}\Big(n_f,\frac{Q^2}{p^2},\mu^2\Big)\Biggr] \,.
\end{eqnarray}
In this equation the charge of the light quark is represented by $e_k$
and $n_f$ denotes the number of light flavours. Further $\otimes$ denotes the 
convolution symbol and, for convenience, the dependence of all 
the above quantities on the hadronic
scaling variable $x$ and the partonic scaling variable $z$
is suppressed. The bare quantities
in (2.1) are indicated by a hat in order to distinguish them from
their finite analogues, which emerge after mass factorization. Starting with
the parton densities, $\hat{\Sigma}(n_f)$ 
and $\hat{G}(n_f)$ denote the singlet
combination of light-quark densities and the gluon density respectively,
with $n_f$ light flavours. The former is given by
\begin{equation}
\hat \Sigma(n_f) = \sum\limits_{l=1}^{n_f} \Big[
\hat f_l(n_f) + \hat f_{\bar l}(n_f) \Big] \,, 
\end{equation}
where $\hat{f}_l(n_f)$ and $\hat{f}_{\bar{l}}(n_f)$ 
stand for the light-quark and light-anti-quark densities respectively. 
The non-singlet combination of light-quark densities is given by
\begin{equation}
\hat \Delta_k(n_f) = 
\hat f_k(n_f) + \hat f_{\bar k}(n_f)- \frac{1}{n_f}
\hat\Sigma(n_f) \,. 
\end{equation}

The QCD  radiative corrections due to the (virtual) photon light-parton  
subprocesses are described by the partonic structure functions 
$\hat{\cal F}_{i,l}(n_f)$
$(i=2,L ; l=q,g)$ where $l$ stands for the parton 
which appears in the initial
state. As with the parton densities they can also be classified into singlet,
non-singlet and gluonic parts. Furthermore 
we assume that coupling-constant renormalization has been performed 
on the partonic (bare) structure functions, which is
indicated by their dependence on the renormalization scale $\mu$. However they 
still contain collinear divergences which, for convenience, are regularized
by taking the external momentum $p$ of the incoming parton 
off-mass-shell $(p^2 < 0)$. Notice that
these divergences do not show up in the final state because
the deep-inelastic structure functions are totally-inclusive quantities. 

The reason we choose off-shell regularization is that it allows us to
distinguish between the collinear divergences coming from the massless partons
and from the heavy quarks. The former divergences  
are the $\ln^i(-\mu^2/p^2)$ terms in the perturbative expansion,
whereas the latter are the $\ln^i(\mu^2/m^2)$ terms. 

We can now also reexpress Eq. (2.1) in finite quantities so that the collinear
divergences are absent. This is achieved via mass factorization which
proceeds as follows
\begin{equation}
\hat{\cal F}_{i,q}^{\rm NS}\Big(n_f,\frac{Q^2}{p^2},\mu^2\Big)
= A_{qq}^{\rm NS}\Big(n_f, \frac{\mu^2}{p^2}\Big)
 \otimes
 {\cal C}_{i,q}^{\rm NS}\Big(n_f, \frac{Q^2}{\mu^2}\Big)\,,
\end{equation}
and
\begin{equation}
\hat{\cal F}_{i,k}^{\rm S}\Big(n_f,\frac{Q^2}{p^2},\mu^2\Big)
= \sum\limits_{l=q,g} A_{lk}^{\rm S}\Big(n_f, \frac{\mu^2}{p^2}\Big)
 \otimes
 {\cal C}_{i,l}^{\rm S}\Big(n_f, \frac{Q^2}{\mu^2}\Big)\,.
\end{equation}
In the above expressions ${\cal C}_{i,k} (i=2,L ; k=q,g)$ denote the 
light-parton coefficient  
functions and the $A_{lk}$ represent the renormalized operator-matrix 
elements (OME's) which are defined by
\begin{equation}
 A_{lk}\Big(n_f, \frac{\mu^2}{p^2}\Big)
= < k(p)| O_l(0)| k(p)> \,, \quad  \quad (l,k=q,g) \,.
\end{equation}
Here $O_l$ are the renormalized operators which appear in the
operator-product expansion of two electromagnetic currents near the light 
cone. The product of these two currents is sandwiched between 
proton states and its Fourier transform into 
momentum space defines the structure functions in (2.1). Like the other
quantities ${\cal C}_{i,k}$ and $A_{lk}$ can 
be divided into singlet and non-singlet parts. The scale $\mu$ 
appearing in Eqs. (2.4)-(2.6) originates from operator
renormalization as well as from coupling-constant renormalization. Notice
that the operator-renormalization scale is identical to the mass-factorization
scale. Using the mass-factorization relations in Eqs. (2.4),(2.5) we can cast
the hadronic structure functions $F_i(n_f,Q^2)$ (2.1) into the form
\begin{eqnarray}
 F_i(n_f,Q^2)\!\! &=& \!\! \frac{1}{n_f} \sum\limits_{k=1}^{n_f} e_k^2 
\Biggl[
\Sigma(n_f,\mu^2) \otimes 
{\cal C}_{i,q}^{\rm S}\Big(n_f,\frac{Q^2}{\mu^2}\Big)
+  G(n_f,\mu^2) \otimes
{\cal C}_{i,g}^{\rm S}\Big(n_f,\frac{Q^2}{\mu^2}\Big)
\nonumber \\ &&
+n_f \Delta_k(n_f, \mu^2) \otimes
{\cal C}_{i,q}^{\rm NS}\Big(n_f,\frac{Q^2}{\mu^2}\Big)\Biggr] \,,
\end{eqnarray}
where the finite (renormalized) parton densities $\Sigma$, $\Delta$ and
$G$ are expressed in the bare ones $\hat \Sigma$, $\hat \Delta$
and $\hat G$ in the following way
\begin{equation}
 \hskip-3.5cm \Delta_k(n_f,\mu^2) = 
A_{qq}^{\rm NS}\Big(n_f, \frac{\mu^2}{p^2}\Big) \otimes \hat \Delta_k(n_f)
 \,, 
\end{equation}
\begin{equation}
 \Sigma(n_f,\mu^2) = 
A_{qq}^{\rm S}\Big(n_f, \frac{\mu^2}{p^2}\Big) \otimes \hat \Sigma(n_f)
+
A_{qg}^{\rm S}\Big(n_f, \frac{\mu^2}{p^2}\Big) \otimes \hat G(n_f)
 \,, 
\end{equation}
and 
\begin{equation}
 G(n_f,\mu^2) = 
A_{gq}^{\rm S}\Big(n_f, \frac{\mu^2}{p^2}\Big) \otimes \hat \Sigma(n_f)
+
A_{gg}^{\rm S}\Big(n_f, \frac{\mu^2}{p^2}\Big) \otimes \hat G(n_f)
 \,. 
\end{equation}

All quantities given above satisfy renormalization group equations (RGE's).
Here we are only interested in those RGE's for the OME's and the parton 
densities.
Define the differential operator $D$ as
\begin{equation}
D = \mu \frac{\partial}{\partial\mu} + \beta(n_f, g)
\frac{\partial}{\partial g} \,,\quad  \quad g \equiv g(n_f,\mu^2)\,,
\end{equation}
where $\alpha_s = g^2/(4\pi)$, then the OME's satisfy the following RGE's
\begin{equation}
D A_{qq}^{\rm NS}\Big(n_f, \frac{\mu^2}{p^2}\Big) = 
- \gamma_{qq}^{\rm NS}(n_f) \otimes 
 A_{qq}^{\rm NS}\Big(n_f, \frac{\mu^2}{p^2}\Big) \,, 
\end{equation}
\begin{equation}
D A_{ij}^{\rm S}\Big(n_f, \frac{\mu^2}{p^2}\Big) = 
- \sum\limits_{k=q,g} \gamma_{ik}^{\rm S}(n_f) \otimes 
 A_{kj}^{\rm S}\Big(n_f, \frac{\mu^2}{p^2}\Big) \,. 
\end{equation}
Here $\gamma_{ij}$ denote the anomalous dimensions corresponding to the 
operators
$O_i$. They can be expanded as a perturbation series in $\alpha_s$. In
Bjorken $x$-space there exists a relation between the anomalous dimensions
$\gamma_{ij}$ and the DGLAP splitting functions, denoted by $P_{ij}$, which
is given by
\begin{equation}
 \gamma_{ij}(n_f)  = - P_{ij}(n_f)\,. 
\end{equation}
Notice that this relation only holds for twist-two operators. From the above
equation one infers that the $\gamma_{ij}$, which are the residues of the
ultraviolet divergences in the unrenormalized OME's, have just the opposite
signs to those of the $P_{ij}$. The latter show up in the partonic quantities 
$\hat{{\cal F}}_{i,l}(n_f)$ in (2.1) and they represent the residues of the  
collinear divergences. 
We will return to this relation (2.14) 
when we discuss the heavy-quark OME's.
The finite (renormalized) parton densities satisfy the RGE's
\begin{equation}
\hskip-3.5cm D \Delta_k(n_f, \mu^2) = 
- \gamma_{qq}^{\rm NS}(n_f) \otimes 
 \Delta_k(n_f, \mu^2) \,, 
\end{equation}
\begin{equation}
D \Sigma(n_f, \mu^2) = 
- \gamma_{qq}^{\rm S}(n_f) \otimes \Sigma(n_f, \mu^2) 
- \gamma_{qg}^{\rm S}(n_f) \otimes G(n_f, \mu^2) 
\,, 
\end{equation}
and
\begin{equation}
D G(n_f, \mu^2) = 
- \gamma_{gq}^{\rm S}(n_f) \otimes \Sigma(n_f, \mu^2) 
- \gamma_{gg}^{\rm S}(n_f) \otimes G(n_f, \mu^2) 
\,. 
\end{equation} From these equations we can derive the 
Altarelli-Parisi equations.

Before we add the heavy-quark contributions to the deep-inelastic structure 
functions (2.7) it is convenient to split the singlet quantities 
$\hat{\cal F}_{i,q}^{\rm S}$, ${\cal C}_{i,q}^{\rm S}$ and $A_{qq}^{\rm S}$ 
into non-singlet and purely-singlet parts, namely
\begin{equation}
\hat{\cal F}_{i,q}^{\rm S}  = \hat{\cal F}_{i,q}^{\rm NS}   
+ \hat{\cal F}_{i,q}^{\rm PS}   
\,, 
\end{equation}
\begin{equation}
{\cal C}_{i,q}^{\rm S}  = {\cal C}_{i,q}^{\rm NS}   
+ {\cal C}_{i,q}^{\rm PS}   
\,, 
\end{equation}
and
\begin{equation}
A_{qq}^{\rm S}  = A_{qq}^{\rm NS}   
+ A_{qq}^{\rm PS}   
\,. 
\end{equation}
This decomposition facilitates the mass factorization of the 
heavy-quark coefficient functions and can be explained as follows. 
If we calculate the diagrams
contributing to the parton subprocesses with a quark in the initial state, 
the resulting expressions have to be projected on the singlet and non-singlet
channels with respect to the flavour group. The latter projection leads to
$\hat{\cal F}_{i,q}^{\rm NS}$. 
However the singlet part i.e. $\hat{{\cal F}}_{i,q}^{\rm S}$
can be split into two types of contributions. The first one is equal to
$\hat{{\cal F}}_{i,q}^{\rm NS}$ whereas the second one is represented by
$\hat{{\cal F}}_{i,q}^{\rm PS}$. The purely-singlet partonic 
structure function arises from the Feynman graphs 
where the projection on the non-singlet channel yields zero,
so that only singlet contributions remain. They are characterized by 
those Feynman graphs in which only gluons are exchanged in the $t$-channel. 
Such graphs show up for 
the first time in two-loop order. An example is given in Fig. 1. The same 
characteristics also hold for $A_{qq}^{\rm PS}$ (see Fig. 2) and the resulting
coefficient function ${\cal C}_{i,q}^{\rm PS}$. 
Another important feature is that the
purely-singlet quantities are proportional to the number of 
light flavours $n_f$.
This property is shared by the gluonic quantities 
$\hat{{\cal F}}_{i,g}^{\rm S}$,
${\cal C}_{i,g}^{\rm S}$ and $A_{qg}^{\rm S}$. The proportionality to $n_f$ 
can be traced back to the fact that in the case of 
$\hat{{\cal F}}_{i,q}^{\rm PS}$, $\hat{{\cal F}}_{i,g}^{\rm S}$ the virtual 
photon is attached to the light-quark loop (see e.g. Fig. 1) and one has to 
sum over all light flavours. In the case of $A_{qq}^{\rm PS}$ and 
$A_{qg}^{\rm S}$ this is due to the insertion of the operator vertex into the
light-quark loop where also a sum over all light flavours 
has to be carried out.
Because of the mass-factorization relation (2.5) the proportionality to $n_f$ 
is transferred to the coefficient functions ${\cal C}_{i,q}^{\rm PS}$,
${\cal C}_{i,g}^{\rm S}$ and the anomalous dimensions $\gamma_{qq}^{\rm PS}$,
$\gamma_{qg}^{\rm S}$. To facilitate the mass factorization of the heavy-quark
coefficient functions it is very convenient to extract this overall factor of 
$n_f$ from the quantities above so we define
\begin{equation}
T_{i,k}  = n_f \tilde T_{i,k} \,, 
\end{equation}
where
\begin{equation}
T_{i,q}  =  \hat {\cal F}_{i,q}^{\rm PS},
 {\cal C}_{i,q}^{\rm PS}\,; \qquad 
T_{i,g}  =  \hat {\cal F}_{i,g}^{\rm S}, 
{\cal C}_{i,g}^{\rm S} \,; 
\end{equation}
and
\begin{equation}
R_{ij}  = n_f \tilde R_{ij} \,, 
\end{equation}
where
\begin{equation}
R_{qq}  =  A_{qq}^{\rm PS} , \gamma_{qq}^{\rm PS} \,; 
\qquad R_{qg}  =  A_{qg}^{\rm S} \,,  \gamma_{qg}^{\rm S} \,. 
\end{equation}
Besides the overall dependence on $n_f$, which we have extracted from the
quantities defined by $T_{i,k}$ and $R_{ij}$ above, there still remains a 
residual dependence on $n_f$ in $\tilde T_{i,k}$ and $\tilde R_{ij}$. The
latter dependence originates from internal light-flavour loops which are
neither attached to the virtual photon nor to the operator-vertex insertions.
Since there are more contributions to $\tilde T_{i,k}$ and $\tilde R_{ij}$
which are not due to these light flavour-loops it is impossible to extract
an overall factor $n_f$ anymore from the quantities indicated by a tilde.
Therefore the dependence of the latter on $n_f$ means that it can be only 
attributed to internal light-flavour loops. 
Using the above definitions one can now rewrite Eqs. (2.1) and (2.7) and the
results become
\begin{eqnarray}
 F_i(n_f,Q^2) \!\! &=& \!\!  \sum\limits_{k=1}^{n_f} e_k^2 
\Biggl[
\hat\Sigma(n_f) \otimes 
\tilde{\hat{\cal F}}_{i,q}^{\rm PS}\Big(n_f,\frac{Q^2}{p^2},\mu^2\Big)
+ \hat G(n_f) \otimes
\tilde{\hat{\cal F}}_{i,g}^{\rm S}\Big(n_f,\frac{Q^2}{p^2},\mu^2\Big) 
\nonumber\\
&&+ \Big\{\hat f_k(n_f) + \hat f_{\bar k}(n_f) \Big\} \otimes
\hat{\cal F}_{i,q}^{\rm NS}\Big(n_f,\frac{Q^2}{p^2},\mu^2\Big)\Biggr] \,,
\end{eqnarray}
\begin{eqnarray}
 F_i(n_f,Q^2) \!\! &=& \!\!  \sum\limits_{k=1}^{n_f} e_k^2 
\Biggl[
\Sigma(n_f,\mu^2) \otimes 
\tilde{\cal C}_{i,q}^{\rm PS}\Big(n_f,\frac{Q^2}{\mu^2}\Big)
+ G(n_f,\mu^2) \otimes
\tilde{\cal C}_{i,g}^{\rm S}\Big(n_f,\frac{Q^2}{\mu^2}\Big)
\nonumber\\
&&+ \Big\{f_k(n_f,\mu^2) + f_{\bar k}(n_f,\mu^2) \Big\} \otimes
{\cal C}_{i,q}^{\rm NS}\Big(n_f,\frac{Q^2}{\mu^2}\Big)\Biggr] \,,
\end{eqnarray}
with the relation 
\begin{eqnarray}
 f_k(n_f,\mu^2) +  f_{\bar k}(n_f,\mu^2) &=&
A_{qq}^{\rm NS}\Big(n_f,\frac{\mu^2}{p^2}\Big) \otimes 
 \Big\{\hat f_k(n_f) + \hat f_{\bar k}(n_f) \Big\}
\nonumber \\ &&
+ \tilde A_{qq}^{\rm PS}\Big(n_f,\frac{\mu^2}{p^2}\Big) \otimes
\hat \Sigma(n_f) 
+ \tilde A_{qg}^{\rm S}\Big(n_f,\frac{\mu^2}{p^2}\Big) \otimes \hat G(n_f)
 \,.
\nonumber \\
\end{eqnarray}
Using Eqs. (2.5),(2.9) and (2.10) one can now write the mass factorization
relations for the quantities indicated by $\tilde T_{i,k}$ and $\tilde R_{ij}$
in (2.21)-(2.24). The same can be done for the RGE's which can be derived from
(2.13),(2.16) and (2.17). Since this derivation is easy it is left to
the reader. Here we only want to report the RGE for the left-hand-side of
(2.27) which follows from (2.15) and (2.16). It is given by
\begin{eqnarray}
 D[ f_k(n_f,\mu^2) + f_{\bar{k}}(n_f,\mu^2)] &=& -\gamma_{qq}^{NS}(n_f) 
\otimes  
\Big[ f_k(n_f,\mu^2) + f_{\bar{k}}(n_f,\mu^2)\Big] 
\nonumber \\ &&
- \tilde{\gamma}_{qq}^{\rm PS}(n_f) \otimes
\Sigma(n_f,\mu^2) -\tilde{\gamma}_{qg}^{\rm S}(n_f) \otimes G(n_f,\mu^2)\,.
\nonumber \\
\end{eqnarray}
After having presented the formulae needed for the mass factorization of  
the light-parton structure functions $\hat{\cal F}_{i,l}$ with the 
corresponding RGE's we want to deal in a similar way with the asymptotic
heavy-quark coefficient functions where the large logarithmic terms 
depending on the heavy-quark mass $m$ have to be removed. For that purpose we
have to add the heavy-quark contribution to the deep-inelastic structure
function $F_i(n_f,Q^2)$ (2.26) which is equal to
\begin{eqnarray}
\lefteqn{F_{i,H}(n_f,Q^2,m^2) =}
\nonumber\\
&& \sum\limits_{k=1}^{n_f} e_k^2 
\Biggl[
\Sigma(n_f,\mu^2) \otimes 
\tilde L_{i,q}^{\rm PS}\Big(n_f,\frac{Q^2}{m^2},\frac{m^2}{\mu^2}\Big)
+ G(n_f,\mu^2) \otimes
\tilde L_{i,g}^{\rm S}\Big(n_f,\frac{Q^2}{m^2},\frac{m^2}{\mu^2}\Big)
\nonumber\\
&&+\Big\{ f_k(n_f,\mu^2) + f_{\bar k}(n_f,\mu^2)\Big\}
\otimes
L_{i,q}^{\rm NS}\Big(n_f,\frac{Q^2}{m^2}, \frac{m^2}{\mu^2}\Big) \Biggr]
\nonumber\\
&&+ e_H^2 \Biggl[ \Sigma(n_f,\mu^2) \otimes 
H_{i,q}^{\rm PS}\Big(n_f,\frac{Q^2}{m^2},\frac{m^2}{\mu^2}\Big)
+  G(n_f,\mu^2) \otimes 
H_{i,g}^{\rm S}\Big(n_f,\frac{Q^2}{m^2},\frac{m^2}{\mu^2}\Big)
\Biggr] \,,
\nonumber\\
\end{eqnarray}
where $e_H$ stands for the charge of the heavy quark denoted by H. Further
$L_{i,k}$ and $H_{i,k} (i=2,L ; k=q,g )$ represent the heavy-quark coefficient
functions. Like in the case of the light-parton coefficient functions
${\cal C}_{i,k}$ they can be split into singlet and non-singlet parts.
The former can be decomposed into non-singlet and purely-singlet pieces
in a similar way as given in (2.19). 
The distinction between $L_{i,k}$
and $H_{i,k}$ can be traced back to the different (virtual) photon-parton
heavy-quark production mechanisms from which they originate. 
The functions $L_{i,k}$
are attributed to the reactions where the virtual photon couples to the light
quarks (u, d, and s), whereas the $H_{i,k}$ describe the 
interactions between the virtual photon and
the heavy quark. Hence $L_{i,k}$ 
and $H_{i,k}$ in (2.29) are multiplied by $e_k^2$ 
and $e_H^2$ respectively. Moreover, when the reaction where the photon couples
to the heavy quark contains a light quark in the initial 
state, then it can only 
proceed via the exchange of a gluon in the $t$-channel (see Fig. 1). Therefore
$H_{i,q}$ is purely-singlet and a non-singlet contribution does not exist.
This is in contrast with $L_{i,q}$ 
which has both purely-singlet and non-singlet contributions. 

For $Q^2 \gg m^2$ the heavy-quark coefficient functions take their 
asymptotic forms, which are dominated by the large
logarithms mentioned at the beginning of this section. If these terms
become too large they vitiate the perturbation series so that a
resummation via the RGE is necessary. Before this resummation can be
carried out we have first to perform mass factorization to remove the 
mass-singular terms $\ln^i(\mu^2/m^2)$ from 
the asymptotic heavy-quark coefficient functions. This
can be done in a similar way as shown for the partonic structure 
functions in Eqs. (2.4),(2.5), where now $\ln^i(-\mu^2/p^2)$ are replaced by
$\ln^i(\mu^2/m^2)$. In the VFNS approach we require that 
the following relation holds
\begin{equation}
F_i(n_f,Q^2) + \lim_{Q^2 \gg m^2} \Big[F_{i,H}(n_f,Q^2,m^2)\Big] 
= F_i^{\rm VFNS}(n_f+1,Q^2)\,.  
\end{equation}
The above formula implies that, at large $Q^2$, 
the left-hand-side can be written
as a structure function containing densities and coefficient functions
corresponding to only light partons, in which the number of light flavours 
is enhanced by one. In $F_i^{\rm VFNS}(n_f+1,Q^2)$ the light-parton densities
are modified with respect to those appearing in $F_i(n_f,Q^2)$ 
in (2.26) and the
heavy-flavour component $F_{i,H}(Q^2,m^2)$ in (2.29)
because the former have absorbed
the logarithms $\ln^i(\mu^2/m^2)$ coming from the heavy-quark coefficient
functions. Moreover a new parton density appears in $F_i^{\rm VFNS}$  
corresponding to the heavy quark H which is now treated as a light quark.

If relation (2.30) holds one has to impose the following mass-factorization
relations for the heavy-quark coefficient functions.
In the case of the coefficient functions $L_{i,k}$ we get for $Q^2\gg m^2$
\begin{eqnarray}
\hskip-4.5cm {\cal C}_{i,q}^{\rm NS}(n_f) + L_{i,q}^{\rm NS}(n_f) = 
A_{qq,H}^{\rm NS}(n_f) \otimes {\cal C}_{i,q}^{\rm NS}(n_f+1)
\,, 
\end{eqnarray}
\begin{eqnarray}
\tilde {\cal C}_{i,q}^{\rm PS}(n_f) + \tilde L_{i,q}^{\rm PS}(n_f)& = &
\Big[ 
A_{qq,H}^{\rm NS}(n_f) + 
n_f \tilde A_{qq,H}^{\rm PS}(n_f) + 
\tilde A_{Hq}^{\rm PS}(n_f)\Big]  
\otimes \tilde {\cal C}_{i,q}^{\rm PS}(n_f+1)
\nonumber \\ 
&+& \tilde A_{qq,H}^{\rm PS}(n_f) \otimes
 {\cal C}_{i,q}^{\rm NS}(n_f+1)
+ A_{gq,H}^{\rm S}(n_f) \otimes
 \tilde {\cal C}_{i,g}^{\rm S}(n_f+1)
\,, 
\nonumber \\
\end{eqnarray}
and
\begin{eqnarray}
 \tilde {\cal C}_{i,g}^{\rm S}(n_f) + \tilde L_{i,g}^{\rm S}(n_f) & = & 
\tilde A_{qg,H}^{\rm S}(n_f) \otimes
  {\cal C}_{i,q}^{\rm NS}(n_f+1)
+  A_{gg,H}^{\rm S}(n_f) \otimes
 \tilde {\cal C}_{i,g}^{\rm S}(n_f+1)
\nonumber \\ &&
+\Big[n_f \tilde A_{qg,H}^{\rm S}(n_f)+\tilde A_{Hg}^{\rm S}(n_f)\Big] \otimes 
 \tilde {\cal C}_{i,q}^{\rm PS}(n_f+1)
\,. 
\end{eqnarray}
For the heavy quark coefficient functions $H_{i,k}$ we obtain for 
$Q^2 \gg m^2$
\begin{eqnarray}
\hskip-2cm H_{i,q}^{\rm PS}(n_f)  & = &  \tilde A_{Hq}^{\rm PS}(n_f)
\otimes \Big[ 
{\cal C}_{i,q}^{\rm NS}(n_f+1) + 
\tilde {\cal C}_{i,q}^{\rm PS}(n_f+1) \Big] 
\nonumber \\  &&
+ \Big[ A_{qq,H}^{\rm NS}(n_f) 
+ n_f \tilde A_{qq,H}^{\rm PS}(n_f)\Big] \otimes
 \tilde {\cal C}_{i,q}^{\rm PS}(n_f+1)
\nonumber \\  &&
+  A_{gq,H}^{\rm S}(n_f)  
\otimes \tilde {\cal C}_{i,g}^{\rm S}(n_f+1)
\,, 
\end{eqnarray}
and
\begin{eqnarray}
H_{i,g}^{\rm S}(n_f)  & = &
  A_{gg,H}^{\rm S}(n_f)  
\otimes \tilde {\cal C}_{i,g}^{\rm S}(n_f+1)
+ n_f \tilde A_{qg,H}^{\rm S}(n_f) \otimes
 \tilde {\cal C}_{i,q}^{\rm PS}(n_f+1)
\nonumber \\ &&
+ \tilde A_{Hg}^{\rm S}(n_f)
\otimes \Big[
{\cal C}_{i,q}^{\rm NS}(n_f+1) +
\tilde {\cal C}_{i,q}^{\rm PS}(n_f+1) \Big]
\,. 
\end{eqnarray}
In the above equations $A_{Hk}$ denotes the heavy-quark OME defined by
\begin{equation}
 A_{Hk}\Big(n_f, \frac{\mu^2}{m^2}\Big)
= < k(p)| O_H(0)| k(p)> \,, 
\end{equation}
which is the analogue of the light-quark OME's in (2.6). The quantities
$A_{lk,H}(n_f,\mu^2/m^2)$, which also appear above, represent 
the heavy-quark-loop contributions
to the light-quark and gluon OME's defined in (2.6).

Although there exists some similarity between the mass factorization of
$\hat{\cal F}_{i,l}$ in (2.4),(2.5) and the ones given for the heavy-quark
coefficient functions above we also observe a striking difference which makes
the proof of Eqs. (2.31)-(2.35) and consequently of Eq. (2.30) much harder.
The main difference is that the light partons appear in the initial as well as 
in the final state of the subprocesses contributing to $\hat{\cal F}_{i,l}$
whereas the heavy quark only shows up in the final state of the reactions
leading to $L_{i,l}$,$H_{i,l}$ (2.29). This implies that there is no analogue
for $\hat{\cal F}_{i,H}$ in Eq. (2.29) meaning that the coefficient functions
$L_{i,H}$ and $H_{i,H}$ do not appear in the latter equation. The same holds
for the bare heavy-quark density $\hat{f}_H$ which has no counter part in
(2.26) and (2.27) either. Hence the mass factorization 
relations in (2.31)-(2.35) are much more cumbersome 
than those presented in (2.4),(2.5). 
Therefore the proofs of the former relations and implicitly of Eq. (2.30)
become more involved. We have explicitly checked that the above relations
hold up to order $\alpha_s^2$ using the asymptotic heavy-quark 
coefficient functions
in \cite{bmsmn} and the heavy quark OME's in Appendix B.

Yet another complication arises when we consider the OME's
$A_{Hk}$ and $A_{lk,H}$ defined above. Contrary to the light-parton OME's in
(2.6) which only depend, apart from $\mu^2$, on one mass scale $p^2$ the 
former also depend on the mass scale $m^2$ due to the presence of the 
heavy quark. Therefore
the unrenormalized expressions of $A_{Hk}$ and $A_{lk,H}$ contain besides
ultraviolet (UV) divergences two types of collinear divergences which are
represented by $\ln^i(\mu^2/p^2)$ (light partons) and $\ln^i(\mu^2/m^2)$
(heavy quark) respectively. The singularity at $p^2=0$ shows up because 
external massless lines represented by $k=q,g$ in (2.36) are 
coupled to internal massless
quanta. This phenomenon shows up for the first time in order $\alpha_s^2$
see \cite{bmsmn}. Notice that the singularities at $p^2=0$ also appear in the
partonic quantities leading to the heavy-quark coefficient functions before
they are removed by mass factorization as outlined in the beginning of this
section. Therefore we also have to subtract the collinear singularities in
$p^2=0$ from the unrenormalized OME's $A_{Hk}$,$A_{lk,H}$ in addition to the
UV divergences. This twofold subtraction leads to two different scales
in the renormalized OME's in (2.31)-(2.36) which are called the 
operator-renormalization scale and mass-factorization 
scale respectively. Usually
these two scales are set to be equal and they are denoted by one parameter
$\mu^2$. However the appearance of these two scales 
results in more complicated
RGE's for the heavy-quark OME's in comparison with those presented for
the light-parton OME's in (2.12),(2.13) as we will see below. In the actual
calculations of $A_{Hk}$,$A_{lk,H}$ in \cite{bmsmn} we have put $p^2=0$ 
in (2.36) because we did the same for the partonic quantities computed in
\cite{lrsn} leading to the heavy-quark coefficient functions $L_{i,l}$,
$H_{i,l}$. Hence we had to adopt $n$-dimensional regularization for the UV
as well as the collinear singularities in $p^2=0$. In this way both are 
represented by pole terms of the type $1/(n-4)^j$. If the latter are removed
in the $\overline {\rm MS}$-scheme 
the OME's $A_{Hk}$ and $A_{lk,H}$ automatically
depend on one scale $\mu$.

Another comment we want to make is that $\alpha_s$ appearing in the 
heavy-quark coefficient functions is renormalized at $n_f$ flavours. 
This means that the
renormalization of the coupling constant is carried out in such a way that
all quarks equal to H and heavier than H decouple in the quark loops 
contributing to $\alpha_s$. In the right-hand-side of Eqs. (2.31)-(2.35) the
decoupling of the heavy quark H has been undone so that here $\alpha_s$ 
depends on $n_f+1$ flavours. This decoupling gives rise to additional
logarithms of the type $\ln^i(\mu^2/m^2)$, which are cancelled by the OME's
of the type $A_{kl,H}$. However explicit indication of
this procedure further 
complicates our mass factorization relations above, which we want 
to avoid, so we have to bear in 
mind that this decoupling is implicitly understood.

If we now substitute Eqs. (2.31)-(2.35) into the left-hand-side of (2.29)
and rearrange some terms then we obtain the 
expression for $F_k^{\rm VFNS}$ as presented in (2.26), 
where $n_f\rightarrow n_f+1$. This implies that the new parton
densities taken at $n_f+1$ light flavours can be expressed in 
terms of those given for $n_f$ flavours. 
The original $n_f$ light-flavour densities get modified so that for $k=1,
\dots ,n_f$
\begin{eqnarray}
f_k(n_f\!+\!1, \mu^2) + f_{\bar k}(n_f\!+\!1, \mu^2)\!\! &=& \!\!
 A_{qq,H}^{\rm NS}\Big(n_f, \frac{\mu^2}{m^2}\Big)\otimes
\Big[ f_k(n_f, \mu^2) + f_{\bar k}(n_f, \mu^2) \Big]
\nonumber \\  &&
+ \tilde A_{qq,H}^{\rm PS}\Big(n_f, \frac{\mu^2}{m^2}\Big) \otimes
\Sigma(n_f, \mu^2)
\nonumber \\ &&
+ \tilde A_{qg,H}^{\rm S}\Big(n_f, \frac{\mu^2}{m^2}\Big) \otimes
G(n_f, \mu^2)\,,
\end{eqnarray}
whereas the parton density of the heavy quark can be expressed in the
original light flavours in the following way
\begin{eqnarray}
f_{H+\bar H}(n_f+1, \mu^2) &\equiv&  f_{n_f+1}(n_f+1, \mu^2) 
+ f_{\overline {n_f+1}}(n_f+1, \mu^2) 
\nonumber \\ 
&=& \tilde A_{Hq}^{\rm PS}\Big(n_f, \frac{\mu^2}{m^2}\Big)\otimes
\Sigma(n_f, \mu^2)
+ \tilde A_{Hg}^{\rm S}\Big(n_f, \frac{\mu^2}{m^2}\Big) \otimes
G(n_f, \mu^2)\,. 
\nonumber\\
\end{eqnarray}
Comparing the above expression with Eq. (2.27) we observe that the first
term in (2.27) has no counter part in (2.38). This is because we have no
bare heavy-quark density unless one assumes that there already exists an
intrinsic heavy-quark component of the proton wave function.
The singlet combination of the quark densities becomes
\begin{eqnarray}
\lefteqn{\Sigma(n_f+1, \mu^2) =  \sum\limits_{k=1}^{n_f+1}\Big[
f_k(n_f+1, \mu^2) + f_{\bar k}(n_f+1, \mu^2)\Big]} 
\nonumber \\  
&&=  \Biggl[ A_{qq,H}^{\rm NS}\Big(n_f, \frac{\mu^2}{m^2}\Big)
+ n_f \tilde A_{qq,H}^{\rm PS}\Big(n_f, \frac{\mu^2}{m^2}\Big)
+\tilde A_{Hq}^{\rm PS}\Big(n_f,\frac{\mu^2}{m^2}\Big)\Biggr]
\otimes \Sigma(n_f, \mu^2)
\nonumber\\
&&+ \Biggl[n_f \tilde A_{qg,H}^{\rm S}\Big(n_f, \frac{\mu^2}{m^2}\Big)
+ \tilde A_{Hg}^{\rm S}\Big(n_f,\frac {\mu}{m^2}\Big) \Biggr]
 \otimes
G(n_f, \mu^2)\,. 
\end{eqnarray}
The non-singlet combination $\Delta_k(n_f+1)$ is defined in an analogous
way as in (2.3) and it reads for $k=1,\dots ,n_f+1$
\begin{eqnarray}
\Delta_k(n_f+1,\mu^2)=
f_k(n_f+1, \mu^2) + f_{\bar k}(n_f+1, \mu^2)
- \frac{1}{n_f+1}\Sigma(n_f+1,\mu^2)\,.
\nonumber \\ && 
\end{eqnarray}
Finally the gluon density for $n_f+1$ light flavours is 
\begin{eqnarray}
 G(n_f+1, \mu^2) &=&  A_{gq,H}^{\rm S}(n_f,\mu^2)
\otimes \Sigma(n_f,\mu^2)
\nonumber \\ &&
+ A_{gg,H}^S(n_f,\mu^2)\otimes G(n_f,\mu^2)\,. 
\end{eqnarray}

The old as well as the new parton densities have to satisfy the momentum sum 
rule
\begin{eqnarray}
\int_0^1 \, dx \, x \Big[ \Sigma(n_f,x, \mu^2)
+ G(n_f,x,\mu^2) \Big] = 1 \,,
\end{eqnarray}
for any $n_f$. This implies that the  OME's $A_{Hk}$, $A_{kl,H}$
have to satisfy two relations
\begin{eqnarray}
&& \int_0^1 \, dx \, x \Biggl[ 
A_{qq,H}^{\rm NS}\Big(n_f,x,\frac{\mu^2}{m^2}\Big)
+n_f \tilde A_{qq,H}^{\rm PS}\Big(n_f,x,\frac{\mu^2}{m^2}\Big)
\nonumber \\ &&
+ \tilde A_{Hq}^{\rm PS}\Big(n_f,x,\frac{\mu^2}{m^2}\Big)
+  A_{gq,H}^{\rm S}\Big(n_f,x,\frac{\mu^2}{m^2}\Big)
 \Biggr] = 1 \,,
\end{eqnarray}
and
\begin{eqnarray}
&&\int_0^1 \, dx \, x \Biggl[ 
n_f \tilde A_{qg,H}^{\rm S}\Big(n_f,x,\frac{\mu^2}{m^2}\Big)
+ \tilde A_{Hg}^{\rm S}\Big(n_f,x,\frac{\mu^2}{m^2}\Big)
\nonumber \\ &&
+  A_{gg,H}^{\rm S}\Big(n_f,x,\frac{\mu^2}{m^2}\Big)
 \Biggr] = 1 \,,
\end{eqnarray}
which can be checked up to second order using the results in Appendix B.

The OME's $A_{Hk}$ and $A_{kl,H}$ satisfy the following RGE's
\begin{eqnarray}
\hskip-1cm D \tilde A_{Hq}^{\rm PS} &  =  & 
\Big( \gamma_{qq}^{\rm NS} + n_f \tilde\gamma_{qq}^{\rm PS}\Big) 
\otimes \tilde A_{Hq}^{\rm PS}
+ \gamma_{gq}^{\rm S}\otimes \tilde A_{Hg}^{\rm S}
\nonumber \\ &&
- (\gamma_{HH}^{\rm NS}+\tilde\gamma_{HH}^{\rm PS} )
\otimes \tilde A_{Hq}^{\rm PS} 
- \tilde \gamma_{Hg}^{\rm S}\otimes  A_{gq,H}^{\rm S} 
\nonumber \\ &&
- \tilde \gamma_{Hq}^{\rm PS}\otimes \Big( A_{qq,H}^{\rm NS}
+ n_f \tilde A_{qq,H}^{\rm PS} \Big)
  \,,
\end{eqnarray}
\begin{eqnarray}
\hskip+1cm  D \tilde A_{Hg}^{\rm S} & = &  
 \gamma_{gg}^{\rm S} \otimes \tilde A_{Hg}^{\rm S}
+ n_f \tilde\gamma_{qg}^{\rm S} 
\otimes \tilde A_{Hq}^{\rm PS}
- (\gamma_{HH}^{\rm NS}
+\tilde\gamma_{HH}^{\rm PS})\otimes\tilde A_{Hg}^{\rm S} 
\nonumber \\&&
- \tilde \gamma_{Hg}^{\rm S}\otimes  A_{gg,H}^{\rm S} 
- n_f \tilde \gamma_{Hq}^{\rm PS}\otimes \tilde A_{qg,H}^{\rm S} \,,
\end{eqnarray}
\begin{eqnarray}
\hskip-4cm D A_{qq,H}^{\rm NS} =  
-  \gamma_{qq,H}^{\rm NS}\otimes  A_{qq,H}^{\rm NS} \,,
\end{eqnarray}
\begin{eqnarray}
 D \tilde A_{qq,H}^{\rm PS} & =  & 
 \gamma_{gq}^{\rm S}\otimes \tilde A_{qg,H}^{\rm S}  
- \Big( \gamma_{qq,H}^{\rm NS} + n_f \tilde\gamma_{qq,H}^{\rm PS}\Big) 
\otimes \tilde A_{qq,H}^{\rm PS}
\nonumber \\ &&
- \tilde \gamma_{qq,H}^{\rm PS}\otimes  A_{qq,H}^{\rm NS} 
- \Big(\tilde \gamma_{qg,H}^{\rm S} + \tilde\gamma_{qg}^{\rm S} \Big)
\otimes  A_{gq,H}^{\rm S} 
\nonumber \\ &&
-\tilde\gamma_{qH}^{\rm PS}\otimes \tilde A_{Hq}^{\rm PS}\,,
\end{eqnarray}
\begin{eqnarray}
 D \tilde A_{qg,H}^{\rm S} & = &  
 \tilde\gamma_{qg}^{\rm S}\otimes \Big(A_{qq,H}^{\rm NS}  
+ n_f \tilde A_{qq,H}^{\rm PS} \Big)
+ \gamma_{gg}^{\rm S} \otimes \tilde A_{qg,H}^{\rm S}
\nonumber \\ &&
- \Big( \gamma_{qq,H}^{\rm NS} + n_f \tilde\gamma_{qq,H}^{\rm PS}
+ \gamma_{qq}^{\rm NS} + n_f \tilde\gamma_{qq}^{\rm PS} \Big) 
\otimes \tilde A_{qg,H}^{\rm S}
\nonumber \\ &&
- \Big(\tilde \gamma_{qg,H}^{\rm S} + \tilde\gamma_{qg}^{\rm S}\Big)
\otimes  A_{gg,H}^{\rm S} 
-\tilde\gamma_{qH}^{\rm PS}\otimes \tilde A_{Hg}^{\rm S}\,,
\end{eqnarray}
\begin{eqnarray}
 \hskip-1cm D  A_{gq,H}^{\rm S} &  = &   
 \Big( \gamma_{qq}^{\rm NS} + n_f \tilde\gamma_{qq}^{\rm PS}\Big) 
\otimes  A_{gq,H}^{\rm S}
+\gamma_{gq}^{\rm S} \otimes A_{gg,H}^{\rm S}
\nonumber \\ && 
- \Big(\gamma_{gq,H}^{\rm S}+ \gamma_{gq}^{\rm S} \Big)
\otimes \Big( A_{qq,H}^{\rm NS} + n_f \tilde A_{qq,H}^{\rm PS}
\Big)
\nonumber \\ &&
- \Big(\gamma_{gg,H}^{\rm S} + \gamma_{gg}^{\rm S}\Big)
\otimes  A_{gq,H}^{\rm S} 
-\gamma_{gH}^{\rm S}\otimes \tilde A_{Hq}^{\rm PS}\,,
\end{eqnarray}
and
\begin{eqnarray}
\hskip-4cm D  A_{gg,H}^{\rm S} & = &  
 n_f \tilde\gamma_{qg}^{\rm S}\otimes A_{gq,H}^{\rm S}  
- \gamma_{gg,H}^{\rm S} \otimes  A_{gg,H}^{\rm S}
\nonumber \\  &&
- n_f \Big( \gamma_{gq,H}^{\rm S}+
\gamma_{gq}^{\rm S} \Big) \otimes \tilde A_{qg,H}^{\rm S} 
-\gamma_{gH}^{\rm S}\otimes \tilde A_{Hg}^{\rm S}\,.
\end{eqnarray}

The above RGE's are much more complicated than those written for $A_{kl}$ in
Eqs. (2.12),(2.13). This is due to the fact already mentioned below (2.36)
that $\mu$ represents the operator-renormalization scale as well as the 
mass-factorization scale. In Eqs. (2.12),(2.13) it only stands for 
the operator-renormalization scale. The anomalous dimensions coming from 
operator renormalization carry a minus sign whereas those associated with mass 
factorization have a plus sign in front of them. This 
difference in sign can be traced back to Eq. (2.14) where it was 
stated that the residues of the ultraviolet
divergences are just the opposite of the ones corresponding to the collinear
divergences. Furthermore one has to bear in mind that the residues of the
ultraviolet and collinear divergences are equal to the anomalous dimensions
coming from operator renormalization and mass factorization respectively.
In Eqs. (2.45)-(2.51) the anomalous dimensions $\gamma_{kl}$ 
have a plus sign on account
of the collinear divergences occurring in $A_{kl}$, where the partons
indicated by $k,l$ are massless. However the anomalous 
dimensions $\gamma_{HH}$,
$\gamma_{Hl}$, $\gamma_{lH}$ and $\gamma_{kl,H}$ 
carry a minus sign because the mass of the
heavy quark $m$ prevents these OME's from
being collinearly divergent so that we
only have to deal with ultraviolet singularities.

Using the above equations and (2.15)-(2.17) one can derive the RGE's for
the new parton densities appearing in $F_i^{\rm VFNS}$ (2.30). 
For $k=1,\dots ,n_f+1$, including the heavy-quark flavour, the RGE reads
\begin{eqnarray}
\lefteqn{D~\Big[f_k(n_f+1, \mu^2) + f_{\bar k}(n_f+1, \mu^2)\Big] =} 
\nonumber\\
&& -\gamma_{qq}^{\rm NS}(n_f+1)\otimes
\Big[ f_k(n_f+1, \mu^2) + f_{\bar k}(n_f+1, \mu^2) \Big]
\nonumber\\ &&
- \tilde \gamma_{qq}^{\rm PS}(n_f+1) \otimes
\Sigma(n_f+1, \mu^2)
- \tilde \gamma_{qg}^{\rm S}(n_f+1) \otimes
G(n_f+1, \mu^2)\,.
\end{eqnarray}
The non-singlet combination of the quark densities satisfies the RGE
\begin{eqnarray}
& D \Delta_k (n_f+1, \mu^2) =  
 - \gamma_{qq}^{\rm NS} (n_f+1)\otimes \Delta_k(n_f+1 , \mu^2)\,.  
\end{eqnarray}
The singlet combination of the quark densities satisfies the RGE
\begin{eqnarray}
 D \Sigma (n_f+1, \mu^2) & = &  
 - \gamma_{qq}^{\rm S} (n_f+1) 
\otimes \Sigma(n_f+1 , \mu^2)
\nonumber \\ && 
-  \gamma_{qg}^{\rm S}(n_f +1) \otimes 
G(n_f+1, \mu^2) \,,
\end{eqnarray}
and the gluon density is given by
\begin{eqnarray}
 D G(n_f+1, \mu^2) & = &  
 - \gamma_{gq}^{\rm S} (n_f+1)
\otimes \Sigma(n_f+1 , \mu^2)
\nonumber \\ && 
- \gamma_{gg}^{\rm S}(n_f +1) \otimes 
G(n_f+1, \mu^2) \,.
\end{eqnarray}
In the above equations we have used the identities
\begin{eqnarray}
& \gamma_{ij}(n_f) + \gamma_{ij,H}(n_f) =  
  \gamma_{ij} (n_f+1)\,,
\end{eqnarray}
\begin{eqnarray}
& \gamma_{HH}^{\rm NS}(n_f+1)  =  
  \gamma_{qq}^{\rm NS} (n_f+1)\,; \quad
\quad \tilde \gamma_{HH}^{\rm PS}(n_f+1)  =
\tilde  \gamma_{qq}^{\rm PS} (n_f+1)\,,
\end{eqnarray}
\begin{eqnarray}
& \tilde\gamma_{qH}^{\rm PS}(n_f+1) = \tilde\gamma_{Hq}^{\rm PS}(n_f+1) =
  \tilde \gamma_{qq}^{\rm PS} (n_f+1)\,,
\end{eqnarray}
\begin{eqnarray}
& \gamma_{gH}^{\rm S}(n_f+1)  =  
  \gamma_{gq}^{\rm S} (n_f+1)\,; \quad 
\quad  \tilde \gamma_{Hg}^{\rm S}(n_f+1)  =  
  \tilde \gamma_{qg}^{\rm S} (n_f+1)\,,
\end{eqnarray}
because the anomalous dimensions do not depend on the mass $m$ of the 
heavy-quark H.
A comparison of the above RGE's with those presented for the light-parton
densities in Eqs. (2.15)-(2.17) reveals that they are exactly the same
in spite of the fact that there is no counterpart of the bare heavy-quark 
density in the derivation of Eqs. (2.52)-(2.55).

All perturbative quantities which appear in this section are now 
available up to $O(\alpha_s^2)$. This 
holds for the anomalous dimensions
$\gamma_{kl}$ \cite{cfp}, the massless-parton 
coefficient functions ${\cal C}_{i,k}$ \cite{zn1}
and the heavy-quark coefficient functions $L_{i,k}$, $H_{i,k}$ \cite{lrsn}. 
The OME's
$\tilde A_{Hq}^{\rm PS}$, $\tilde A_{Hg}^{\rm S}$ 
and $A_{qq,H}^{\rm NS}$ are computed up to $O(\alpha_s^2)$ 
and listed in unrenormalized form in
Appendix C of \cite{bmsmn}. 
We still require $A_{gq,H}^{\rm S}$ (Fig. 3) and $A_{gg,H}^{\rm S}$ (Fig. 4)
which are 
calculated up to the same order in this paper. Exact expressions 
for the unrenormalized OME's can be found 
in Appendix A. Notice that up to second order in $\alpha_s$
both $\tilde A_{qg,H}^{\rm S}$ 
and $\tilde A_{qq,H}^{\rm PS}$ are zero. 
The renormalized (finite) expressions for all these OME's, which we will 
use in the next section, are presented in Appendix B.

After having found the representation of the heavy-quark density 
$f_{H + \bar H}$ in (2.38) and the RGE in (2.52) which determines its scale
evolution we can write down the charm component of the deep-inelastic 
structure function in the VFNS representation. The latter is given by
\begin{eqnarray}
F_{i,H}^{\rm VFNS} (n_f+1, Q^2) &  = &  e_H^2 \Big[
f_{H +\bar H}(n_f+1,\mu^2) \otimes 
{\cal C}_{i,q}^{\rm NS}\Big(n_f+1,\frac{Q^2}{\mu^2}\Big)
\nonumber\\
&& + \Sigma(n_f+1,\mu^2)\otimes\tilde{\cal C}_{i,q}^{\rm PS}
\Big(n_f+1,\frac{Q^2}{\mu^2}\Big)
\nonumber \\ &&
 + G(n_f+1,\mu^2)\otimes \tilde{\cal C}_{i,g}^{\rm S}
\Big(n_f+1,\frac{Q^2}{\mu^2}\Big)
\Big]
\,.
\end{eqnarray}
The expression above satisfies $D F_{i,H}^{\rm VFNS}=0$
(see (2.11)), so it is renormalization group invariant, 
which means that it is scheme independent and becomes a 
physical quantity. Notice that even though the form 
of $F_{i,H}^{\rm VFNS}$ is the same
as the one presented for intrinsic heavy-quark production, their
origins are completely different. The VFNS result, Eq. (2.60) 
is derived from FOPT when the exact heavy-quark coefficient 
functions are replaced by their 
asymptotic expressions taken at $Q^2\gg m^2$ (see (2.30)). 
Then we have performed mass factorization and absorbed the mass 
singularities with respect to the heavy-quark mass $m$ into 
the light-parton densities taken at the original $n_f$
light flavours (see Eqs. (2.37)-(2.41)). This procedure leads to relations
between the parton densities in the $n_f$ and $n_f+1$ flavour schemes.
In particular this applies to the
heavy-quark density $f_{H + \bar H}$ in (2.38). For intrinsic heavy-quark
production the origin of Eq. (2.60) is completely different. In this case
it is not 
derived from perturbation theory and therefore there does not 
exist any relation
between the heavy-quark density and the light-parton densities like in 
Eq. (2.38). Actually the intrinsic-charm density is determined by a simple fit
to experimental data.

Furthermore we want to emphasize that the VFNS 
approach is only valid for totally
inclusive quantities like structure functions since the 
logarithmic terms $\ln^i(Q^2/m^2)\ln^j(\mu^2/m^2)$ only 
appear in the asymptotic form of the heavy-quark coefficient functions
in this case. Hence expression (2.60) is
just an alternative description for FOPT when the production of the heavy
quarks occurs far above threshold where $Q^2\gg m^2$. The only
difference between the FOPT and the VFNS descriptions
is that the large logarithms have been resummed in the latter approach
so that one gets an improved expression with respect to normal 
perturbation theory in the large $Q^2$-region. However on the
level of differential distributions the large logarithms of the kind given 
above do not show up in the perturbation series. Therefore in this case
the VFNS approach cannot be applied. As has been mentioned in the introduction
one can only distinguish between intrinsic and extrinsic heavy-quark 
production.

Finally we want to comment about the work in \cite{acot} where one has proposed
the idea of the VFNS approach. In particular we want to 
make some remarks about equation (9) in \cite{acot}(ACOT)
which is similar to our equation (2.60).
Using the notations
in the latter reference this equation reads
\begin{eqnarray}
&& \sum_{\lambda} W_{BN}^{\lambda} = f_N^Q\otimes \sum_{\lambda}
\omega_{BQ}^{\lambda(0)}-f_N^g \otimes f_g^{Q(1)}\otimes\sum_{\lambda}
\omega_{BQ}^{\lambda(0)}+f_N^g \otimes \sum_{\lambda}
\omega_{Bg}^{\lambda(1)} \,,
\nonumber\\&&
\end{eqnarray}
where we have summed over all helicities of the virtual photon denoted by
$\lambda$. Further we have corrected a misprint because $f_g^{Q(0)}$ in (9)
should read $f_g^{Q(1)}$. To translate (2.61) into our language we have to
make the following replacements
\begin{eqnarray}
 B \rightarrow \gamma^* \,;\quad Q \rightarrow H \,; \quad
\sum_{\lambda} W_{BN}^{\lambda} \rightarrow F_{i,H}^{\rm ACOT}\,,
\end{eqnarray}
\begin{eqnarray}
 \sum_{\lambda} \omega_{BQ}^{\lambda(0)} \rightarrow 
e_H^2~{\cal C}_{i,q}^{\rm NS,(0)}=e_H^2\delta(1-z) \,;\quad 
\sum_{\lambda}\omega_{Bg}^{\lambda(1)} \rightarrow e_H^2 H_{i,g}^{\rm S,(1)}\,,
\end{eqnarray}
\begin{eqnarray}
 f_g^{Q(1)} \rightarrow \tilde A_{Hg}^{\rm S,(1)} \,,
\end{eqnarray}
and
\begin{eqnarray}
 f_N^Q \rightarrow f_{H+ \bar H} \,; \quad f_N^g \rightarrow G\,.
\end{eqnarray}
Hence Eq. (2.61) can be written in our notation as
\begin{eqnarray}
\lefteqn{F_{i,H}^{\rm ACOT}
=e_H^2\Big[ f_{H + \bar H} \otimes {\cal C}_{i,q}^{\rm NS,(0)} 
+ G\otimes\tilde {\cal C}_{i,g}^{\rm S,(1)}\Big]}
\nonumber\\
&&+ e_H^2 \;\Big[G\otimes\{ H_{i,g}^{\rm S,(1)}
-\tilde {\cal C}_{i,g}^{\rm S,(1)}-\tilde A_{Hg}^{\rm S,(1)} \otimes
{\cal C}_{i,q}^{\rm NS,(0)}\}\Big] \,.
\end{eqnarray}
Comparing this expression with ours in (2.60) we notice the 
following differences. First Eq. (2.61) and therefore Eq. (2.66) in \cite{acot}
was only derived in lowest order (LO), whereas Eq. (2.60) is valid in all
orders of perturbation theory. Second Eqs. (2.61) and (2.66) are not valid in 
next-to-leading order (NLO) but this can be repaired by replacing 
${\cal C}_{i,q}^{\rm NS,(0)}$ by ${\cal C}_{i,q}^{\rm NS,(0)}+
(\alpha_s/4\pi)\,{\cal C}_{i,q}^{\rm NS,(1)}$ in (2.66). 
However the most important
difference is that the last term between the square brackets in (2.66) has
no counter part in our expression (2.60). This is because we took the limit
$Q^2\rightarrow \infty$ and dropped all terms proportional to 
$(m^2/Q^2)^l$ in the asymptotic heavy-quark coefficient functions. 
This is revealed by the
mass-factorization relation (2.35) for $H_{i,g}^{\rm S}$, 
which only holds in the limit $Q^2 \gg m^2$. 
However the exact expressions for $H_{i,g}^{\rm S,(1)}$,
$\tilde {\cal C}_{i,g}^{\rm S,(1)}$
and $\tilde A_{Hg}^{\rm S,(1)}$ can be found in the literature 
(see \cite{lrsn}, \cite{zn1} and \cite{bmsmn}). 
Expanding the latter in powers of $m^2/Q^2$ we obtain
\begin{eqnarray}
\lefteqn{H_{i,g}^{\rm S,(1)}
-\tilde {\cal C}_{i,g}^{\rm S,(1)}-\tilde A_{Hg}^{\rm S,(1)} \otimes
{\cal C}_{i,q}^{\rm NS,(0)}}
\nonumber\\
&&= \alpha_s(\mu^2) \sum\limits_{l=1}^{\infty} (\frac{m^2}{Q^2})^l \Big[
a_i^{(l)} \ln(\frac{Q^2}{m^2}) + b_i^{(l)}\Big]  \,.
\end{eqnarray}
When $Q^2 \rightarrow \infty$ (2.67) vanishes, as was already expected 
from (2.35).
The motivation for the above expression, included in \cite{acot}, was to get
a better stability 
of $F_{i,H}^{\rm VFNS}$ in the threshold region with respect  
to variations in the factorization scale. However we have dropped these type
of contributions as shown in Eq. (2.67) in our
representation for $F_{i,H}^{\rm VFNS}$ (2.60) for theoretical as well as
practical reasons. The theoretical argument is that mass factorization
does not apply to terms proportional to $(m^2/Q^2)^l$ for $l\geq 1$ so that 
Eq. (2.66) or (9) in \cite{acot} cannot be generalized to higher orders in
$\alpha_s$. The practical reason is that the inclusion of higher order
corrections automatically improves the stability of the perturbation series
with respect to variations with respect to the factorization scale 
as is shown in \cite{grs}, \cite{Vogt}, \cite{or}.

Summarizing this section we have presented expressions for the total structure 
functions $F_i^{\rm VFNS}(n_f+1,Q^2)$ (2.30) as well as for its heavy-quark
component $F_{i,H}^{\rm VFNS}(n_f+1,Q^2)$ (2.60) in the context 
of the variable-flavour-number scheme (VFNS),
which are valid in all orders of perturbation theory.
Starting from the structure functions $F_i(n_f,Q^2)$, expressed in 
the light-parton densities and coefficient functions, we could show 
that by adding the heavy-flavour contribution the sum can be written 
as $F_i^{\rm VFNS}(n_f+1,Q^2)$,
provided $Q^2 \gg m^2$ (see Eq. (2.30)). 
This procedure imposes relations between
the parton densities appearing in $F_i(n_f,Q^2)$ and the new ones showing up
in $F_i^{\rm VFNS}(n_f+1,Q^2)$. 
These relations, given by (2.37)-(2.41), have to be 
satisfied in the VFNS approach. Unfortunately the 
existing parton-density sets in the literature do not satisfy 
these requirements. They do however satisfy the
RGE's in (2.52)-(2.55). The consequences of the relations between the new 
parton densities taken at $n_f+1$ flavours and the old ones presented at
$n_f$ flavours will be discussed for charm production ($n_f=3$) in the next
section.

\mysection{Validity of FOPT and the VFNS}
In this section we apply the findings of the last section to charm production.
Here we want to investigate which of the two approaches
i.e., FOPT or VFNS, is the most appropriate to describe the total structure
function $F_{i}(x,Q^2)$ and its charm component $F_{i,c}(x,Q^2,m^2)$ in the
different kinematical regimes.
To do this we will study the following issues. The first one concerns
the question at which values of $x$ and $Q^2$ the large logarithmic terms
$\ln^i(Q^2/m_c^2)\ln^j(\mu^2/m_c^2)$ in the heavy-quark coefficient functions
constitute the bulk of the radiative corrections to the charm component of
the structure functions.
Further in the spirit of VFNS one should use the parton 
densities defined in (2.37)-(2.41) 
instead of the usual ones which are available
in the literature. In the former there exist direct relations between the
parton densities taken at $n_f=3$ 
(no charm-quark density) and the parton densities
at $n_f=4$ (charm-quark density included). In the literature this relation
is broken because in all parametrizations the charm-quark density is given
by an arbitrary function in which the parameters are fitted to the data and 
there exists no direct relation between the charm density and the remaining
u,d,s and g densities except that they have to satisfy the momentum sum
rule. The effect of this difference on the charm structure 
function $F_{2,c}(x,Q^2)$ will be investigated. Finally we study the 
difference in the behaviour of this function 
which occurs when going from the FOPT to the VFNS descriptions. 
In particular we focus on
the visibility of the charm threshold which appears in FOPT but is absent in 
the case of the VFNS. We also study the rate of convergence of the perturbation
series over a wide range of $Q^2$-values which will be different for these
two approaches. 

Before presenting our results we want to mention that all
perturbative quantities like the operator-matrix elements (OME's), 
heavy-quark coefficient functions and light-parton coefficient functions are
presented in the ${\overline {\rm MS}}$-scheme. 
Therefore we have to use parton densities parametrized in the same scheme.
Furthermore in FOPT the number of light flavours in the running coupling
constant and the coefficient functions has to be equal to three  
($\Lambda_3 = 232$ MeV (LO); $\Lambda_3=248$ MeV (NLO)) 
whereas in the VFNS this number should be
equal to four ($\Lambda_4 = 200$ MeV (LO and NLO)). 
For the mass of the charm quark
and the factorization (renormalization) scale we have chosen 
$m_c = 1.5$ ${\rm GeV}/c$ and $\mu^2=Q^2$ respectively
because the same scale is usually adopted for the light-parton components
of the structure functions. Notice that in the literature \cite{Adl}, 
\cite{acot}, \cite{Vogt} a different scale
is chosen for the charm structure function in FOPT. However the results in
NLO are rather independent of the scale as is shown in \cite{grs},\cite{Vogt}
\cite{or}.

Starting with the first question about the dominance of the large logarithms
we define the charm component in the FOPT approach by the function
$F_{i,c}(x,Q^2,m_c^2)$. The latter is computed from (2.29) and contains 
the heavy-quark coefficient functions $L_{i,k}$ 
and $H_{i,k}$, where $i=2,L$ and $k=q,g$.
The above coefficient functions are exactly calculated up to $O(\alpha_s^2)$
in \cite{lrsn}. Their asymptotic expressions, which contain the large 
logarithms above, as well as constant terms, are presented in \cite{bmsmn}. 
The latter are strictly speaking only valid when $Q^2 \gg m_c^2$. 
Further let us denote by $F_{i,c}^{\rm exact}$ and 
$F_{i,c}^{\rm asymp}$ the charm 
structure functions computed by
using the exact and asymptotic forms of the heavy-quark coefficient functions 
respectively. In order to determine the $Q^2$ value above 
which $F_{i,c}^{\rm exact}$ 
and $F_{i,c}^{\rm asymp}$ coincide we plot the following ratios
\begin{equation}
 R_i(x,Q^2,m_c^2)= 
\frac{F_{i,c}^{\rm asymp}(x,Q^2,m_c^2)}{F_{i,c}^{\rm exact}(x,Q^2,m_c^2)} 
\,.
\end{equation}
For these plots we adopt the GRV94HO parton density set \cite{grv94}.
The reason that this set is chosen is because it is obtained from a fit to
the deep-inelastic scattering data performed in the 
spirit of the FOPT approach, which means that the number of active 
flavours is chosen to be three and the charm component of the structure
function is calculated from the photon-gluon fusion process 
and its higher order QCD corrections.

In Fig. 5 we plot $R_2(NLO)$ as a function of $Q^2$ for 
four different 
values of $x$ i.e. $x=10^{-1}, 10^{-2}, 10^{-3}$ and $10^{-4}$. From this 
figure we infer that $R_2$ 
tends to unity at $Q^2$ values which are an order of 
magnitude larger than $m_c^2$
( $1 > R_2 > 0.9$ for $Q^2 > 20$ $({\rm GeV}/c)^2$). 
This holds provided $x < 10^{-2}$, where
the limit $R_2 = 1$ is almost always approached from below. 
However for $x > 10^{-1}$ the limit is 
reached from above and at a much bigger value of 
$Q^2$ $(1.1 > R_2  > 1$ for $Q^2 > 300$ $({\rm GeV}/c)^2$),
which is two orders of magnitude larger than $m_c^2$. The
fact that the rate of convergence to $R_2=1$ is much slower 
at large $x$ is more clearly shown in Fig. 6. Here we plot $R_2(NLO)$ 
as a function of $x$
at $Q^2=10,50$ and $100$ $({\rm GeV}/c)^2$. At very large $x$ 
all curves strongly deviate from
$R_2 = 1$. When $Q^2$ decreases the deviations occur at smaller $x$. 
This fact that $R_2 \not= 1$ at large $x$ and small $Q^2$ can 
be wholly attributed to threshold terms, which are present in 
the exact heavy-quark coefficient functions, but which are absent 
in their asymptotic expressions.

In Fig. 7 we show the same plot for $R_L(NLO)$ as 
presented for $R_2(NLO)$ in Fig. 5. 
Here the approach to $R_L=1$ starts at a much larger value of $Q^2$ 
than has been observed for $R_2$. 
One sees that $0.9 <R_L <1.1$ when $Q^2 > 10^3$ $({\rm GeV}/c)^2$.
Again the rate of convergence is slower as $x$ increases, which is the region
where threshold effects become important. This becomes even more visible
in Fig. 8. The reason that threshold effects are dominant 
at large $x$ and small $Q^2$ 
can be explained when one looks at the convolution
\begin{equation}
\int_x^{z_{\rm th}} \, \frac{dz}{z} f_k\Big(\frac{x}{z},\mu^2\Big)
H_{i,k}\Big(z,\frac{Q^2}{m^2},\frac{m_c^2}{\mu^2}\Big)\,,
\end{equation}
with a similar expression when $H_{i,k}$ is replaced by $L_{i,k}$.
Here $f_k$ and $H_{i,k}$,$L_{i,k}$ denote
the parton densities and the heavy-quark coefficient
functions respectively. The threshold value is given by
\begin{equation}
z_{\rm th} = \frac{Q^2}{Q^2 + 4~m_c^2}
\end{equation}
in the expression for the structure function (2.29). From the above 
equations one infers that when $x$ is very 
large and $Q^2$ is very small $ x \rightarrow z_{\rm th}$ so that 
only threshold terms can contribute to the integral (3.2).
We also would like to comment on the phenomenon that the 
approach to $R_i(NLO)=1$ is much slower for $i=L$ 
than for $i=2$. It originates from the fact that the power of the
large logarithms appearing in the heavy-quark coefficient functions 
in the case of $i=L$ is one unit smaller than that for $i=2$. 
This phenomenon was
also observed  for heavy-flavour production in the Drell-Yan process 
\cite{rn}.
It appears that the $Q^2$ value for which the exact and asymptotic expressions
of the physical quantities coincide is smaller when the powers of the large
logarithmic terms increase. Notice that the Born contribution to the
longitudinal coefficient function does not contain logarithms in the limit
$Q^2 \gg m_c^2$ so that it is independent of $Q^2$ and $m_c^2$. 
In this case, as well as
in some interference terms in the Drell-Yan process, the convergence to the
asymptotic expressions takes place at an extremely large value of $Q^2$.

We have also studied $R_i(LO)$ in the 
Born approximation to the charm structure 
functions in (3.1).  
Here it turns out that $R_i(LO)\ge 1$ for all
$x$ and $Q^2$ so that the limit is always approached from above. This behaviour
is different from the one observed in NLO (see Figs. 5-8). 
Further $R_i(LO)$ is closer to unity at small $Q^2$ 
and large $x$ than in the case of $R_i(NLO)$ which implies
that threshold effects in leading order are smaller than in next-to-leading 
order. However the $Q^2$ value at which $R_i$ becomes equal 
to one is essentially the same in LO and NLO. 
In Section 5 of \cite{bmsmn} we 
made the same study of $R_i$ in (3.1) but on the level of the 
heavy-quark coefficient functions themselves.
A comparison with the results from \cite{bmsmn} reveals that the $Q^2$ value 
for which the asymptotic and exact heavy-quark coefficient 
functions coincide is the same as the one obtained for the charm structure
functions in (3.1). 
Hence the convolution with the parton densities in (3.2) hardly 
affects the $Q^2$ value at which the exact and asymptotic expressions coincide.

Next we discuss the parton densities which emerge from the VFNS  
according to Eqs. (2.37)-(2.41). The most interesting among them is 
the charm-quark density which appears in the four-flavour scheme. 
It is derived from the formula in Eq. (2.38) where we choose $n_f=3$.
Up to $O(\alpha_s^2)$ Eq. (2.38) becomes equal to
\begin{eqnarray}
\lefteqn{f^{\rm VFNS}_{c + \bar c}(4,x,\mu^2) \equiv f_4(4,x,\mu^2) + 
f_{\bar 4}(4,x,\mu^2)}
\nonumber\\
&&= \Big( \frac{\alpha_s(\mu^2)}{4\pi}\Big)^2
\int_x^1 \frac{dz}{z} \Sigma\Big(3, \frac{x}{z} ,\mu^2\Big)
\tilde A_{cq}^{\rm PS,(2)}\Big(z, \frac{\mu^2}{m_c^2} \Big)        
\nonumber\\ 
&&+ \int_x^1 \, \frac{dz}{z} G\Big(3, \frac{x}{z}, \mu^2\Big)
\Big[  \Big( \frac{\alpha_s(\mu^2)}{4\pi}\Big)
\tilde A_{cg}^{\rm S,(1)}\Big(z, \frac{\mu^2}{m_c^2} \Big)
 +   \Big( \frac{\alpha_s(\mu^2)}{4\pi}\Big)^2
\tilde A_{cg}^{\rm S,(2)}\Big(z, \frac{\mu^2}{m_c^2} \Big) \Big]
\,,
\nonumber\\
\end{eqnarray}
where $\tilde A_{Hq}^{{\rm PS},(2)}$, $\tilde A_{Hg}^{{\rm S},(i)} (i=1,2)$ 
with $H=c$ are the OME's presented in (B.1)-(B.3) which are renormalized
in such a way that $\alpha_s$ in (3.4) depends on four flavours 
($\Lambda_4=200$ MeV). Further one has to put $n_f=3$ in the OME's above
which up to order $\alpha_s^2$ are independent of the number of flavours.
The quantities $\Sigma(3,\mu^2)$ and $G(3,\mu^2)$
represent the singlet combination of parton densities and the gluon density 
in the three-flavour scheme respectively. Notice that the $O(\alpha_s)$ term
was already introduced in Eq. (10) of \cite{acot}. 

In order to make a comparison with the existing charm-quark densities present
in the literature we now choose the GRV92HO set \cite{grv92} to compute
$\Sigma(3,\mu^2)$ and $G(3,\mu^2)$ in (3.4). 
The reason is that the GRV92 set contains a charm-quark density 
which is not included in the GRV94 set \cite{grv94}.
For a comparison between the charm-quark density $f^{\rm VFNS}_{c + \bar c}$ 
and the one presented by the GRV92 set we have to impose the same boundary  
conditions. Therefore we require $f^{\rm VFNS}_{c + \bar c}(4,x,m_c^2)= 0$  
(3.4) since the same has been done for the charm-quark density
in \cite{grv92}.
To that order the non-logarithmic terms in 
$\tilde A_{cq}^{\rm PS,(2)}$ (B.1) and $\tilde A_{cg}^{\rm S,(2)}$ (B.3) 
have to be removed, which is
not needed for $\tilde A_{cg}^{\rm S,(1)}$ (B.2) because the latter 
already vanishes at $\mu^2=m_c^2$.

In Fig. 9 we plot the ratio
\begin{equation}
  R_{\rm ch} = \frac{f^{\rm VFNS}_{c + \bar c}(4,x,\mu^2)}
{f^{\rm PDF}_{c + \bar c}(4,x,\mu^2)}      
\,,
\end{equation}
as a function of $\mu^2$ for the values $x = 10^{-1}, 10^{-2}, 10^{-3}$
and $10^{-4}$. 
Here $f^{\rm VFNS}_{c + \bar c}$ is computed from (3.4) by choosing one of the 
parton density sets in the literature (here GRV92HO) for the determination of
$\Sigma(3,\mu^2)$ and $G(3,\mu^2)$ while
$f^{\rm PDF}_{c + \bar c}$ is the charm-quark density belonging to the
same set.
For $x < 10^{-2}$ it turns out that $1 > R_{\rm ch} > 0.9$.
This is surprising because (3.4) is calculated up to finite
order in perturbation theory
whereas the GRV density is the NLO solution of  
RGE's in which all leading and next-to-leading logarithms are
resummed. Hence we should expect a larger difference between the scale
evolutions. However from the results in \cite{zn2}, \cite{brnv} 
we anticipate that this 
difference is not so dramatic as long as $\mu^2/m_c^2 < 10^3$. 
Apparently the leading logarithms beyond $O(\alpha_s^2)$, which 
are neglected in (3.4), but not in the GRV charm density,
do not play an important role provided $\mu^2$ is not chosen to be too large.
What is really striking is
that the charm density depends on four-light-parton densities 
in the VFNS, whereas in the case of GRV this density is unconstrained 
except that it has to satisfy the momentum sum rule. Therefore it is
unexpected that the ratio $R_{\rm ch}$ is so close to unity. 

Besides the appearance of the charm-quark density, the four other light-parton
densities representing u,d,s and g are modified while going from three to four
active flavours. The modification for the light-quark densities is given 
by Eq. (2.37). 
Summing the latter over the three light flavours (u,d and s) one obtains
the singlet density corrected up to $O(\alpha_s^2)$ which is given by 
\begin{eqnarray}
\lefteqn{ \Sigma'(4,x,\mu^2)  =
\int_x^1 \, \frac{dz}{z}  \Sigma\Big(3,\frac{x}{z},\mu^2\Big)}
\nonumber\\ 
&&\times \Big[ \delta(1-z) 
 +   \Big( \frac{\alpha_s(\mu^2)}{4\pi}\Big)^2
 A_{qq,c}^{\rm NS,(2)}\Big(z, \frac{\mu^2}{m_c^2} \Big) \Big]
\,,
\end{eqnarray}
where $A_{qq,H}^{\rm NS,(2)}$ 
is given in (B.4) with $H = c $. Notice that 
$\Sigma(4,x,\mu^2)=\Sigma'(4,x,\mu^2)+f_{c + \bar c}(4,x,\mu^2)$
in Eq. (2.39).
Up to $O(\alpha_s^2)$ the gluon density in the four-flavour scheme 
(see Eq. (2.41)) is
\begin{eqnarray}
\lefteqn{G(4,x,\mu^2) =
   \Big( \frac{\alpha_s(\mu^2)}{4\pi}\Big)^2
\int_x^1 \, \frac{dz}{z} \Sigma\Big(3,\frac{x}{z},\mu^2\Big)
 A_{gq,c}^{\rm S,(2)}\Big(z, \frac{\mu^2}{m_c^2}\Big)} 
\nonumber\\ 
&&+ \int_x^1 \, \frac{dz}{z} G\Big(3,\frac{x}{z},\mu^2\Big)
\Big[
\delta(1-z) +
   \Big( \frac{\alpha_s(\mu^2)}{4\pi}\Big)
A_{gg,c}^{\rm S,(1)}\Big( z, \frac{\mu^2}{m_c^2}\Big)
\nonumber\\
&&+ \Big( \frac{\alpha_s(\mu^2)}{4\pi}\Big)^2
A_{gg,c}^{\rm S,(2)}\Big( z, \frac{\mu^2}{m_c^2}\Big)
\Big]
\,.
\end{eqnarray}
The functions
$A_{gq,H}^{\rm S,(2)}$ and $A_{gg,H}^{{\rm S},(i)}$ ($i=1,2$) with $H = c$ are 
presented in Eqs. (B.5)-(B.7). Notice that up to $O(\alpha_s^2)$ the 
above OME's are independent of the number of internal flavours $n_f$. 

In the literature all parton density sets are presented either in the 
three-flavour scheme ($n_f=3$) e.g. GRV94 \cite{grv94}, GRV92 \cite{grv92} 
or in the four-flavour scheme ($n_f=4$) e.g. CTEQ \cite{cteq} and 
MRS \cite{mrs}. 
This implies that $n_f$, which occurs in the anomalous dimensions and in
the coefficient functions, is kept fixed above and below the charm threshold.
In the case of GRV92 it means that even when $Q^2 \gg m_c^2$ 
the light-parton densities evolve 
according to $n_f = 3$ whereas in principle $n_f = 4$ 
has to be chosen. 
In Fig. 10 we show the singlet 
combinations of parton densities $\Sigma(3,\mu^2)$ 
and $\Sigma'(4,\mu^2)$ (3.6) and, as the difference between them 
is $O(\alpha_s^2)$, it is essentially invisible. 
Therefore the error made in choosing $\Sigma(3,\mu^2)$
instead of $\Sigma'(4,\mu^2)$ above the charm threshold 
is extremely small. In Fig. 11
we make the same study for the gluon density where we 
compare $G(3,\mu^2)$ with $G(4,\mu^2)$. Here the latter function yields
the higher curves. In this case the 
difference is a larger than for the singlet-quark combination (3.6).
This is no surprise because the difference already starts
in $O(\alpha_s)$ in the gluon case (see (3.7)). 
We also made a comparison
with the parton densities of \cite{cteq} (CTEQ) and  
\cite{mrs} (MRS). In this case the parton-density sets are 
presented in the four-flavour scheme. Inverting Eqs. (3.6) and (3.7) 
one can easily get the corresponding densities for the three-flavour 
scheme and recompute $f^{\rm VFNS}_{c + \bar c}$ in (3.4). 
Like in the GRV92 case the fitted charm-quark density 
$f^{\rm PDF}_{c + \bar c}$ 
and the computed charm-quark density $f^{\rm VFNS}_{c + \bar c}$ 
show a remarkable agreement for each parton-density set separately.
Since $f^{\rm VFNS}_{c + \bar c}$ is mainly determined by the gluon
density in (3.4) the mutual differences between the charm-quark densities 
in the various parton-density sets may be attributed to the different 
parametrizations for the gluon.

Summarizing our findings above we can conclude that for the 
actual computation
of the charm structure function $F^{\rm VFNS}_{2,c}$ in (2.60), it 
does not makes much difference 
when the light-flavour densities in the the VFNS, as given by 
Eqs. (3.4),(3.6) and (3.7) are replaced by those obtained from the 
various parton-density sets available in the literature. 

Now we want to study the differences between the FOPT and VFNS 
approaches for the description of the charm structure function $F_{2,c}$.
Using the set GRV92HO \cite{grv92} we present plots 
for $F_{2,c}^{\rm exact}$ (2.29) in Figs. 12a-e (see also above (3.1)), which
represents the FOPT.
We do the same for $F_{2,c}^{\rm VFNS}$, which is given by Eq. (2.60), using
the densities in Eqs. (3.4),(3.6) and (3.7). Both are calculated in NLO.
For $Q^2 =3$ $({\rm GeV}/c)^2$ (see Fig. 12a) we observe the
threshold effect in $F_{2,c}^{\rm exact}$ at large $x$. 
Here this function vanishes
for $x\geq 0.1$. This effect does not show up in $F_{2,c}^{\rm VFNS}$
although the latter is so small in the large $x$-region that, in view of
the low statistics of the data in that region, any distinction between the
FOPT and the VFNS approaches will be invisible. 
The difference becomes even
smaller when $Q^2$ increases (see Figs. 12b-e). However at small $x$ the
difference between $F_{2,c}^{\rm exact}$ and $F_{2,c}^{\rm VFNS}$ becomes more 
conspicuous. It is in this region that data will appear 
from the HERA collider experiments. 
Although the difference is not that large 
at $Q^2=3$ $({\rm GeV}/c)^2$ 
it becomes big for $5<Q^2<10$ $({\rm GeV}/c)^2$ 
(see Figs. 12b,c). Here $F_{2,c}^{\rm VFNS}$
exceeds $F_{2,c}^{\rm exact}$ by more than $60\%$ with respect to the latter 
quantity at $x=10^{-4}$. The difference becomes less when $Q^2$ increases, e.g.
at $Q^2=100$ $({\rm GeV}/c)^2$ (Fig. 12e) it is about $25\%$. 
The H1-collaboration has published data in \cite{Adl} 
for $Q^2=12,25$ and $45$ $({\rm GeV}/c)^2$ 
which lie in the region $x \le 10^{-2}$. From their 
analysis we infer that the predictions from FOPT lie below the data which
means that, in view of our findings above, the predictions from VFNS are in 
better agreement with experiment. 
However one has to be cautious to draw too premature conclusions from
the observations made above. First one has to bear in mind that
$F_{2,c}^{\rm VFNS}$ is derived from $F_{2,c}^{\rm asymp}$ via mass 
factorization (see Section 2). Therefore the former is only reliable when
the large logarithms mentioned in the beginning of this section dominate
the corrections to the charm structure function. From Figs. 5,6 we infer that
this happens for $Q^2>20$ $({\rm GeV}/c)^2$. Another criterion is the
perturbative stability of the various approaches, 
which will be discussed shortly, and from which we 
infer that the VFNS is better than FOPT 
for $Q^2>10$ $({\rm GeV}/c)^2$. 
This implies that 
$F_{2,c}^{\rm VFNS}$ gives a more reliable description of the data than 
$F_{2,c}^{\rm exact}$ for $Q^2>20$ $({\rm GeV}/c)^2$.

Further we have also plotted $F_{2,c}^{\rm asymp}$ (see above (3.1)). As
expected from Figs. 5,6 this function approaches $F_{2,c}^{\rm exact}$
when $Q^2$ gets very large. We did not plot $F_{2,c}^{\rm asymp}$ at
$Q^2=3$ $({\rm GeV}/c)^2$ (Fig. 12a) because here it becomes negative for
$x<0.1$. From Figs. 12b-e one observes a noticeable difference between
$F_{2,c}^{\rm asymp}$ and $F_{2,c}^{\rm VFNS}$, which is remarkable since
the latter is derived from the former via mass factorization; a
procedure which holds in all orders of perturbation theory. In our actual
computations $F_{2,c}^{\rm asymp}$ has been computed up to order $\alpha_s^2$
whereas $F_{2,c}^{\rm VFNS}$ involves $\alpha_s^3$ contributions. The latter
are due to the convolutions in Eq. (2.60) of the order $\alpha_s^2$
corrected densities in (3.4),(3.6) and (3.7) with the 
order $\alpha_s$ corrected
light-parton coefficient functions. Due to the large logarithms 
these order $\alpha_s^3$ contributions
are quite appreciable and they survive in the region $3<Q^2<100$ 
$({\rm GeV}/c)^2$ covered by our plots. The corrections beyond order 
$\alpha_s^2$ present in $F_{2,c}^{\rm VFNS}$ can be resummed using the RGE's in
(2.52)-(2.55). This procedure has been applied to obtain the parton densities
in the literature. If we substitute one of them (here we use GRV92HO 
\cite{grv92}) in the expression for $F_{2,c}^{\rm VFNS}$ in Eq. (2.60)
we call the corresponding charm structure function $F_{2,c}^{\rm PDF}$.
The results for this function are given in Figs. 12a-e. 
As can be expected from our discussions
above $F_{2,c}^{\rm PDF}$ is slightly larger than $F_{2,c}^{\rm VFNS}$. This
is not surprising because the parton densities represent the resummation of
the large logarithms in all orders of perturbation theory. 

We also want to comment on the presentation of the charm structure function in
(2.66) originating from (9) in \cite{acot} (ACOT). From (2.60) and 
the definitions for $F_{i,c}^{\rm exact}$,$F_{i,c}^{\rm asymp}$ 
given above (3.1) one can easily 
derive that Eq. (2.66) is nothing but the $O(\alpha_s)$ approximation to
the following expression
\begin{eqnarray}
F_{i,c}^{\rm ACOT}(x,Q^2,m_c^2)
=& F_{i,c}^{\rm VFNS}(x,Q^2)+F_{i,c}^{\rm exact}(x,Q^2,m_c^2)
 \nonumber \\ & \qquad -F_{i,c}^{\rm asymp}(x,Q^2,m_c^2).
\end{eqnarray}
Up to $O(\alpha_s)$, $F_{i,c}^{\rm VFNS}$ 
and $F_{i,c}^{\rm exact}-F_{i,c}^{\rm asymp}$
correspond with the first and the last part of (2.66) respectively (for the 
last part see also (2.67)). In the previous figures we have seen that for $Q^2
\ge 20$$(\rm GeV/c)^2$, $F_{i,c}^{\rm exact}$ approaches $F_{i,c}^{\rm asymp}$
so that $F_{i,c}^{\rm ACOT}$ coincides with $F_{i,c}^{\rm VFNS}$. Therefore 
$F_{i,c}^{\rm ACOT}$ gives a good description of the charm structure function
at large $Q^2$-values. However at small $Q^2$, $F_{i,c}^{\rm VFNS}\not=
F_{i,c}^{\rm asymp}$ (see Figs .12b-e) 
so that $F_{i,c}^{\rm ACOT}$ is not 
dominated by $F_{i,c}^{\rm exact}$. 
Therefore $F_{i,c}^{\rm ACOT}$ (3.8) does not have
the correct threshold behaviour exhibited by $F_{i,c}^{\rm exact}$. 
The inequality
between $F_{i,c}^{\rm VFNS}$ and $F_{i,c}^{\rm asymp}$ is due 
to the large value of $\alpha_s(Q^2)$ and the different ways 
that the corrections beyond $O(\alpha_s)$
have been included in the latter two functions. 
In our case $F_{i,c}^{\rm asymp}$
is computed up $O(\alpha_s^2)$ whereas $F_{i,c}^{\rm VFNS}$ contains even
$O(\alpha_s^3)$ contributions. If $F_{i,c}^{\rm VFNS}$ is replaced by 
$F_{i,c}^{\rm PDF}$ the above inequality 
becomes even larger since the latter includes
all leading contributions in all orders of $\alpha_s$. Due to
these considerations and the fact that the large logarithms only show
up at large $Q^2$ the expressions for $F_{i,c}^{\rm VFNS}$ 
and $F_{i,c}^{\rm asymp}$
have no physical meaning in the threshold region (low $Q^2$). Hence they
should be dropped in the latter region so that here the charm structure
function is only represented by $F_{i,c}^{\rm exact}$. In fact the 
EMC data on charm production were recently reexamined by \cite{hsv} using
the NLO results for $F_{2,c}^{\rm exact}(x,Q^2,m_c^2)$. 
The theoretical results are in excellent agreement
with the experimental data, except for one data point.

We have also studied the total structure function $F_2(x,Q^2)$,
where the charm component is included. 
Since this structure function is dominated by the light-parton 
(u,d,s and g) contributions the differences between the various
descriptions is much smaller than those observed for the charm component.
At maximum these differences are of the order of $10 \%$, which
occurs in the region $5<Q^2<10$ $({\rm GeV}/c)^2$.

To study the stability of the perturbation 
series for $F_{i,c}$ one can proceed 
in two different ways. The first one is discussed in \cite{acot} and concerns
the behaviour of the charm structure function with respect to variations in the
factorization scale. It was found that near threshold (large $x$ and small 
$Q^2$) $F_{2,c}^{\rm exact}$ shows a better
stability than $F_{2,c}^{\rm VFNS}$
under variations of the factorization scale. 
Far above threshold it turns out
that just the opposite happens, so that at large $Q^2$ it is more preferable 
to use $F_{2,c}^{\rm VFNS}$ instead of $F_{2,c}^{\rm exact}$ (FOPT). 
However in the analysis of \cite{acot} the NLO corrections from \cite{lrsn}
were not taken into account. In \cite{grs},\cite{Vogt} and \cite{or} these
corrections were included and one could show that far away from threshold 
$F_{2,c}^{\rm exact}$ is as stable as $F_{2,c}^{\rm VFNS}$ with respect to
variations in the factorization scale. The second way to study the stability
of the perturbation series is to look at the actual size of the higher order
corrections. They have to decrease when the order in $\alpha_s$ increases.
To be more specific we study the following quantities
\begin{equation}
  K_{i,c}^{(l)} = \frac{F_{i,c}^{(l)}(x,Q^2)}
{F_{i,c}^{(l-1)}(x,Q^2)}
\,,
\end{equation}
where $F_{i,c}^{(l)}$ denotes the $O(\alpha_s^l)$ corrected charm structure
function. Our criterion is that the perturbation series gets more stable
if $K_{i,c}^{(l)}\rightarrow 1$ for increasing $l$. 
Here we want to compare $K_{2,c}^{(1),\rm exact}$ (FOPT) with 
$K_{2,c}^{(l),\rm VFNS}$ ($l=1,2$),
which are derived from $F_{2,c}^{\rm exact}$ and
$F_{2,c}^{\rm VFNS}$ respectively. 
In virtue of the observations made in Figs. 12a-e
we replace the latter by $F_{2,c}^{\rm PDF}$ 
because this does not appreciably change the results which we now present. 
In Figs. 13a-g we plot $K_{2,c}^{(1),\rm exact}$,
$K_{2,c}^{(1),\rm PDF}$ and $K_{2,c}^{(2),\rm PDF}$. Here 
$K_{2,c}^{(1),\rm exact}$ is given by the ratio of the NLO 
over the  LO charm 
structure functions in FOPT using the GRV92-set \cite{grv92}. The same holds
for $K_{2,c}^{(1),\rm PDF}$ which is obtained from the NLO over 
LO approximations to Eq. (2.60). 
Finally $K_{2,c}^{(2),\rm PDF}$ stands for the ratio of the
next-to-next-to-leading-order (NNLO) and NLO corrected 
charm structure functions, where in NNLO only the
light-parton coefficient functions of \cite{zn1} have been included. Notice
that the NNLO parton densities are not known yet because
the three-loop splitting functions (anomalous dimensions) have not been
calculated. From Fig. 13a we infer that for $Q^2=3$ $({\rm GeV}/c)^2$, 
$K_{2,c}^{(1),\rm exact}$ is rather close to unity over the whole $x$-range
contrary to $K_{2,c}^{(l),\rm PDF}$ $(l=1,2)$ which deviate from unity in  
spectacular ways. 
This shows that one should use FOPT instead of
VFNS at small $Q^2$.
At $Q^2=5$ $({\rm GeV}/c)^2$ (see Fig. 13b) the deviation from unity
becomes the same for $K_{2,c}^{(1),\rm exact}$ and $K_{2,c}^{(1),\rm PDF}$
in the small $x$-region. However at large $x$ ($x>0.01$), which represents the 
threshold region, the description by FOPT is still superior to the one given 
by VFNS. This picture changes when $Q^2>10$ $({\rm GeV}/c)^2$ (Fig. 13c).
In the threshold region ($x>0.01$) both approaches are equally bad whereas at
small $x$ the VFNS is better than FOPT. The superiority of the 
VFNS with respect to the
FOPT approach becomes even clearer when one looks at Eq. (3.9) plotted
at still larger $Q^2$-values, which are probably not realizable experimentally,
in Figs. 13d-g. This holds for all $x$-values
including the threshold region. We also notice a considerable improvement
when the order $\alpha_s^2$ corrections are included in $F_{2,c}^{\rm VFNS}$.
In particular $K_{2,c}^{(2),\rm PDF}$ is much closer
to one than $K_{2,c}^{(1),\rm PDF}$ 
at larger $Q^2$-values, indicating the rapid convergence of
the perturbation series in case of VFNS. From the 
observations made above we conclude
that Eq. (3.9) provides us with a better criterion to determine the predictive
power of the perturbation series than the investigation of the stability
of the charm structure function under changes in the factorization scale. It
shows very clearly that for $Q^2>10$ $({\rm GeV}/c)^2$ it is better to use
$F_{2,c}^{\rm VFNS}$ instead of $F_{2,c}^{\rm exact}$ even when $x$ gets large.

The most important results found in this paper can be summarized as follows.
When the charm structure function $F_{2,c}$ is computed 
in FOPT for $Q^2>20$ $({\rm GeV}/c)^2$ 
the results obtained from the exact ($F_{2,c}^{\rm exact}$)
and the asymptotic heavy-quark coefficient functions ($F_{2,c}^{\rm asymp}$)
are indistinguishable. For $F_{L,c}$ the minimal value of $Q^2$ becomes
larger, say around $Q^2=10^3$ $({\rm GeV}/c)^2$. Above this minimal
value the large logarithms dominate the perturbation series and they can be
resummed after having performed mass factorization on the heavy-quark 
coefficient functions. In this way starting from  
$F_{2,c}^{\rm asymp}$ in the three-flavour scheme
one can derive an expression for $F_{2,c}^{\rm VFNS}$ valid
in the four-flavour scheme. This procedure imposes a relation between the
parton density sets parametrized at three and four flavours, which has to be
satisfied in the VFNS approach. In the literature the parton density sets 
(PDF's) do not obey this requirement although in practice this has no serious
consequences for the prediction of $F_{2,c}^{\rm VFNS}$, which is almost
equal to $F_{2,c}^{\rm PDF}$. Further it turns out that $F_{2,c}^{\rm VFNS}
>F_{2,c}^{\rm exact}$ for all $Q^2$-values. This can be attributed to the 
higher-order corrections appearing beyond FOPT which are included in VFNS.
Finally we have shown that comparisons between 
the charm structure functions calculated in different orders in
$\alpha_s$ give better indications 
about the stability of the perturbation series than a variation
in the mass-factorization scale. It turns out that below $Q^2=10$ 
$({\rm GeV}/c)^2$, $F_{2,c}^{\rm exact}$ (FOPT) gives a better description
of the charm structure function whereas above this value it is much better
to use $F_{2,c}^{\rm VFNS}$. In particular this holds in the small $x$-region
under investigation by the HERA experiments.

We stress again that the VFNS is only valid for totally inclusive
quantities. At the level of exclusive distributions the large
logarithms we have been discussing simply do not exist. If
the experimental analysis is carried out 
on the basis of the photon-gluon fusion model in NLO
with three light flavours then we expect that
the observed charm contribution $F_{2,c}(x,Q^2,m^2)$
should agree with the NLO results
$F_{2,c}^{\rm exact}(x,Q^2,m^2)$ for small $Q^2$ 
(say $Q^2 = 5$ $({\rm GeV}/c)^2$) 
and with $F_{2,c}^{\rm VFNS}(x,Q^2,m^2)$ for large $Q^2$
(say $Q^2 = 100$ $({\rm GeV}/c)^2$). 

Acknowledgements.

W.L. van Neerven enjoyed some discussions about the charm-quark density with
A. Vogt. 
The research of J. Smith and Y. Matiounine was partially supported by the 
contract NSF 93-09888.
J. Smith would like to thank the Alexander von Humbolt Stiftung for an
award to allow him to spend his Sabbatical leave at DESY.

\appendix
\mysection*{Appendix A}
\setcounter{section}{1}

In this appendix we present the unrenormalized operator-matrix elements
 $\hat A_{gq,H}^{{\rm S},(2)}$, $\hat A_{gg,H}^{{\rm S},(1)}$ and 
$\hat A_{gg,H}^{{\rm S},(2)}$ where $\hat A_{kl,H}^{{\rm S},(i)}$
denote the coefficients of $(\alpha_s/4\pi)^i$ in the
perturbation series of the operator-matrix elements (OME's).
The corresponding two-loop
Feynman graphs are presented in Fig. 3 and Fig. 4 respectively and the
calculation proceeds in the same way as is outlined for the other
OME's in \cite{bmsmn}. Using $n$-dimensional regularization for the ultraviolet
and collinear divergences the results are given by ($n=4 + \epsilon$)

\begin{eqnarray}
&& \hskip-0.80cm \hat A_{gq,H}^{{\rm S},(2)}\Big(\frac{m^2}{\mu^2},
\epsilon\Big)
=S_{\epsilon}^2\Big(\frac{m^2}
{\mu^2}\Big)^{\epsilon} C_FT_f \Biggl\{ \frac{1}{\epsilon^2} \Biggl[
\frac{64}{3 z} - \frac{64}{3} + \frac{32}{3} z \Biggr]
\nonumber \\ && \qquad
+ \frac{1}{\epsilon} \Biggl[\frac{160}{9 z} - \frac{160}{9} + \frac{128}{9} z
+\Biggl (\frac{32}{3 z} - \frac{32}{3} + \frac{16}{3} z \Biggr)  \ln(1 - z) \Biggr]
\nonumber \\ && \qquad
+ \frac{4}{3} \Biggl(\frac{2}{z} - 2 + z \Biggr) \ln^2(1 - z)
+ \frac{8}{9} \Biggl( \frac{10}{z} - 10 + 8 z \Biggr) \ln(1 - z)
\nonumber \\ && \qquad
+ \frac{8}{3} \Biggl(\frac{2}{z} - 2 + z \Biggr) \zeta(2)
+ \frac{1}{27} \Biggl(\frac{448}{z} - 448 + 344 z \Biggr )
\Biggr\} \, .
\end{eqnarray}
Here $S_\epsilon$ denotes the spherical factor which is given by
\begin{eqnarray}
S_\epsilon = \exp \Big\{ \frac{\epsilon}{2} (\gamma_E - \ln 4\pi) \Big\}\,,
\end{eqnarray}
where $\gamma_E$ is the Euler constant. 
\begin{eqnarray}
 \hskip-3cm \hat A_{gg,H}^{{\rm S},(1)} \Big(\frac{m^2}{\mu^2},\epsilon\Big)
= S_{\epsilon}\Big(\frac{m^2}{\mu^2}\Big)^{\frac{\epsilon}{2}}
\Biggl[\frac{1}{\epsilon}T_f\Biggl(\frac{8}{3}\delta(1-z) \Biggr) \Biggr] \, ,
\end{eqnarray}

\begin{eqnarray}
&&\hat A_{gg,H}^{{\rm S},(2)}\Big(\frac{m^2}{\mu^2},\epsilon\Big)
=S_{\epsilon}^2\Big(\frac{m^2}
{\mu^2}\Big)^{\epsilon}
\Biggl[\frac{1}{\epsilon^2}\Biggl\{C_FT_f\Biggl[32 (1+z)\ln z
+\frac{64}{3 z} + 16
\nonumber \\ && \qquad
- 16 z - \frac{64}{3} z^2  \Biggr]
+C_AT_f\Biggl[
\frac{32}{3}\Biggl(\frac{1}{1-z}\Biggr)_+
+ \frac{32}{3 z} - \frac{64}{3} + \frac{32}{3} z - \frac{32}{3} z^2
\Biggr]\Biggr\}
\nonumber \\ && \qquad
+\frac{1}{\epsilon}\Biggl\{C_FT_f \Biggl[
8 (1 + z) \ln^2 z + (24 + 40 z) \ln z - \frac{16}{3 z} + 64
- 32 z
\nonumber \\ && \qquad
- \frac{80}{3} z^2  + 4 \delta(1-z)\Biggr]
+ C_AT_f\Biggl[
\frac{16}{3} (1 + z) \ln z + \frac{80}{9}\Biggl(\frac{1}{1-z}\Biggr)_+
\nonumber \\ && \qquad
+ \frac{184}{9 z} - \frac{232}{9} + \frac{152}{9} z
- \frac{184}{9} z^2 + \frac{16}{3} \delta(1- z)
 \Biggr]\Biggr\}
+a^{(2)}_{gg,H}(z)\Biggr] \,,
\nonumber \\ && \qquad
+\sum_{f = H}^t S_{\epsilon}^2\Big(
\frac{m_f^2}{\mu^2}\Big)^{\epsilon/2} \Big( \frac{m^2}{\mu^2}\Big)^{\epsilon/2}
\Biggl[\frac{1}{\epsilon^2}T_f^2\Biggl\{
\frac{64}{9}\Big(1 + \frac{\epsilon^2}{4}\zeta(2)\Big)\delta(1 - z)\Biggr\} 
\Biggr] \,,
\end{eqnarray}

with
\begin{eqnarray}
&&a^{(2)}_{gg,H}(z)=C_FT_f\Biggl\{
\frac{4}{3}(1 + z)\ln^3 z
+(6 + 10 z) \ln^2 z + (32 + 48 z) \ln z
\nonumber \\ && \qquad
+ 8 (1 + z) \zeta(2) \ln z
+\Biggl(\frac{16}{3 z} + 4 - 4 z - \frac{16}{3} z^2 \Biggr)\zeta(2)
\nonumber \\ && \qquad
- \frac{8}{z} + 80 - 48 z - 24 z^2 - 15 \delta(1 - z)
\Biggr\}
\nonumber \\ && \qquad
+C_AT_f\Biggl\{\frac{4}{3}(1 + z)\ln^2 z
+ \frac{1}{9} (52 + 88 z) \ln z
- \frac{4}{3} z \ln(1-z)
\nonumber \\ && \qquad
+ \frac{8}{3} \Biggl[ \Biggl(\frac{1}{1-z}\Biggr)_+
+ \frac{1}{z} - 2 + z - z^2 \Biggr] \zeta(2)
\nonumber \\ && \qquad
+\frac{1}{27} \Biggl[224 \Biggl(\frac{1}{1-z}\Biggr)_+ + \frac{556}{z}
- 628 + 548 z - 700 z^2 \Biggr]
\nonumber \\ && \qquad
+ \frac{10}{9} \delta(1 - z)
\Biggr\} \,.
\end{eqnarray}
The last term in Eq. (A.4), which is proportional to $T_f^2$, is due to the
one-loop correction to $\hat A_{gg,H}^{{\rm S},(1)}$ in Eq. (A.3). This
correction is represented by the heavy quark ($f$) loop contribution to the
gluon self energy where $f$ represents all heavy flavours starting with
the quark $H$ ($m_H \equiv m$)
and ending with the top quark $t$.
The corresponding graph is not shown in Fig. 4. In the next section
the renormalization of the OME $\hat A_{gg,H}^{\rm S}$ will be chosen in such
a way that the heavy quarks with $m_f > m$ decouple from the running
coupling constant. This implies that the contributions in the sum of Eq. (A.4)
with $f > H$ completely vanish in the renormalized OME 
$\hat A_{gg,H}^{\rm S}$
presented in the next section. However the contribution due to $f=H$
remains in the renormalized expressions like those coming from the $n_f$ light
flavours. This renormalization prescription implies that the running
coupling constant is presented in the ${\overline {\rm MS}}$-scheme and it 
depends on $n_f+1$ light flavours including the heavy quark $H$.

The $1/(1-z)_+$ terms appearing in Eqs. (A.4) and (A.5) have to be 
understood as distributions, namely 
\begin{eqnarray}
%
\int^1_0 \,dz \Biggl(\frac{1}{1-z}\Biggr)_+ f(z)
=\int_0^1dz\frac{1}{1-z}[f(z)-f(1)] \,.
\end{eqnarray}
The colour factors in $SU(N)$ are given by
%
\begin{eqnarray}
C_F=\frac{N^2-1}{2N} \qquad  C_A=N  \qquad T_f=\frac{1}{2} \, ,
\end{eqnarray}
with $N=3$ for QCD.

\mysection*{Appendix B}
\setcounter{section}{2}

Here we present the renormalized operator-matrix elements (OME's)
corresponding to the unrenormalized expressions 
given in Appendix C of \cite{bmsmn} and in Appendix A of this paper. 
All OME'S have been renormalized in the ${\overline {\rm MS}}$-scheme.
In particular the renormalized coupling constant is presented in the
above scheme for $n_f+1$ light flavours. Here the heavy quark $H$ is treated
on the same footing as the light flavours and it is not decoupled
from the running coupling constant in the VFNS approach. 
The $(\alpha_s/4\pi)^2$ coefficient in the heavy-quark OME 
$\tilde A^{\rm PS}_{Hq}$ is given by

\begin{eqnarray}
&& \hskip-0.5cm \tilde A^{{\rm PS},(2)}_{Hq}\Biggl(\frac{m^2}{\mu^2}\Biggr)=
C_FT_f\Biggl\{
\Biggl[-8(1+z)\ln z-\frac{16}{3z}-4
\nonumber \\ && \qquad
+ 4 z +\frac{16}{3}z^2\Biggr] \ln^2\frac{m^2}{\mu^2}
+\Biggl[8(1+z)\ln^2z-\Biggl(8+40z+\frac{64}{3}z^2\Biggr)\ln z
\nonumber \\ && \qquad
-\frac{160}{9z}
+16-48z+\frac{448}{9}z^2\Biggr] \ln\frac{m^2}{\mu^2}
\nonumber \\ && \qquad
+ (1+z)\Biggl[32{\rm S}_{1,2}(1-z)+16\ln z{\rm Li}_2(1-z)
-16\zeta(2)\ln z
\nonumber \\ && \qquad
-\frac{4}{3}\ln^3z\Biggr]
+\Biggl(\frac{32}{3z}+8-8z-\frac{32}{3}z^2\Biggr) {\rm Li}_2(1-z)
\nonumber \\ && \qquad
+ \Biggl( -\frac{32}{3 z}-8+8z+\frac{32}{3} z^2\Biggr)\zeta(2)
+\Biggl(2+10z+\frac{16}{3}z^2\Biggr)  \ln^2z
\nonumber \\ && \qquad
-\Biggl(\frac{56}{3}+\frac{88}{3}z
+\frac{448}{9}z^2\Biggr)\ln z
-\frac{448}{27z} - \frac{4}{3}
-\frac{124}{3}z+\frac{1600}{27}z^2 \Biggr\}  \,,
\end{eqnarray}

The $(\alpha_s/4\pi)$ and the $(\alpha_s/4\pi)^2$ coefficients of the
heavy quark OME $\tilde A^{\rm S}_{Hg}$ are

\begin{eqnarray}
\hskip-3cm \tilde A_{Hg}^{{\rm S},(1)} \Biggl(\frac{m^2}{\mu^2}\Biggr)
= T_f \Biggl[ - 4 ( z^2 + (1 - z)^2)\ln\frac{m^2}{\mu^2}\Biggr] \, ,
\end{eqnarray}
and

\begin{eqnarray}
&& \tilde A_{Hg}^{{\rm S},(2)}\Biggl(\frac{m^2}{\mu^2}\Biggr)
=
\Biggl\{C_FT_f[ (8 -16 z+16 z^2)\ln(1-z)
\nonumber \\ && \qquad
-(4 -8 z+ 16 z^2)\ln z -(2 - 8 z)]
\nonumber \\ && \qquad
+C_AT_f\Biggl[-(8 - 16 z + 16 z^2)\ln(1-z)
-(8 + 32 z)\ln z
\nonumber \\ && \qquad
 -\frac{16}{3z} -4  - 32 z+\frac{124}{3}z^2\Biggr]
+ T_f^2 \Biggl[ - \frac{16}{3} ( z^2 + (1 - z)^2) \Biggr]
\Biggr\} \ln^2\frac{m^2}{\mu^2}
\nonumber \\ && \qquad
+\Biggl\{C_FT_f \Biggl[( 8 - 16 z + 16z^2)[2\ln z\ln(1-z)
-\ln^2(1-z)+2\zeta(2)]
\nonumber \\ && \qquad
-(4 - 8 z +16 z^2)\ln^2z-32z(1-z)\ln(1-z)
\nonumber \\ && \qquad
-(12 - 16 z + 32 z^2)\ln z  - 56+116z -80z^2 \Biggr]
\nonumber \\ && \qquad
+ C_AT_f\Biggl[(16 +32 z +32 z^2)[{\rm Li}_2(-z) + \ln z\ln(1+z) ]
\nonumber \\ && \qquad
+(8 - 16 z + 16 z^2)\ln^2(1-z)
+(8 + 16 z)\ln^2z
\nonumber \\ && \qquad
+32z\zeta(2)+32z(1-z)\ln(1-z)
-\Biggl(8+64z+\frac{352}{3}z^2\Biggr)\ln z
\nonumber \\ && \qquad
-\frac{160}{9z}+16-200z+\frac{1744}{9}z^2 \Biggr]\Biggl\} \ln \frac{m^2}{\mu^2}
\nonumber \\ && \qquad
+ C_FT_f\Big\{(1-2z+2z^2)[8\zeta(3)
+\frac{4}{3}\ln^3(1-z)
\nonumber \\ && \qquad
-8\ln(1-z){\rm Li}_2(1-z)
+8\zeta(2)\ln z
-4\ln z\ln^2(1-z)
\nonumber \\ && \qquad
+\frac{2}{3}\ln^3z
-8\ln z{\rm Li}_2(1-z)
+8{\rm Li}_3(1-z)
-24{\rm S}_{1,2}(1-z)]
\nonumber \\ && \qquad
+z^2\Biggl[-16\zeta(2)\ln z+\frac{4}{3}\ln^3z
+16\ln z{\rm Li}_2(1-z)+32{\rm S}_{1,2}(1-z)\Biggr]
\nonumber \\ && \qquad
-(4+96z-64z^2){\rm Li}_2(1-z)
-(4-48z+40z^2)\zeta(2)
\nonumber \\ && \qquad
-(8+48z-24z^2)\ln z\ln(1-z)
+(4+8z-12z^2)\ln^2(1-z)
\nonumber \\ && \qquad
-(1+12z-20z^2)\ln^2z-(52z-48z^2)\ln(1-z)
\nonumber \\ && \qquad
-(16+18z+48z^2)\ln z
+26-82z+80z^2\Biggr\}
\nonumber \\ && \qquad
+C_AT_f\Biggl\{(1-2z+2z^2) [
-\frac{4}{3} \ln^3(1-z)
\nonumber \\ && \qquad
+8\ln(1-z){\rm Li}_2(1-z)-8{\rm Li}_3(1-z)]
+(1+2z+2z^2)
\nonumber \\ && \qquad
\times [-8\zeta(2)\ln(1+z)
-16\ln(1+z){\rm Li}_2(-z)
-8\ln z\ln^2(1+z)
\nonumber \\ && \qquad
+4\ln^2z\ln(1+z) + 8\ln z{\rm Li}_2(-z)-8{\rm Li}_3(-z)
-16{\rm S}_{1,2}(-z)]
\nonumber \\ && \qquad
+(16+64z)[2{\rm S}_{1,2}(1-z)
+\ln z{\rm Li}_2(1-z)]
-\Biggl(\frac{4}{3} +  \frac{8}{3} z\Biggr)\ln^3z
\nonumber \\ && \qquad
+(8-32z+16z^2)\zeta(3)-(16+64z)\zeta(2)\ln z+(16+16z^2)
\nonumber \\ && \qquad
\times [ {\rm Li}_2(-z) + \ln z\ln(1+z)  ]
+\Biggl(\frac{32}{3z}+12+64z-\frac{272}{3}z^2\Biggr)
{\rm Li}_2(1-z)
\nonumber \\ && \qquad
-\Biggl( 12 + 48 z - \frac{260}{3} z^2+\frac{32}{3 z}\Biggr)\zeta(2)
-4z^2\ln z\ln(1-z)
\nonumber \\ && \qquad
-(2+8z-10z^2)\ln^2(1-z)+\Biggl(2+8z+\frac{46}{3}z^2\Biggr)\ln^2z
\nonumber \\ && \qquad
+(4+16z-16z^2)\ln(1-z)
-\Biggl(\frac{56}{3}+\frac{172}{3}z+\frac{1600}{9}z^2\Biggr)\ln z
\nonumber \\ && \qquad
-\frac{448}{27z}-\frac{4}{3}-\frac{628}{3}z
+\frac{6352}{27}z^2\Biggr\} \, ,
\end{eqnarray}
respectively.
Now we present the renormalized expressions for 
the heavy-quark loop contributions to the light-parton OME's denoted by
$A_{kl,H}$. The coefficients of the $(\alpha_s/4\pi)^2$ terms in 
$A_{qq,H}$ and $A_{gq,H}$ are
\begin{eqnarray}
&&  A^{{\rm NS},(2)}_{qq,H}\Biggl(\frac{m^2}{\mu^2}\Biggr)
=
 C_F T_f \Biggl\{\Biggr[
\frac{8}{3}\Biggl(\frac{1}{1-z}\Biggr)_+
-\frac{4}{3}-\frac{4}{3}z+2\delta(1-z)\Biggr] \ln^2\frac{m^2}{\mu^2}
\nonumber \\ && \qquad
+\Biggl[\frac{80}{9}\Biggl(\frac{1}{1-z}\Biggr)_+ +\frac{8}{3}
\frac{1+z^2}{1-z}\ln z+\frac{8}{9}-\frac{88}{9}z
\nonumber \\ && \qquad
+\delta(1-z)\Biggl(
\frac{16}{3}\zeta(2)+\frac{2}{3}\Biggr)\Biggr] \ln\frac{m^2}{\mu^2}
\nonumber \\ && \qquad
+\frac{1+z^2}{1-z}\Biggl(\frac{2}{3}\ln^2z+\frac{20}{9}\ln z\Biggr)
\nonumber \\ && \qquad
+\frac{8}{3}(1-z)\ln z
+\frac{224}{27}\Biggl(\frac{1}{1-z}\Biggr)_+
+\frac{44}{27}-\frac{268}{27}z
\nonumber \\ && \qquad
+\delta(1-z)\Biggl(-\frac{8}{3}\zeta(3)+\frac{40}{9}
\zeta(2)+\frac{73}{18}\Biggr) \Biggr\} \,,
\end{eqnarray}
and  
\begin{eqnarray}
&&A_{gq,H}^{{\rm S},(2)}\Biggl(\frac{m^2}{\mu^2}\Biggr) =
C_FT_f \Biggl\{ \Biggl[
\frac{16}{3 z} - \frac{16}{3} + \frac{8}{3} z \Biggr] \ln^2 \frac{m^2}{\mu^2}
\nonumber \\ && \qquad
+\Biggl[\frac{160}{9 z} - \frac{160}{9} + \frac{128}{9} z
+ (\frac{32}{3 z} - \frac{32}{3} + \frac{16}{3} z) \ln(1 - z) \Biggr]
\ln\frac{m^2}{\mu^2}
\nonumber \\ && \qquad
+ \frac{4}{3} \Biggl(\frac{2}{z} - 2 + z \Biggr) \ln^2(1 - z)
+ \frac{8}{9}  \Biggl(\frac{10}{z} - 10 + 8 z \Biggr) \ln(1 - z)
\nonumber \\ && \qquad
+ \frac{1}{27} \Biggl(\frac{448}{z} - 448 + 344 z \Biggr )
\Biggr\} \, .
\end{eqnarray}
respectively. The coefficients of the $\alpha_s/4\pi$ and
$(\alpha_s/4\pi)^2$ terms in $A_{gg,H}$ are 
\begin{eqnarray}
\hskip-3cm  A_{gg,H}^{{\rm S},(1)} \Biggl(\frac{m^2}{\mu^2}\Biggr)
=T_f\Biggl[\frac{4}{3}\delta(1-z) 
\ln\frac{m^2}{\mu^2} \Biggr] \, ,
\end{eqnarray}                      
and
\begin{eqnarray}
&&A_{gg,H}^{{\rm S},(2)}\Biggl(\frac{m^2}{\mu^2}\Biggr) =
\Biggl\{C_FT_f\Biggl[8 (1+z)\ln z
+\frac{16}{3 z} + 4
- 4 z - \frac{16}{3} z^2  \Biggr]
\nonumber \\ && \qquad
+C_AT_f \Biggl[
\frac{8}{3}\Biggl(\frac{1}{1-z}\Biggr)_+
+ \frac{8}{3 z} - \frac{16}{3} + \frac{8}{3} z - \frac{8}{3} z^2 \Biggr]
\nonumber \\ && \qquad
+T_f^2 \Biggl[\frac{16}{9}\delta(1-z) \Biggr]
\Biggr\} \ln^2 \frac{m^2}{\mu^2}
\nonumber \\ && \qquad
+\Biggl\{C_FT_f \Biggl[
8 (1 + z) \ln^2 z + (24 + 40 z) \ln z - \frac{16}{3 z} + 64
- 32 z
\nonumber \\ && \qquad
- \frac{80}{3} z^2  + 4 \delta(1-z)\Biggr]
+ C_AT_f\Biggl[
\frac{16}{3} (1 + z) \ln z + \frac{80}{9}\Biggl(\frac{1}{1-z}\Biggr)_+
\nonumber \\ && \qquad
+ \frac{184}{9 z} - \frac{232}{9} + \frac{152}{9} z
- \frac{184}{9} z^2 + \frac{16}{3} \delta(1- z)
 \Biggr]\Biggr\} \ln \frac{m^2}{\mu^2}
\nonumber \\ && \qquad
+ C_F T_f\Biggl\{
\frac{4}{3}(1 + z)\ln^3 z
+(6 + 10 z) \ln^2 z + (32 + 48 z) \ln z
\nonumber \\ && \qquad
\nonumber \\ && \qquad
- \frac{8}{z} + 80 - 48 z - 24 z^2 - 15 \delta(1 - z)
\Biggr\}
\nonumber \\ && \qquad
+C_AT_f\Biggl\{\frac{4}{3}(1 + z)\ln^2 z
+ \frac{1}{9} (52 + 88 z) \ln z
- \frac{4}{3} z \ln(1-z)
\nonumber \\ && \qquad
+\frac{1}{27} \Biggl[224 \Biggl(\frac{1}{1-z}\Biggr)_+ + \frac{556}{z}
- 628 + 548 z - 700 z^2 \Biggr]
\nonumber \\ && \qquad
+ \frac{10}{9} \delta(1 - z)
\Biggr\} \, ,
\end{eqnarray}
respectively.

The definitions for the polylogarithms ${\rm Li}_n(z)$ and the
Nielsen functions ${\rm S}_{n,p}(z)$, which appear in the
above expressions, can be found in \cite{lbmr}.
We have checked that the renormalized OME's given above satisfy the sum rules
presented in Eqs. (2.43) and (2.44). This provides us with a strong check
on our results in \cite{bmsmn} and in this paper.

%

\centerline{\bf \large{Figure Captions}}

\begin{description}
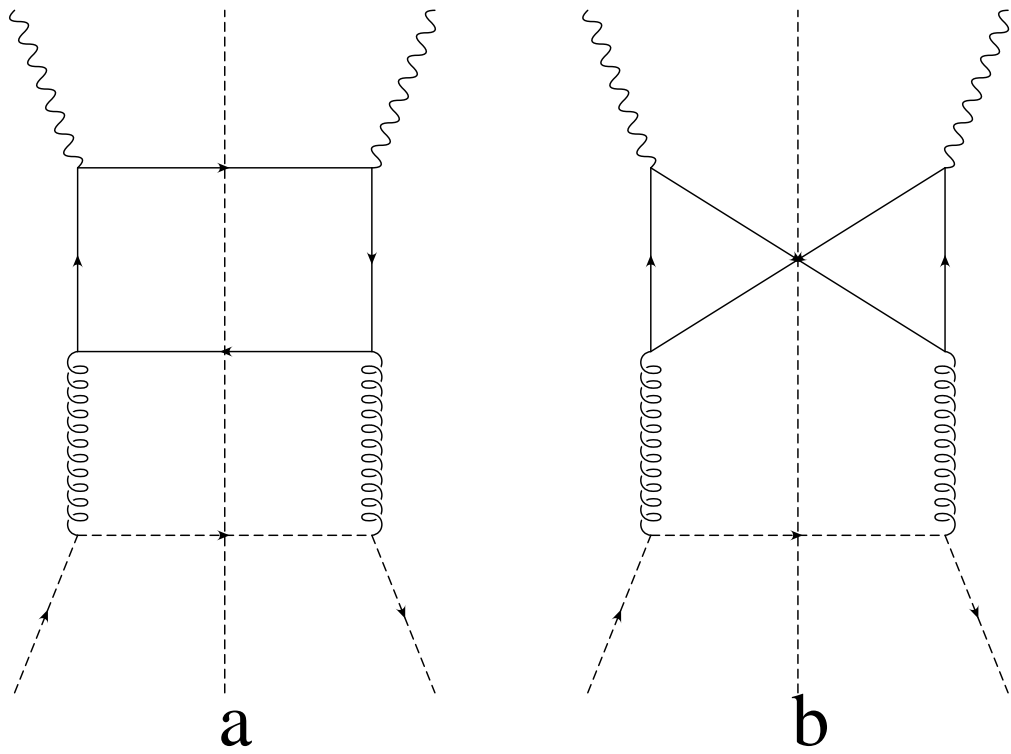
\item[Fig. 1.]
$O(\alpha_s^2)$ contributions to the purely-singlet parton structure function
${\cal F}_{i,q}^{\rm PS}$ 
representing the subprocess $\gamma^{*}$ + $q$ $\rightarrow$
$q$ $+$ $q'$ $+$ $\bar{q}'$. 
Here $q$ and $q'$ are represented by the dashed and solid lines
respectively. In the case of heavy-quark production $q'=H$ and these graphs
contribute to the heavy-quark coefficient function $H_{i,q}^{\rm PS}$.\\ 

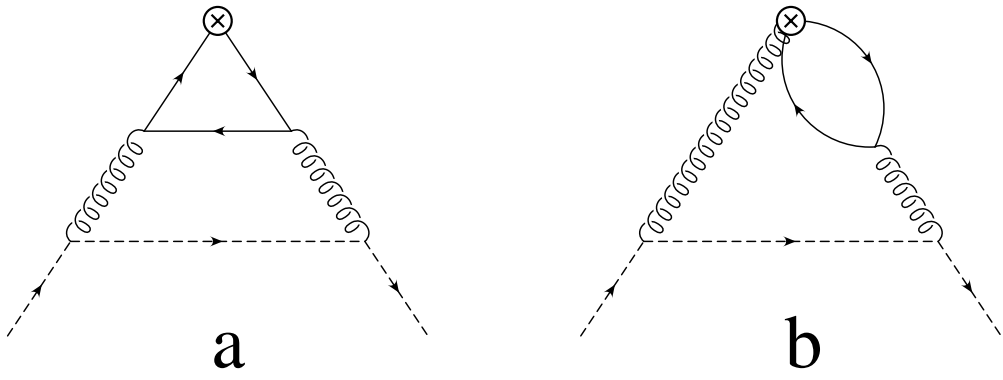
\item[Fig. 2.]
$O(\alpha_s^2)$ contributions to the purely-singlet OME $A_{q'q}^{\rm PS}$.
Here $q$ and $q'$ are represented by the dashed and solid lines
respectively. In the case of $q'=H$ these graphs
contribute to the heavy-quark OME $A_{Hq}^{\rm PS}$.\\

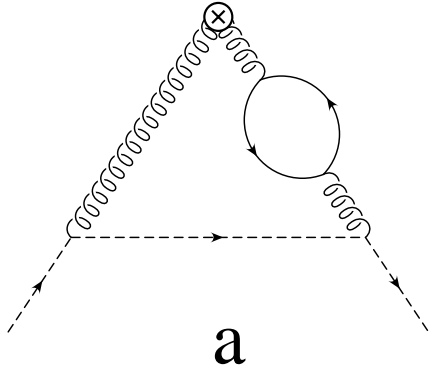
\item[Fig. 3.]
Two-loop contribution to the OME $A_{gq,H}^{\rm S}$. The dashed
and solid lines represent the light quark $q$ and the heavy quark $H$ 
respectively.\\

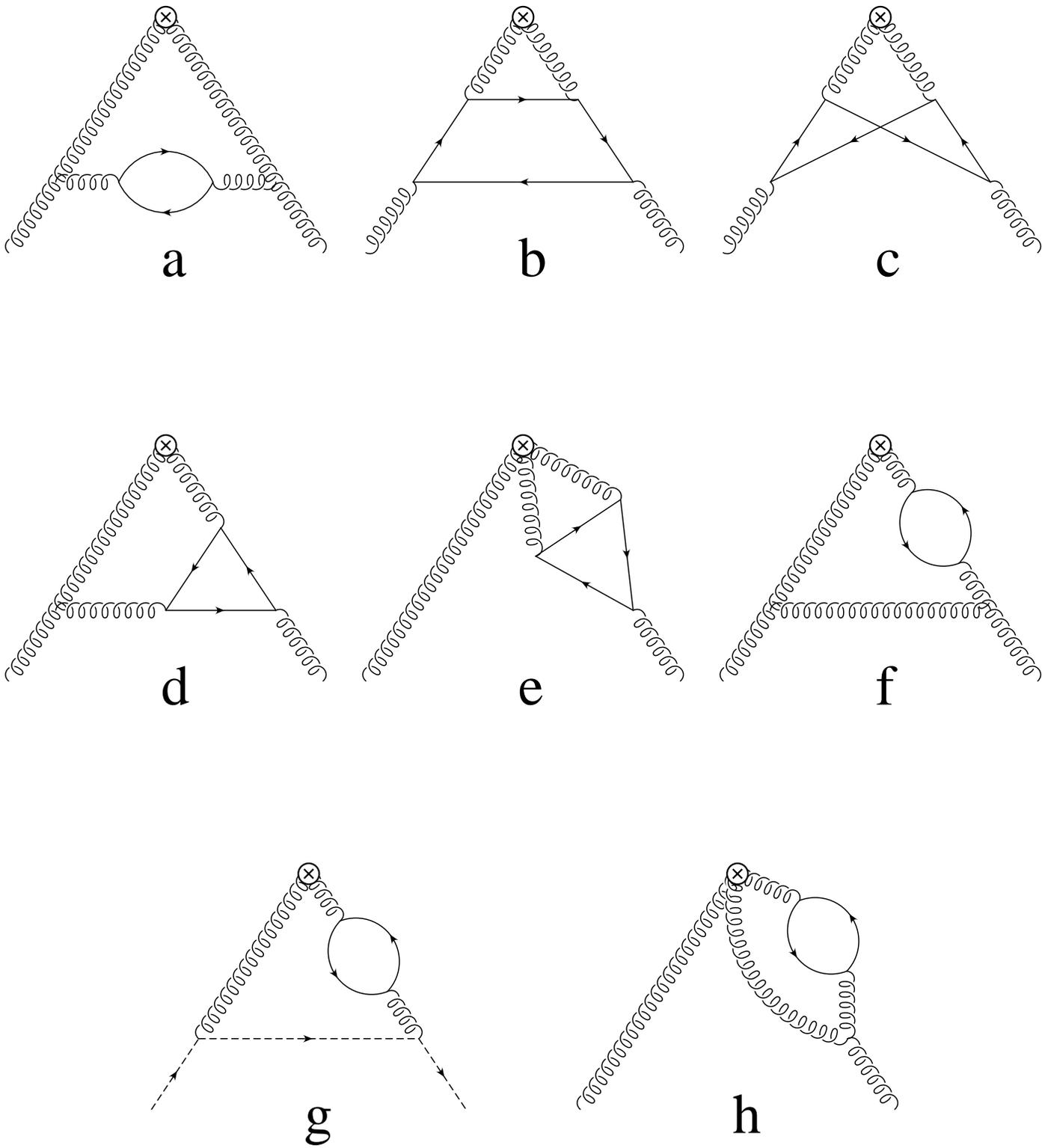
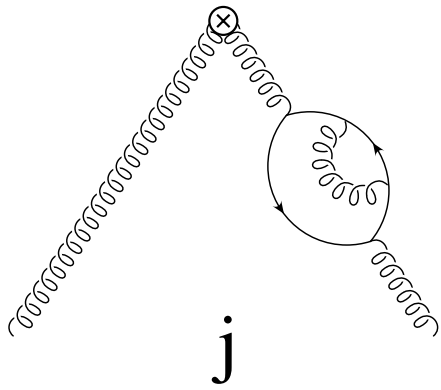
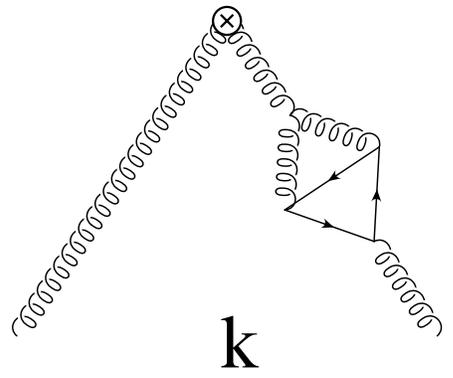
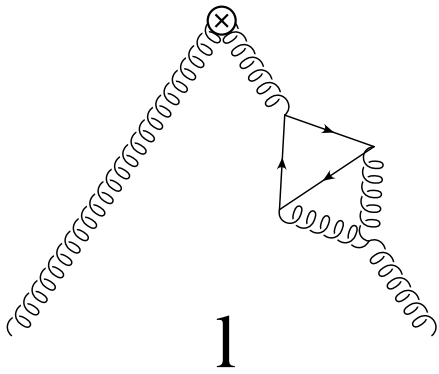
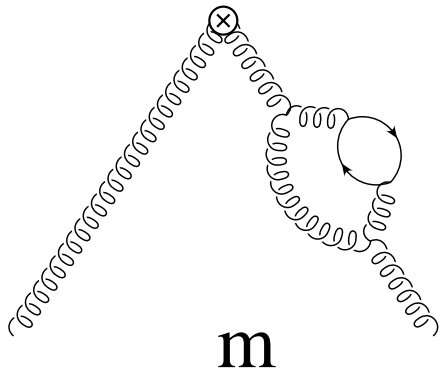
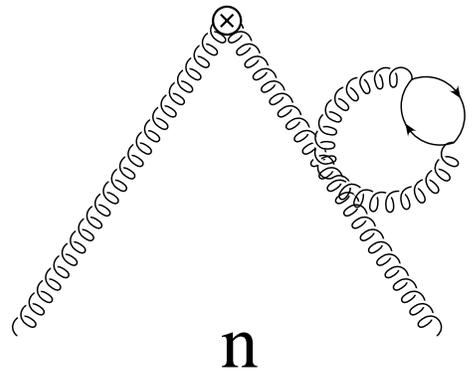
\item[Fig. 4.]
Two-loop graphs contributing to the OME $A_{gg,H}^{\rm S}$. 
The dashed and solid lines
represent the Faddeev-Popov ghost and the heavy quark H respectively.
The graph with the external Faddeev-Popov ghost (fig.4g) has to be included
if the sum over the gluon polarization states involves the contributions 
from unphysical polarizations.

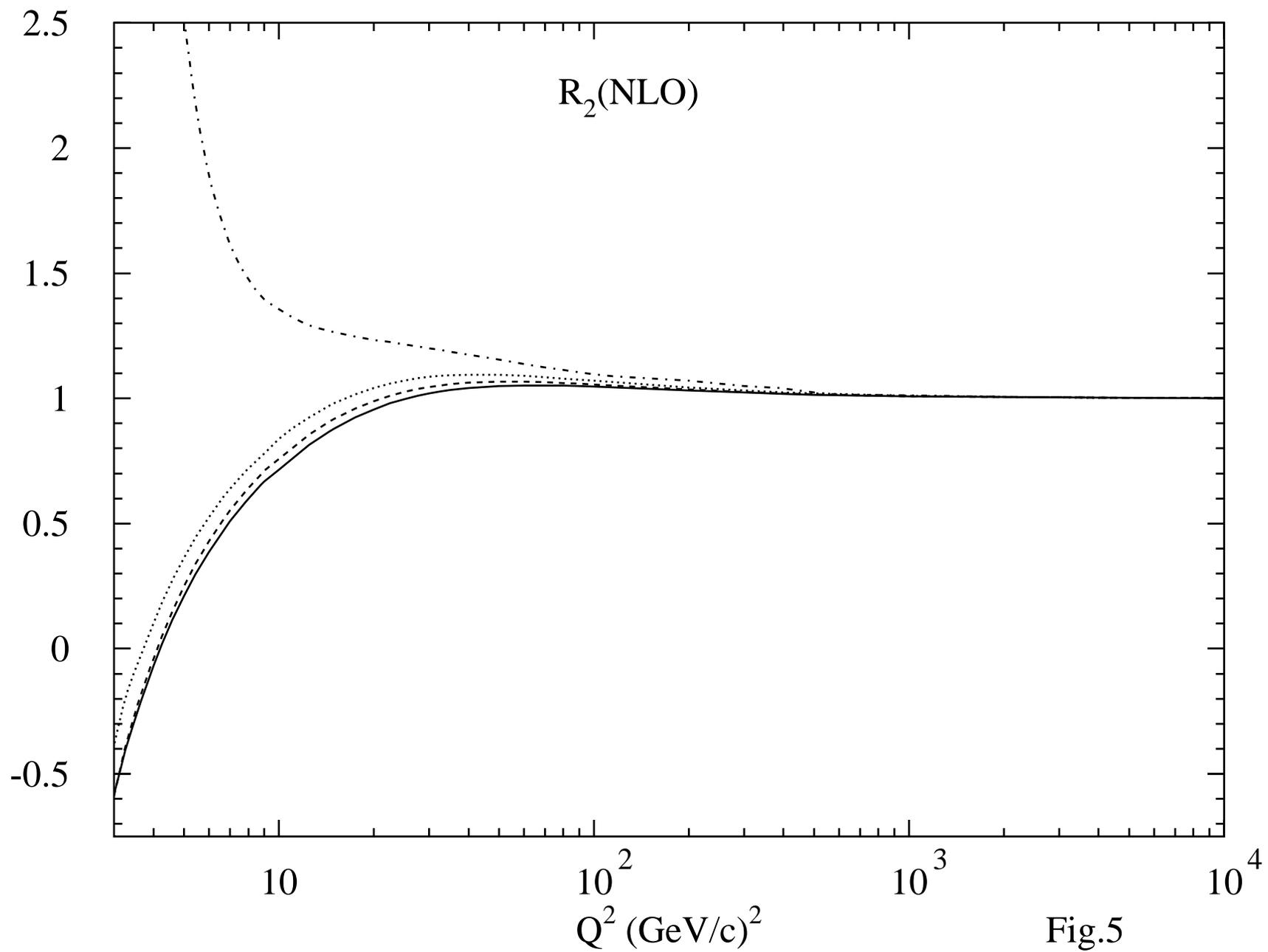
\item[Fig. 5.]
$R_2(NLO)$ (3.1)
plotted as a function of $Q^2$ at fixed $x$;
$x = 10^{-1}$  (dashed-dotted line), $x = 10^{-2}$ (dotted line), 
$x = 10^{-3}$ (dashed line) and $x = 10^{-4}$ (solid line).
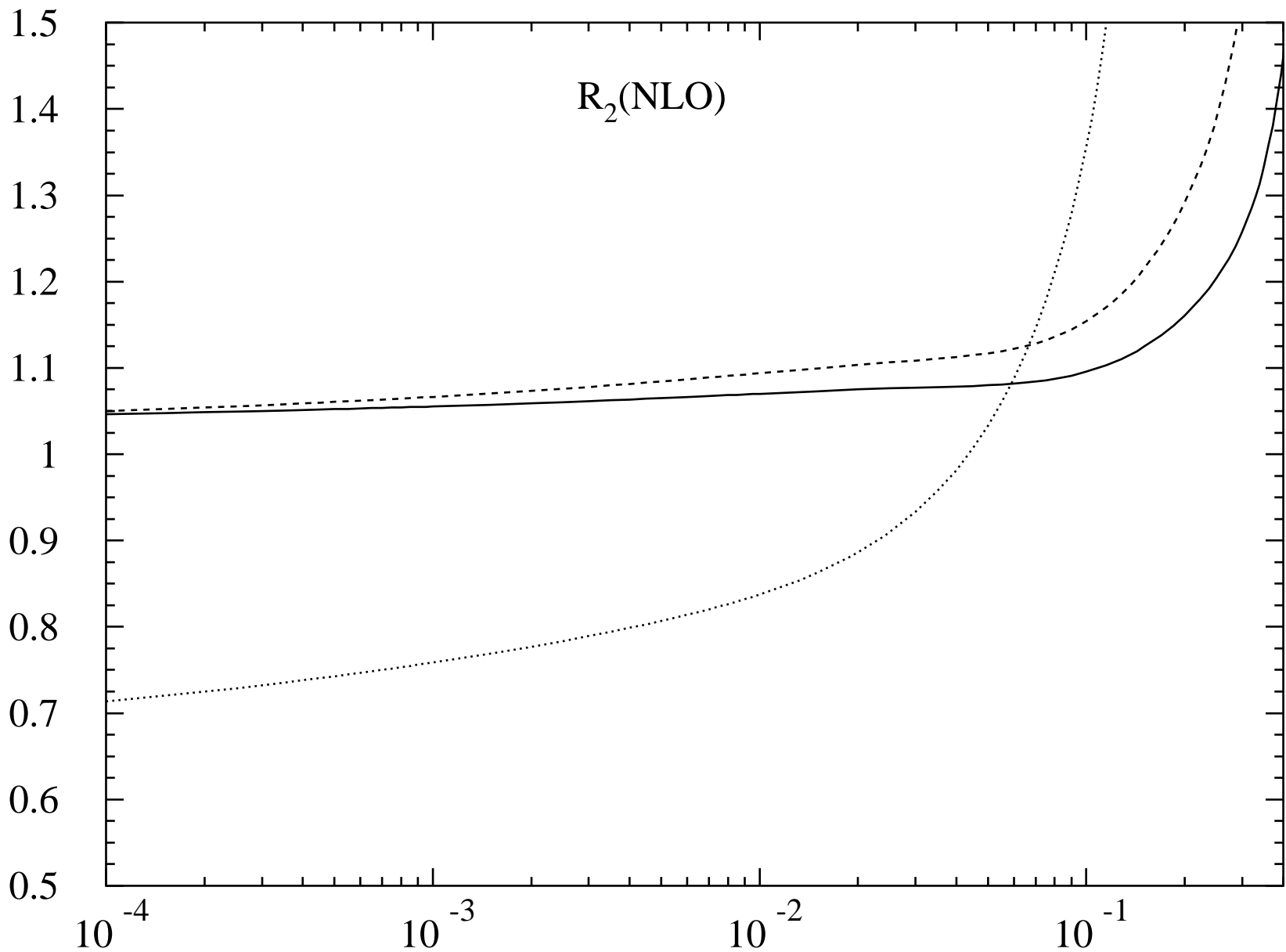
\item[Fig. 6.]
$R_2(NLO)$ (3.1)
plotted as a function of $x$ at fixed $Q^2$;
$Q^2 = 10$ (GeV/$c)^2$ (dotted line), $Q^2 = 50$ (GeV/$c)^2$
(dashed line), $Q^2 = 100$ (GeV/$c)^2$ (solid line).
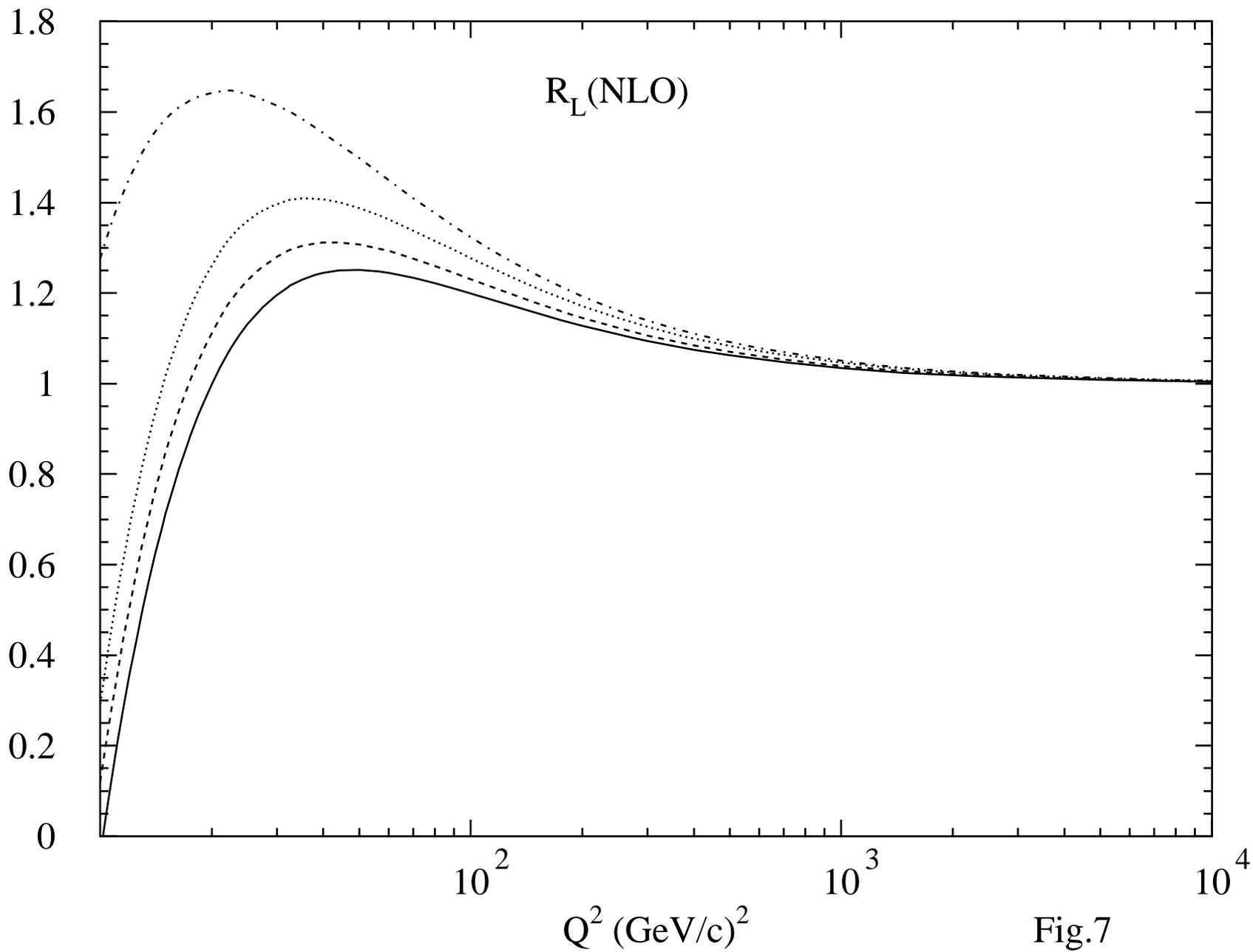
\item[Fig. 7.]
Same as in Fig. 5 but now for $R_L(NLO)$.

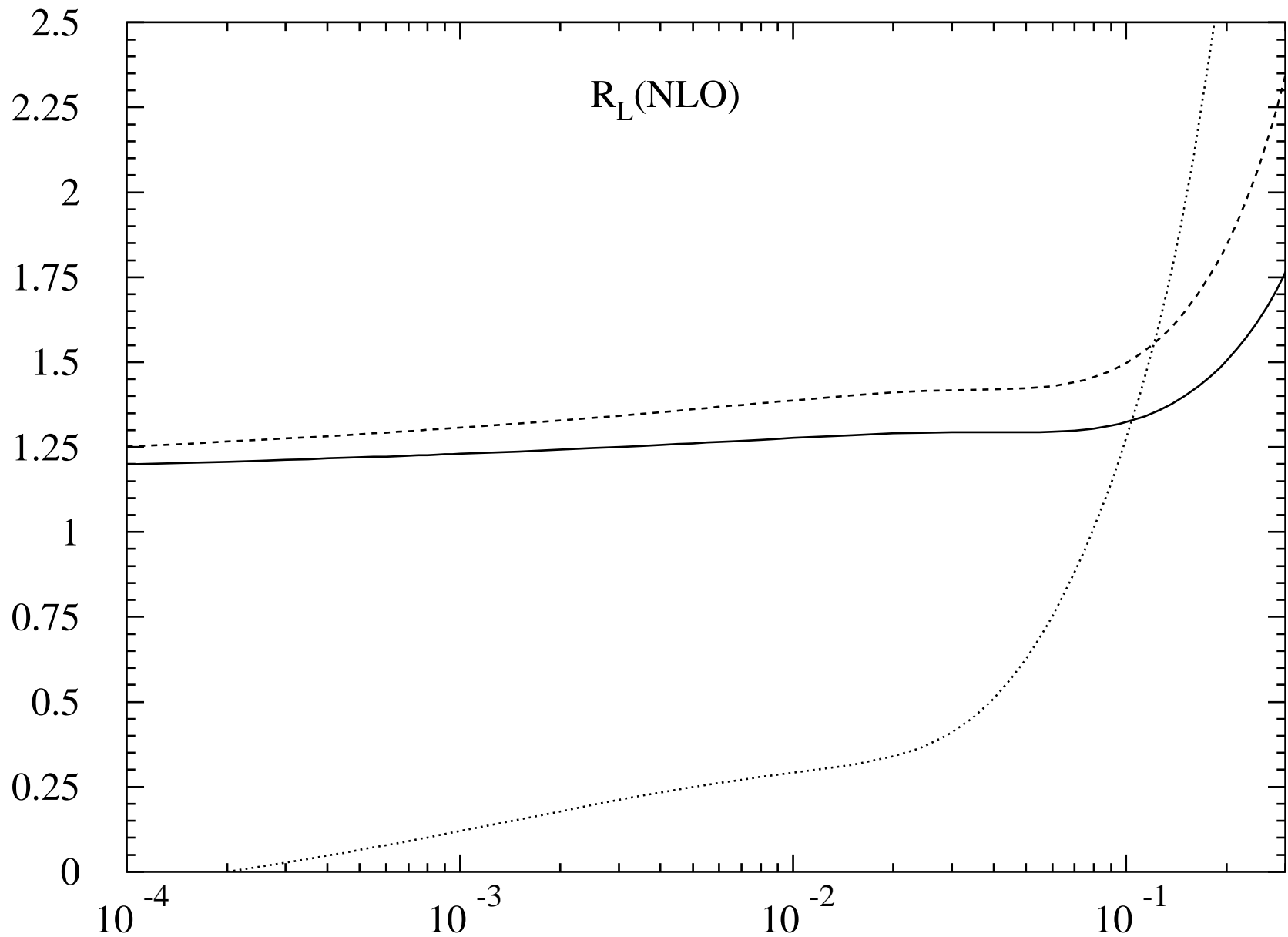
\item[Fig. 8.]
Same as in Fig. 6 but now for $R_L(NLO)$.
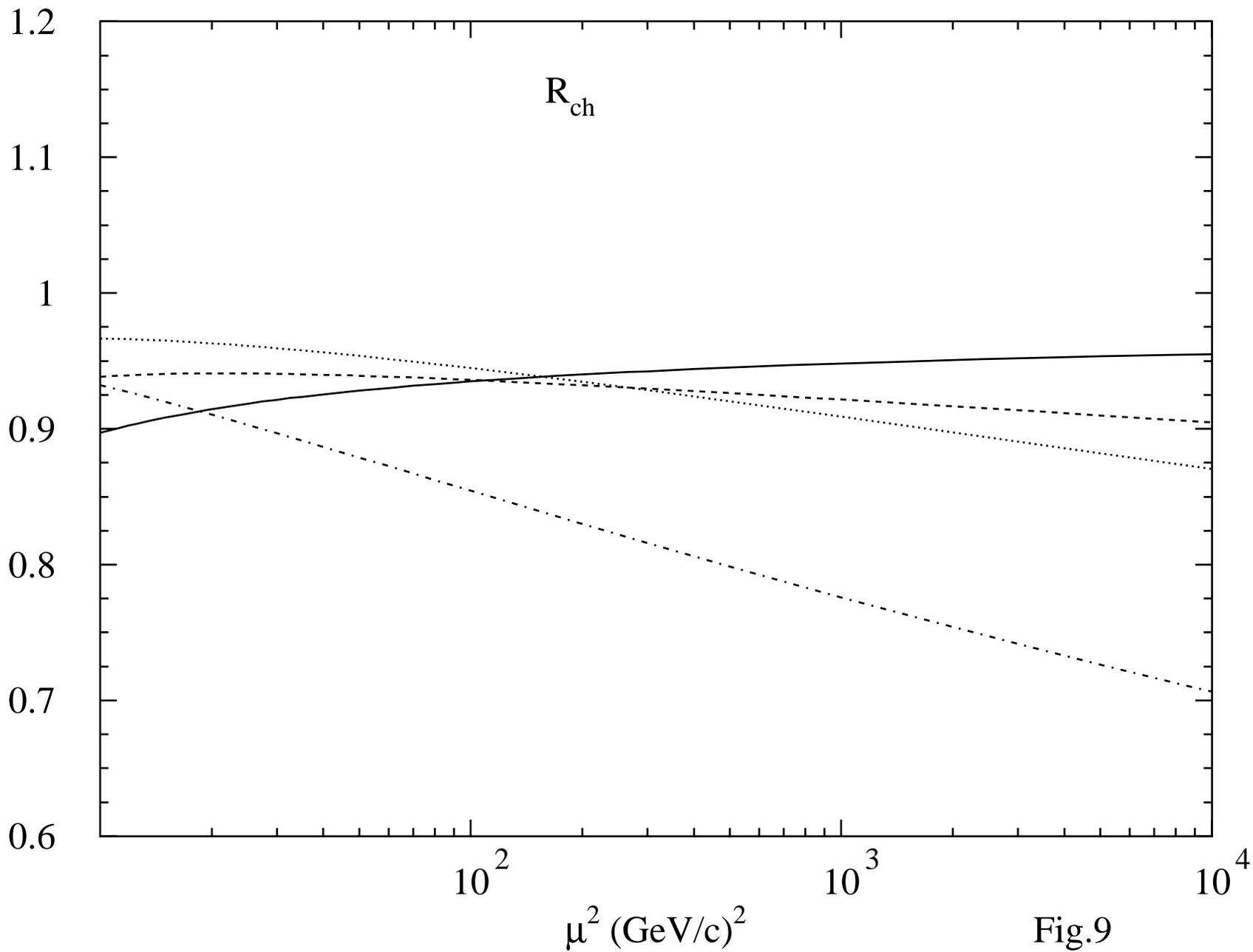
\item[Fig. 9.]
The ratio 
$R_{\rm ch}={f_{c+\bar c}^{\rm VFNS}}/{f_{c+\bar c}^{\rm PDF}}$ (3.5) 
as a function of $\mu^2$ at fixed $x$ in NLO. 
The charm densities $f_{c+\bar c}^{\rm VFNS}$ 
and $f_{c+\bar c}^{\rm PDF}$ are given by (3.4) and
the set GRV92HO \cite{grv92} respectively; 
$x = 10^{-1}$ (dashed-dotted line), $x = 10^{-2}$ (dotted line), 
$x=10^{-3}$ (dashed line), $x=10^{-4}$ (solid line).
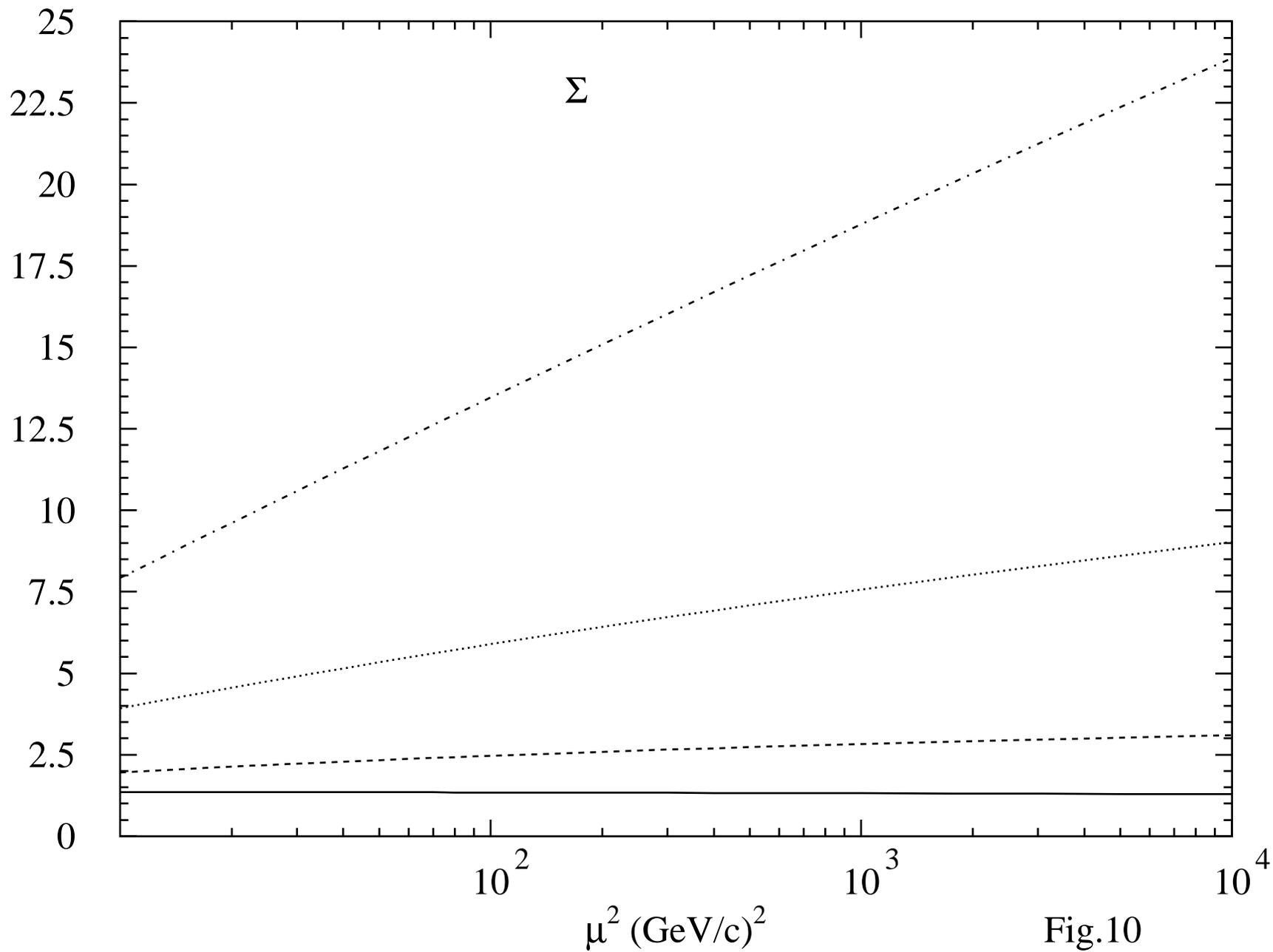
\item[Fig. 10.]
$\Sigma(3,x,\mu^2)$ and $\Sigma'(4,x,\mu^2)$ (the difference between both 
singlet combinations of quark densities is unnoticeable in the figure) 
plotted as functions of $\mu^2$ at
fixed $x$;
$x = 10^{-1}$ (solid line), $x = 10^{-2}$ (dashed line), 
$x = 10^{-3}$ (dotted line), $x = 10^{-4}$ (dashed-dotted line).
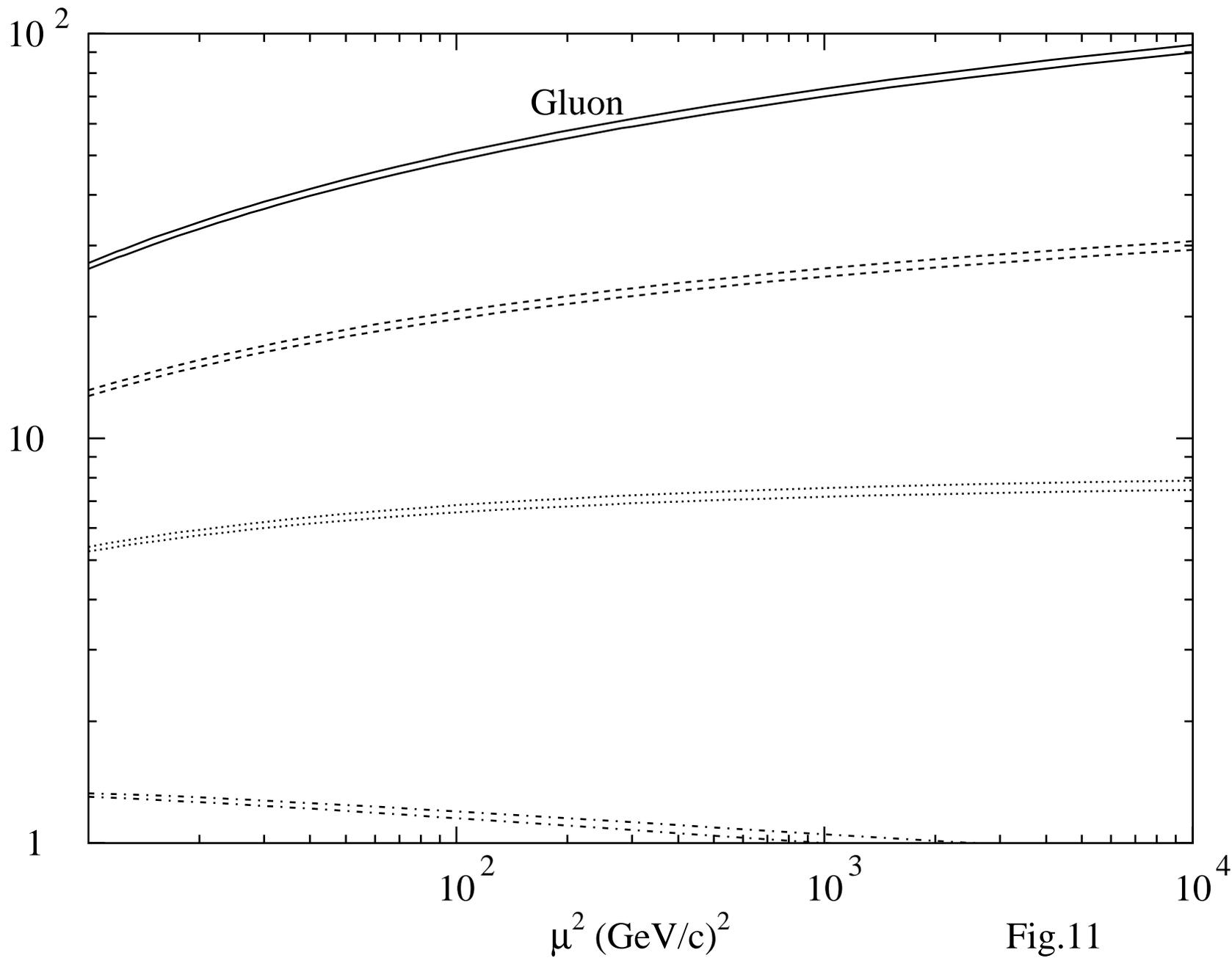
\item[Fig. 11.]
$G(3,x,\mu^2)$ (lower lines) and $G(4,x,\mu^2)$ (upper lines)
plotted as functions of $\mu^2$ at
fixed $x$;
$x = 10^{-1}$  (dashed-dotted line), $x = 10^{-2}$ (dotted line), 
$x = 10^{-3}$ (dashed line), $x = 10^{-4}$ (solid line).
\item[Fig. 12a.]
The charm component of the structure function given by $F_{2,c}$ in NLO
as a function of $x$ at $Q^2 =3 $ (GeV/$c)^2$;
$F_{2,c}^{\rm exact}(x,Q^2,m^2)$ (solid line),
$F_{2,c}^{\rm VFNS}(4,x,Q^2)$ (dashed line),
$F_{2,c}^{\rm PDF}(4,x,Q^2)$ (dotted line) with PDF=GRV92HO. 
\item[Fig. 12b.]
The charm component of the structure function given by $F_{2,c}$ in NLO
as a function of $x$ at $Q^2 =5 $ (GeV/$c)^2$;
$F_{2,c}^{\rm exact}(x,Q^2,m^2)$ (solid line),
$F_{2,c}^{\rm VFNS}(4,x,Q^2)$ (dashed line),
$F_{2,c}^{\rm PDF}(4,x,Q^2)$ (dotted line) with PDF=GRV92HO ,
and $F_{2,c}^{\rm asymp}(x,Q^2,m^2)$ (dashed-dotted line). 
\item[Fig. 12c.] Same as in Fig. 12b but now for $Q^2=10$ (GeV/$c)^2$.
\item[Fig. 12d.] Same as in Fig. 12b but now for $Q^2=50$ (GeV/$c)^2$.
\item[Fig. 12e.] Same as in Fig. 12b but now for $Q^2=100$ (GeV/$c)^2$.
\item[Fig. 13a.]
The ratios $K_{2,c}^{(l)}(x,Q^2)$ (eq.(3.9)) plotted as functions of $x$ at
$Q^2 =3 $ (GeV/$c)^2$;
$K_{2,c}^{(1),\rm exact}$ (solid line),
$K_{2,c}^{(1),\rm PDF}$ (dashed line),
$K_{2,c}^{(2),\rm PDF}$ (dotted line),
with PDF=GRV92HO.
\item[Fig. 13b.] Same as in Fig. 13a but now for $Q^2=5$ (GeV/$c)^2$.
\item[Fig. 13c.] Same as in Fig. 13a but now for $Q^2=10$ (GeV/$c)^2$.
\item[Fig. 13d.] Same as in Fig. 13a but now for $Q^2=50$ (GeV/$c)^2$.
\item[Fig. 13e.] Same as in Fig. 13a but now for $Q^2=100$ (GeV/$c)^2$.
\item[Fig. 13f.] Same as in Fig. 13a but now for $Q^2=10^3$ (GeV/$c)^2$.
\item[Fig. 13g.] Same as in Fig. 13a but now for $Q^2=10^4$ (GeV/$c)^2$.
\end{description}

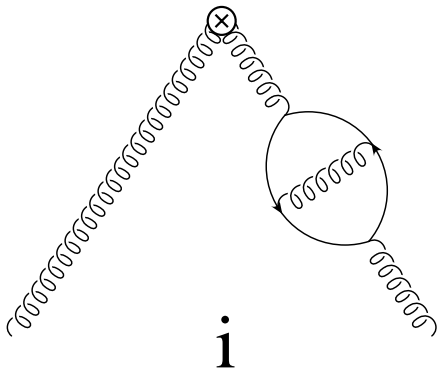
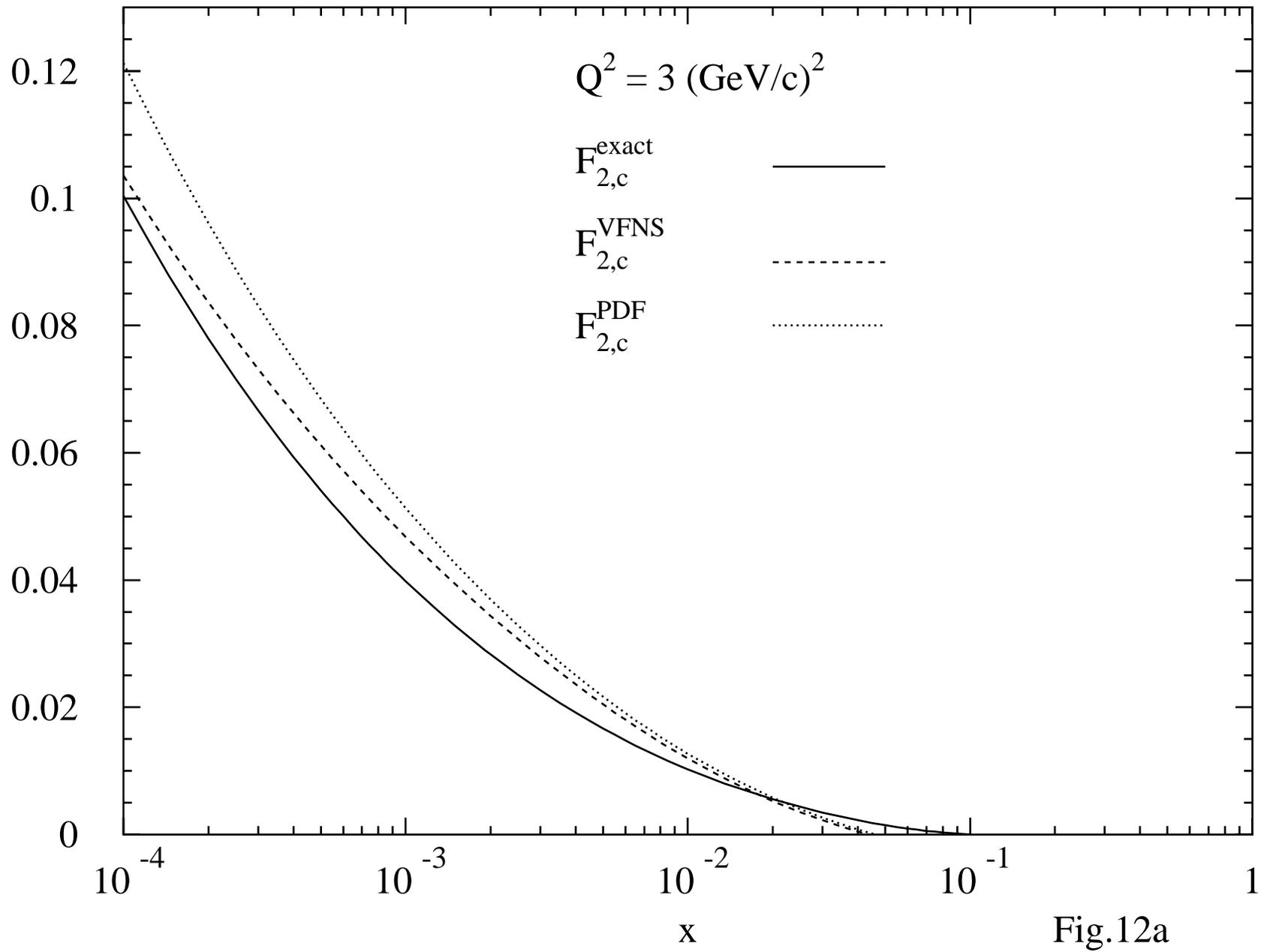
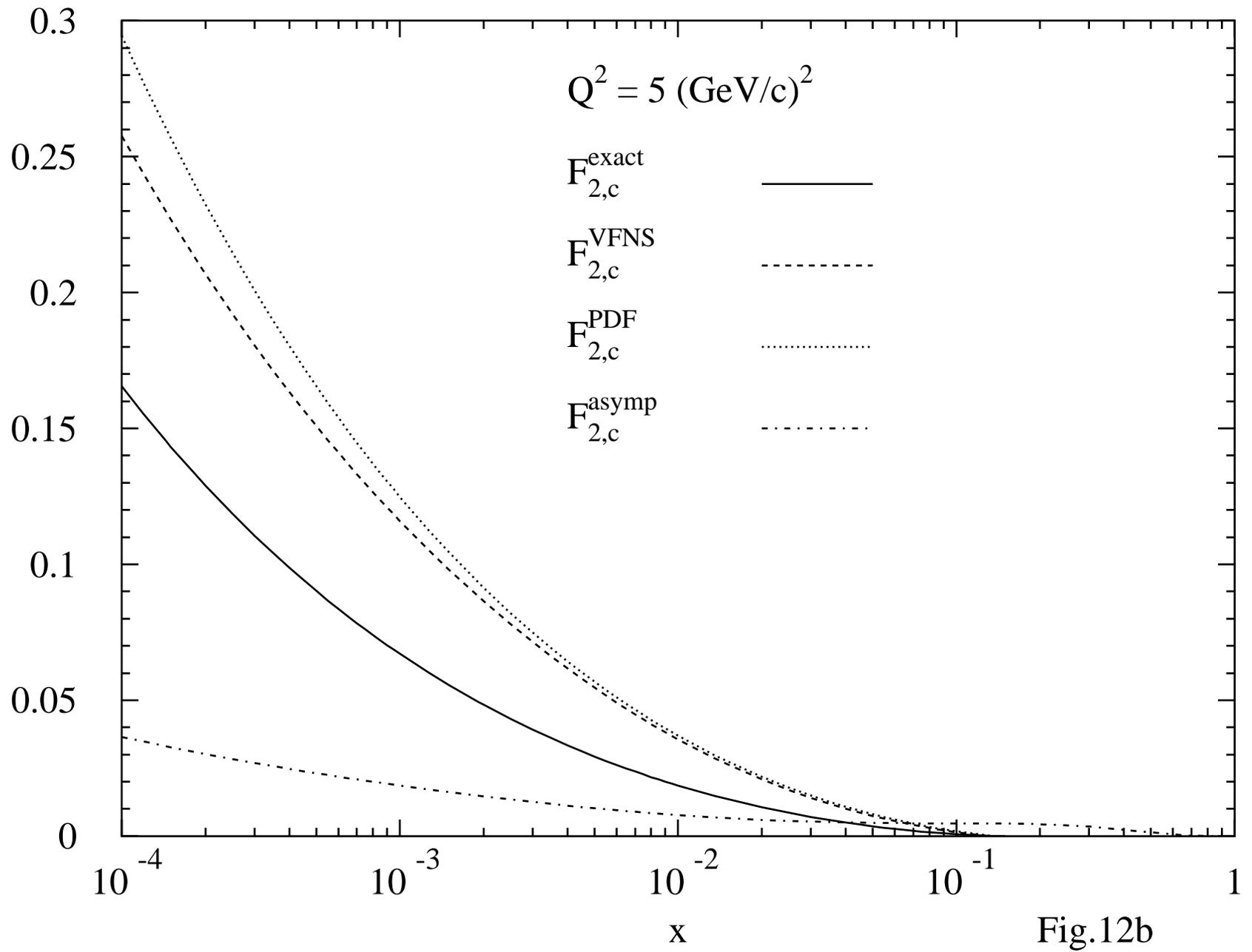
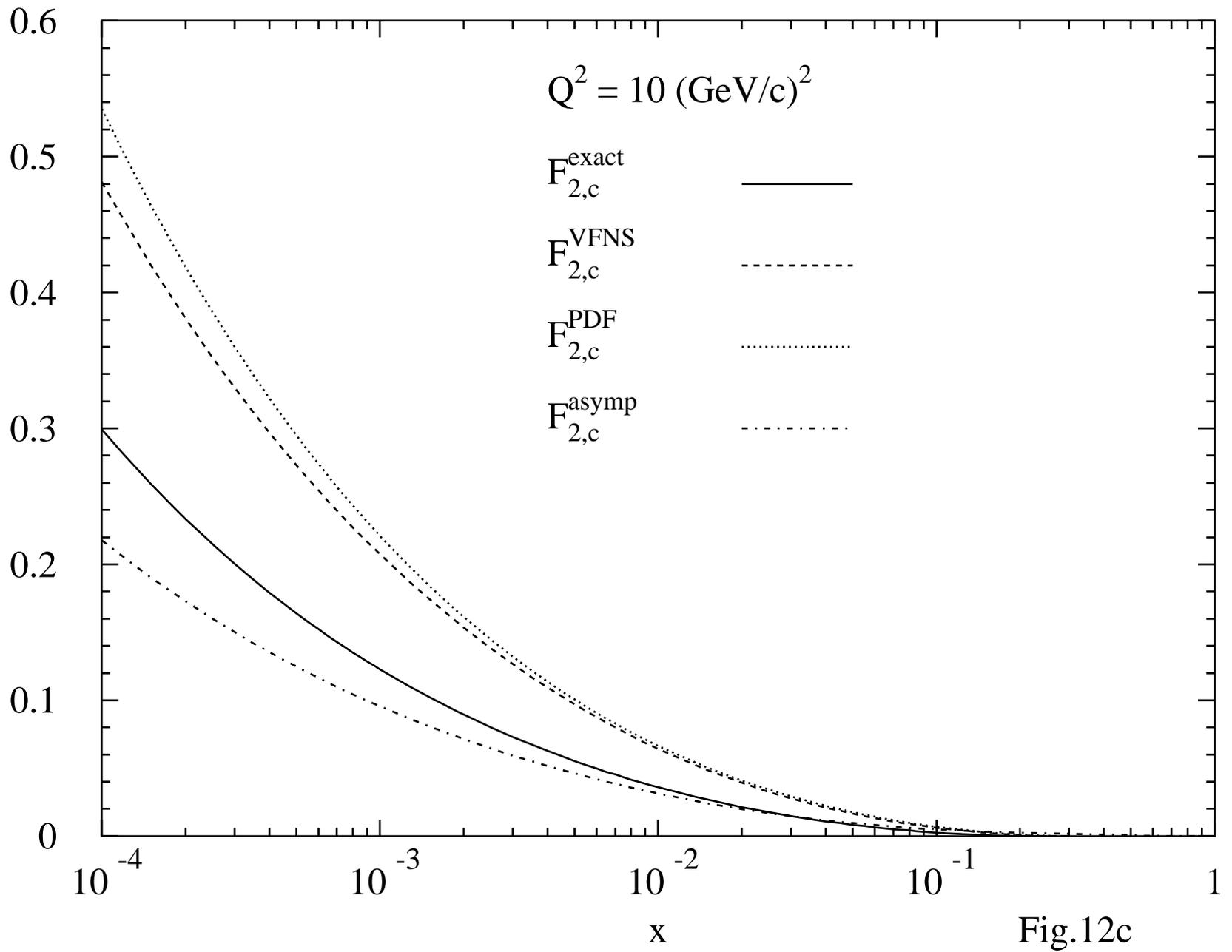
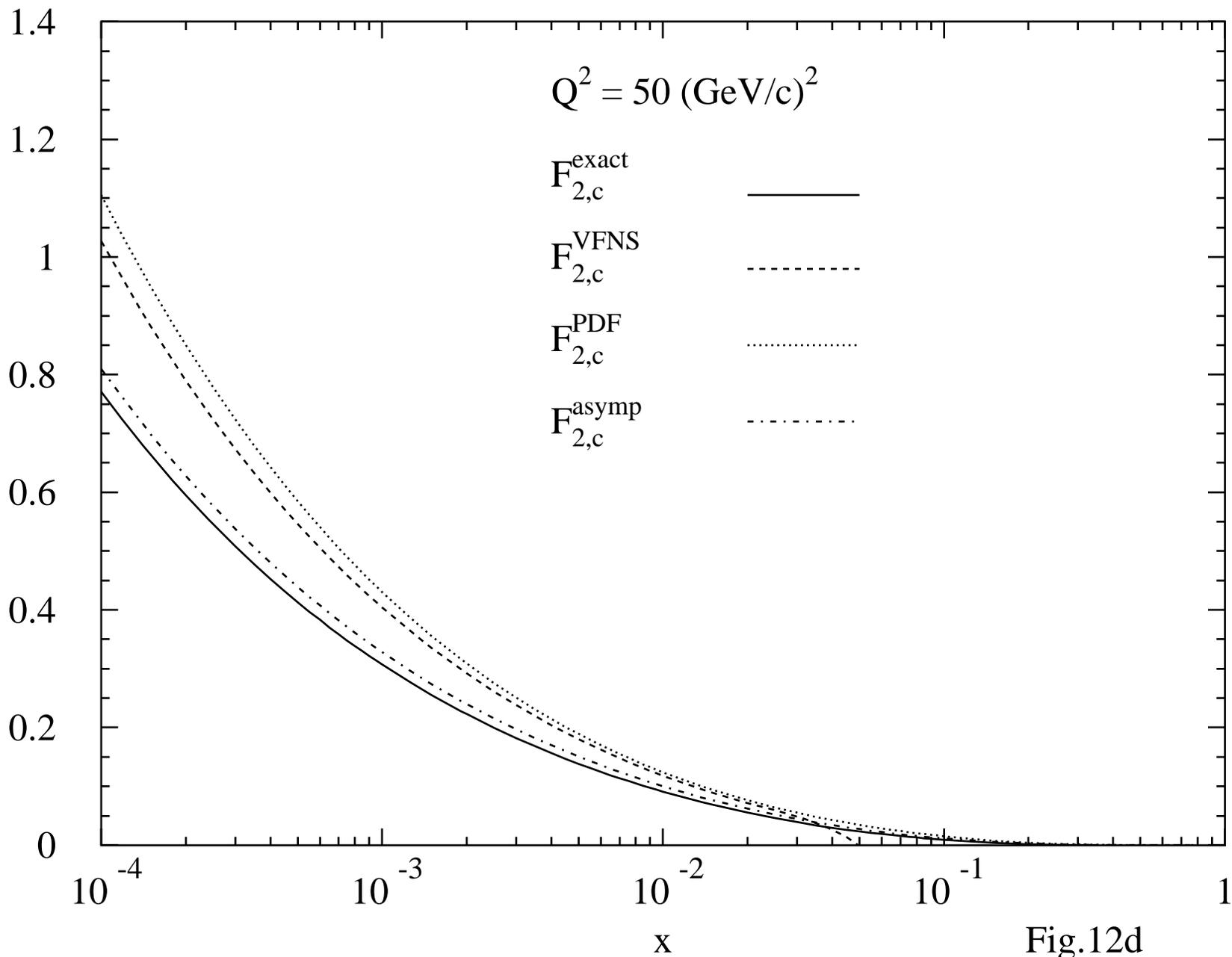
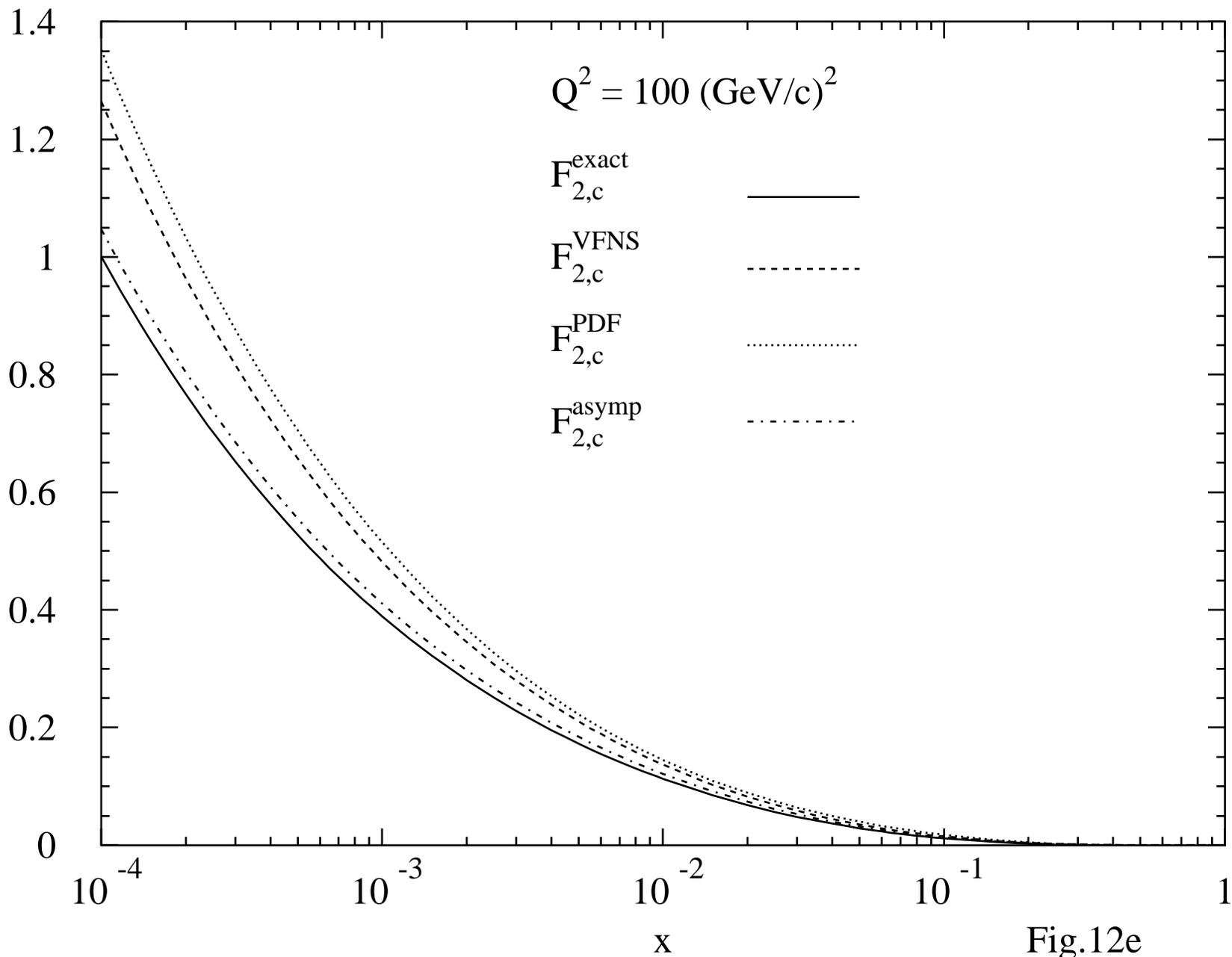
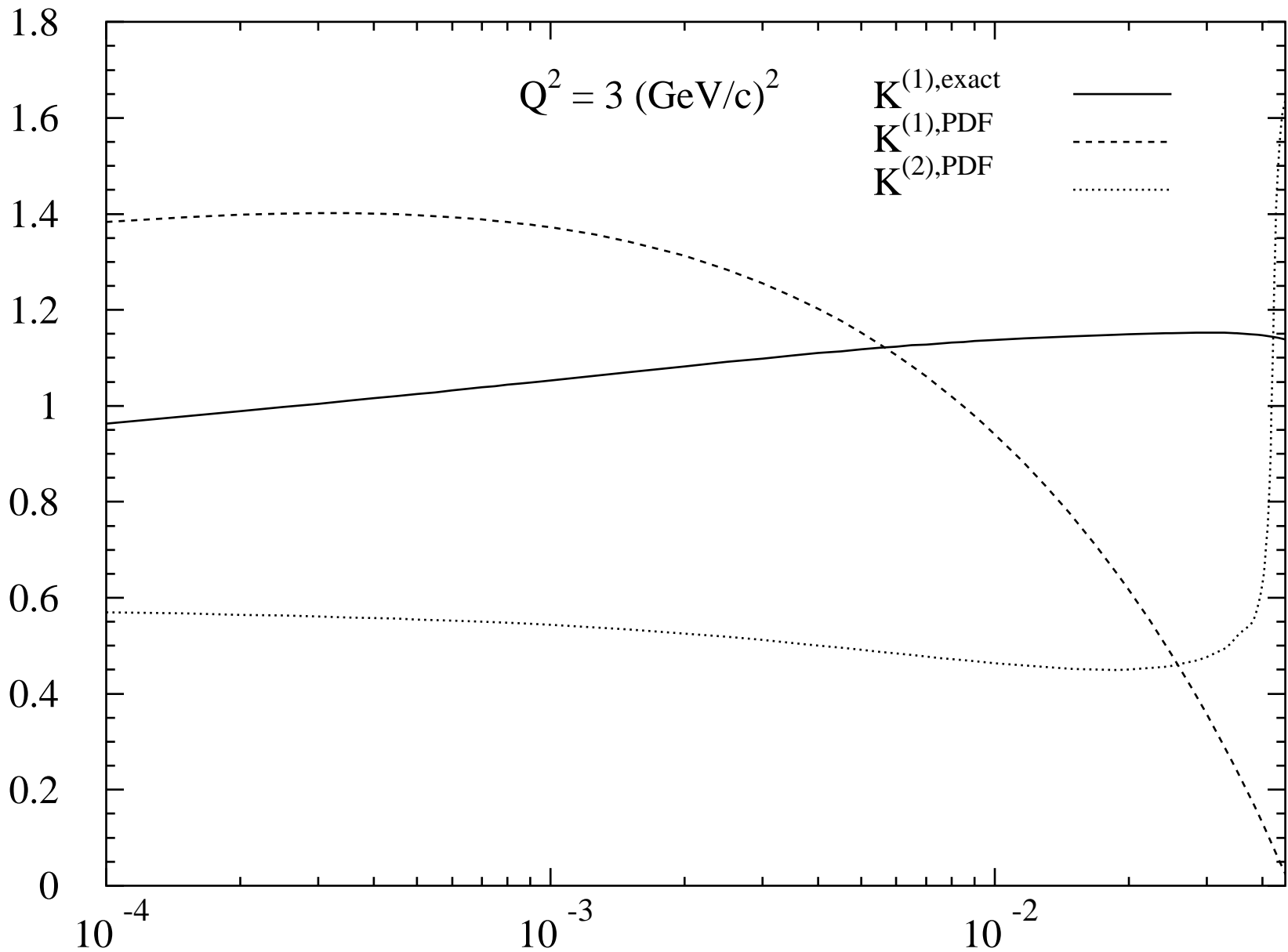
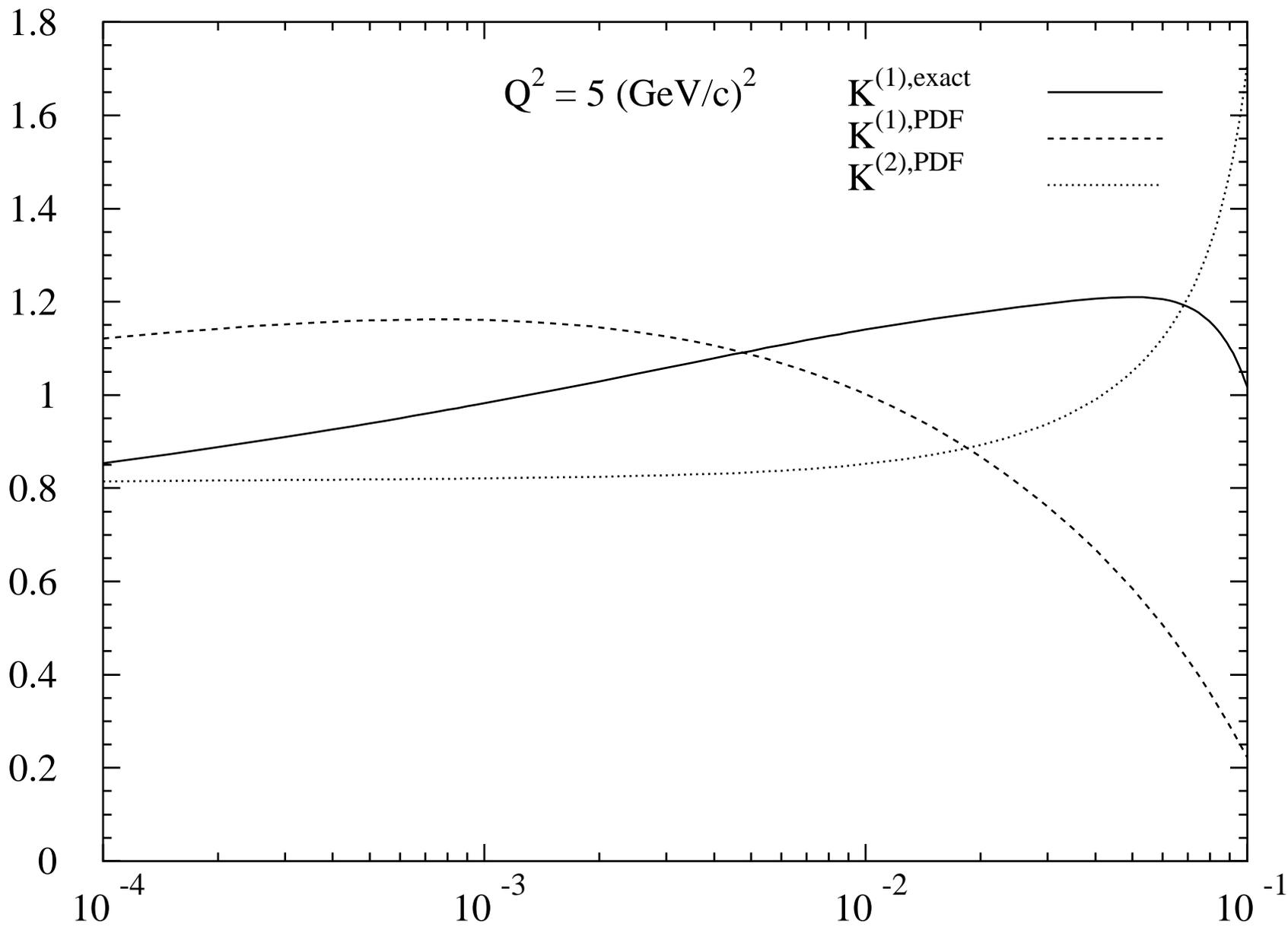
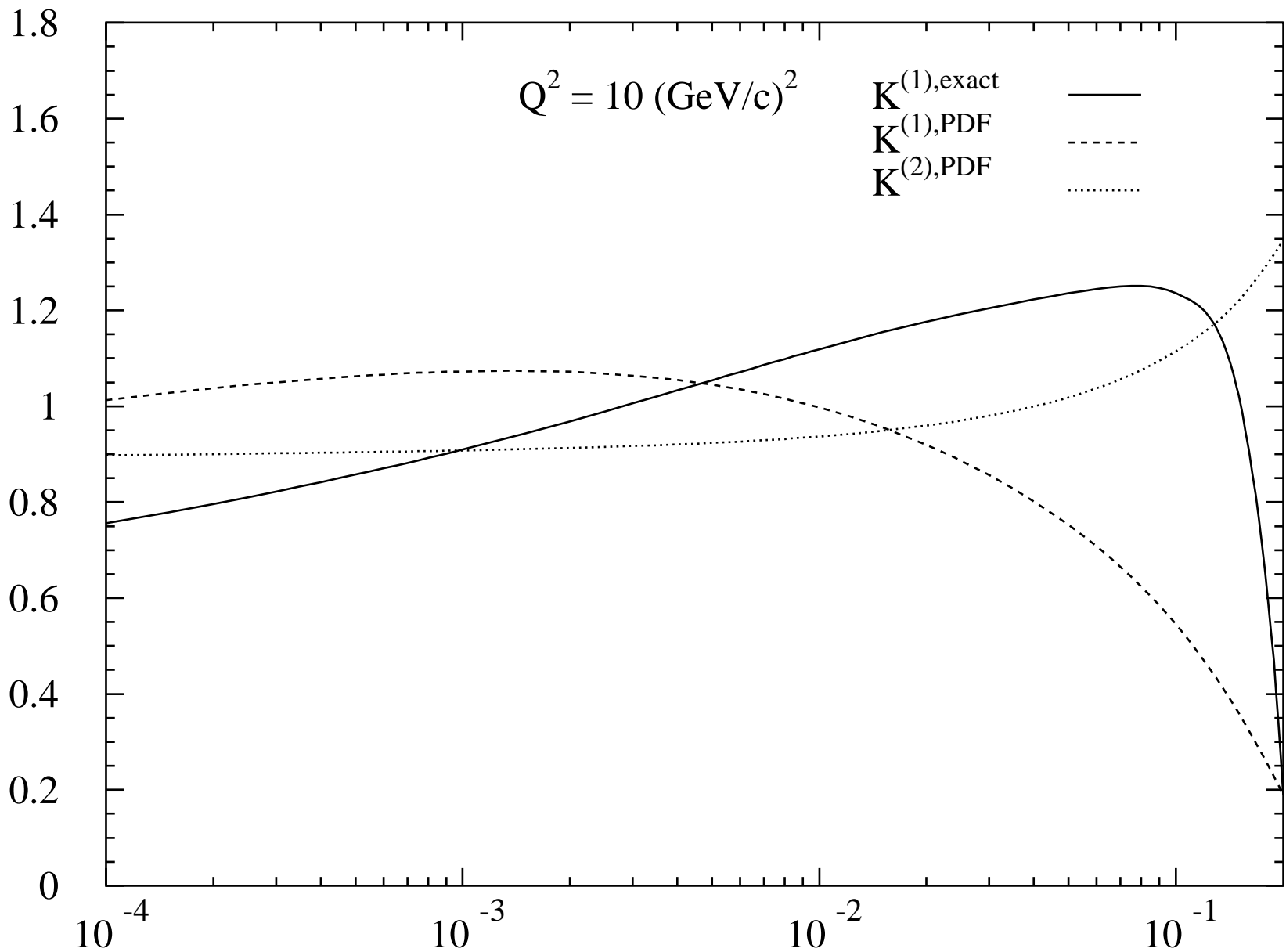
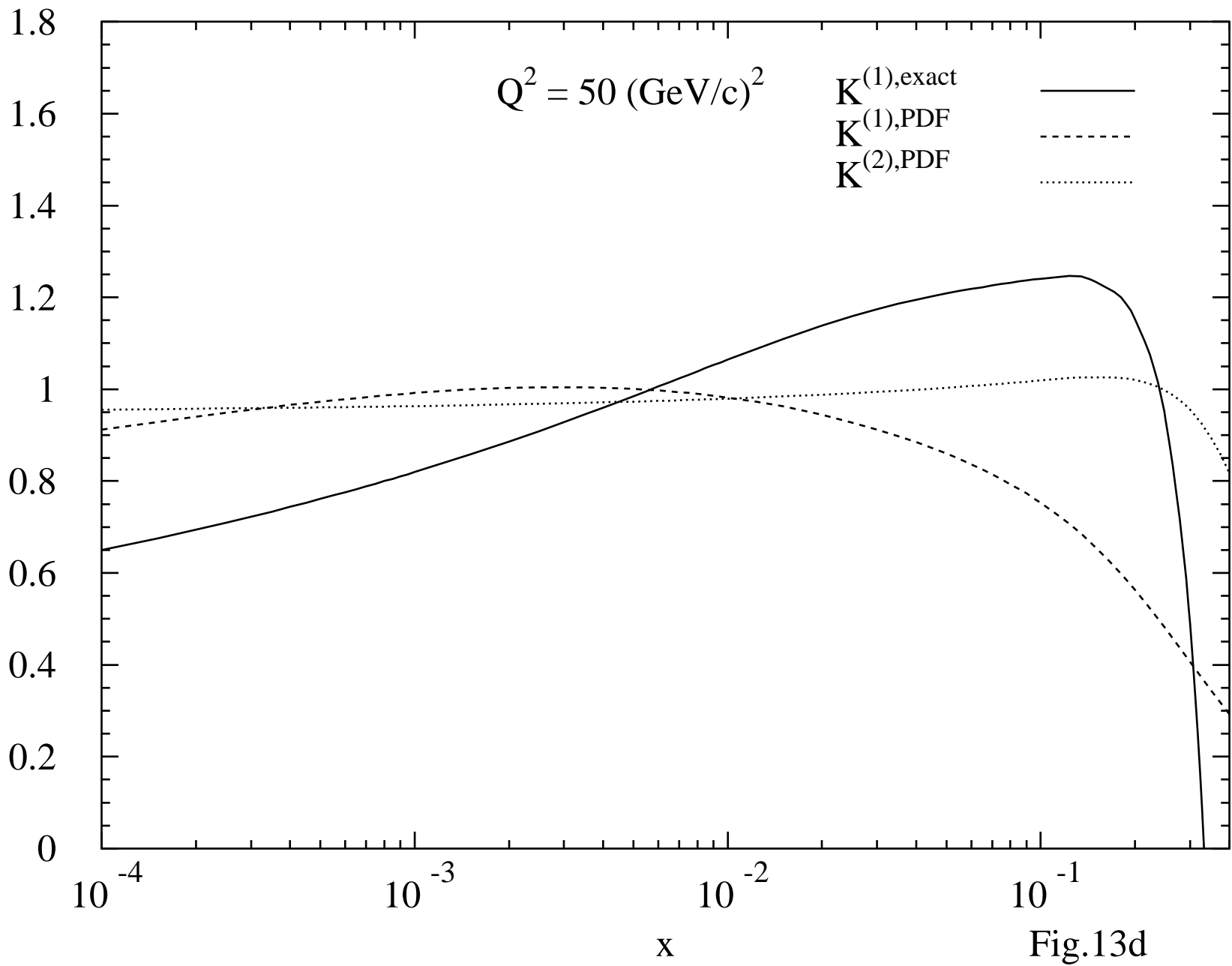
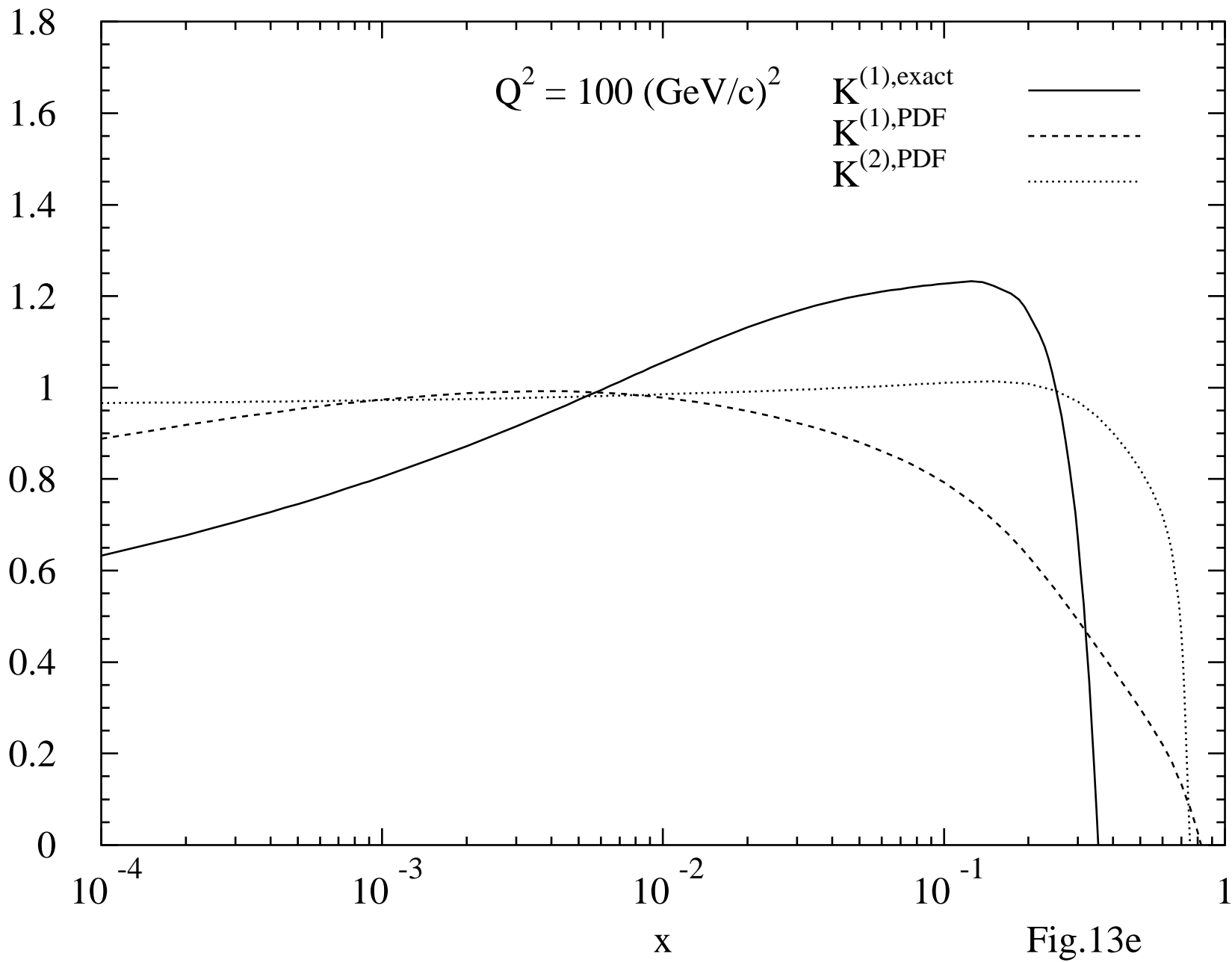
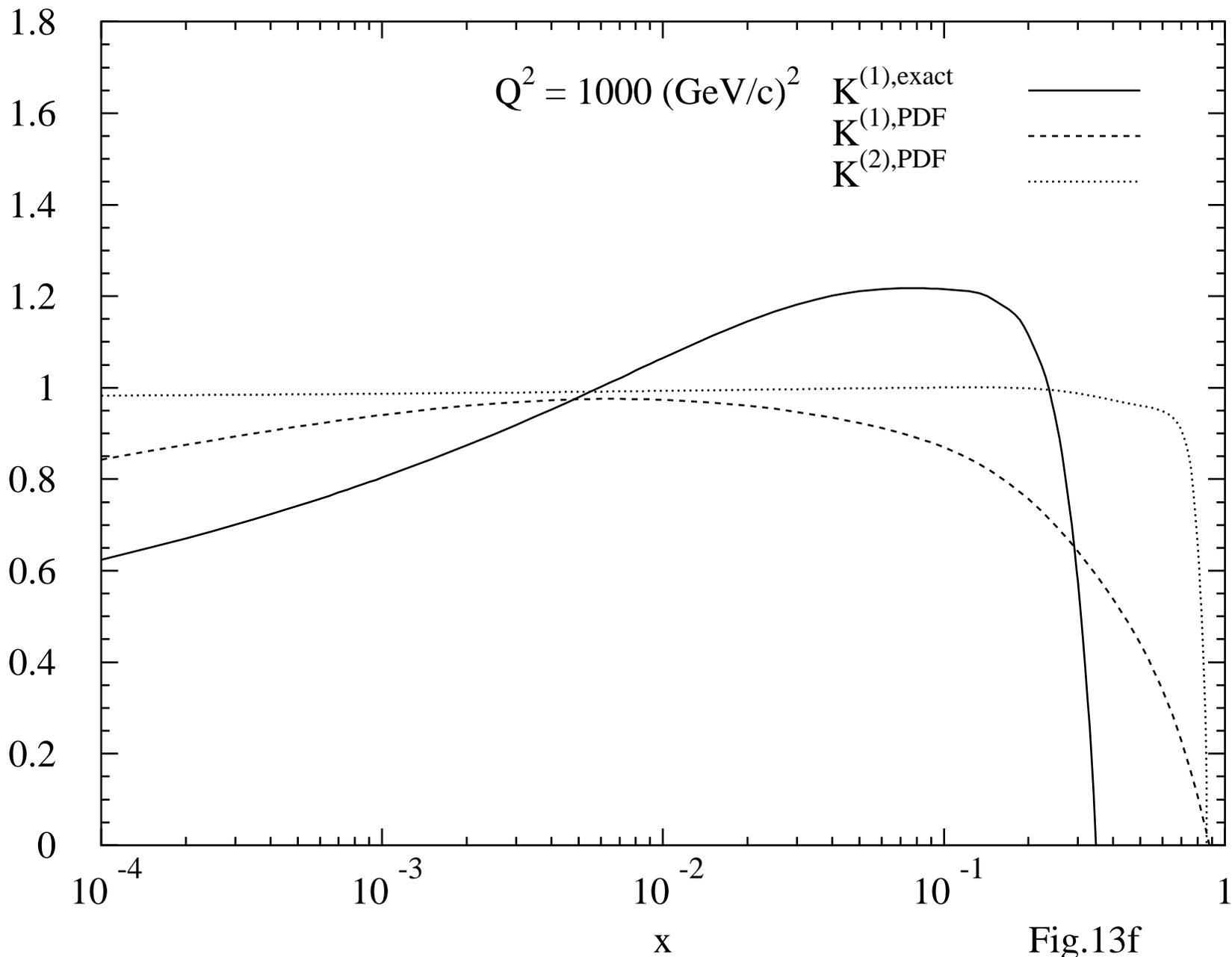
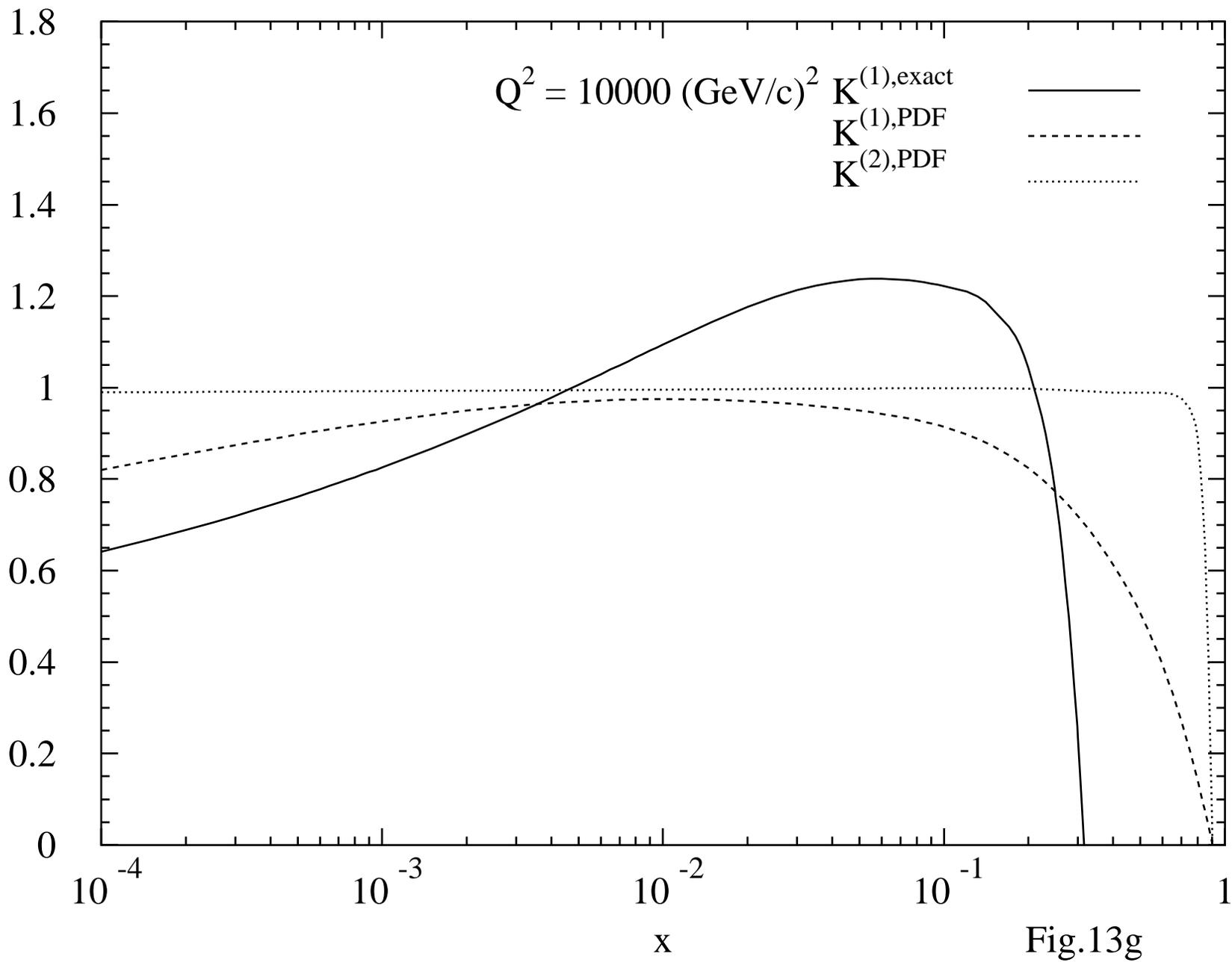
\end{document}